\documentclass[submission,copyright,creativecommons]{eptcs}
\providecommand{\event}{LFMTP 2025}

\usepackage{underscore}
\usepackage[T1]{fontenc}

\usepackage{graphicx}
\usepackage{caption}
\usepackage{amsmath}

\usepackage{etoolbox}

\usepackage{graphicx}
\usepackage{ragged2e}
\usepackage{tikz-cd}
\makeatletter
\@ifundefined{lhs2tex.lhs2tex.sty.read}%
  {\@namedef{lhs2tex.lhs2tex.sty.read}{}%
   \newcommand\SkipToFmtEnd{}%
   \newcommand\EndFmtInput{}%
   \long\def\SkipToFmtEnd#1\EndFmtInput{}%
  }\SkipToFmtEnd

\newcommand\ReadOnlyOnce[1]{\@ifundefined{#1}{\@namedef{#1}{}}\SkipToFmtEnd}
\usepackage{amstext}
\usepackage{amssymb}
\usepackage{stmaryrd}
\DeclareFontFamily{OT1}{cmtex}{}
\DeclareFontShape{OT1}{cmtex}{m}{n}
  {<5><6><7><8>cmtex8
   <9>cmtex9
   <10><10.95><12><14.4><17.28><20.74><24.88>cmtex10}{}
\DeclareFontShape{OT1}{cmtex}{m}{it}
  {<-> ssub * cmtt/m/it}{}

\DeclareFontShape{OT1}{cmtt}{bx}{n}
  {<5><6><7><8>cmtt8
   <9>cmbtt9
   <10><10.95><12><14.4><17.28><20.74><24.88>cmbtt10}{}
\DeclareFontShape{OT1}{cmtex}{bx}{n}
  {<-> ssub * cmtt/bx/n}{}

\newcommand{\Conid}[1]{\mathit{#1}}
\newcommand{\Varid}[1]{\mathit{#1}}
\newcommand{\anonymous}{\kern0.06em \vbox{\hrule\@width.5em}}

\renewcommand{\leq}{\leqslant}

\usepackage{polytable}

\@ifundefined{mathindent}%
  {\newdimen\mathindent\mathindent\leftmargini}%
  {}%

\def\resethooks{%
  \global\let\SaveRestoreHook\empty
  \global\let\ColumnHook\empty}
\newcommand*{\savecolumns}[1][default]%
  {\g@addto@macro\SaveRestoreHook{\savecolumns[#1]}}
\newcommand*{\restorecolumns}[1][default]%
  {\g@addto@macro\SaveRestoreHook{\restorecolumns[#1]}}
\newcommand*{\aligncolumn}[2]%
  {\g@addto@macro\ColumnHook{\column{#1}{#2}}}

\resethooks

\newcommand{\onelinecommentchars}{\quad-{}- }
\newcommand{\commentbeginchars}{\enskip\{-}
\newcommand{\commentendchars}{-\}\enskip}

\newcommand{\visiblecomments}{%
  \let\onelinecomment=\onelinecommentchars
  \let\commentbegin=\commentbeginchars
  \let\commentend=\commentendchars}

\newcommand{\invisiblecomments}{%
  \let\onelinecomment=\empty
  \let\commentbegin=\empty
  \let\commentend=\empty}

\visiblecomments

\newlength{\blanklineskip}
\newcommand{\hsindent}[1]{\quad}
\let\hspre\empty
\let\hspost\empty

\EndFmtInput
\makeatother
\ReadOnlyOnce{polycode.fmt}%
\makeatletter

\newcommand{\hsnewpar}[1]%
  {{\parskip=0pt\parindent=0pt\par\vskip #1\noindent}}

\newcommand{\hscodestyle}{}

\newcommand{\sethscode}[1]%
  {\expandafter\let\expandafter\hscode\csname #1\endcsname
   \expandafter\let\expandafter\endhscode\csname end#1\endcsname}

  {\par\noindent
   \advance\leftskip\mathindent
   \hscodestyle
   \let\\=\@normalcr
   \let\hspre\(\let\hspost\)%
   \pboxed}%
  {\endpboxed\)%
   \par\noindent
   \ignorespacesafterend}

  {\hsnewpar\abovedisplayskip
   \advance\leftskip\mathindent
   \hscodestyle
   \let\hspre\(\let\hspost\)%
   \pboxed}%
  {\endpboxed%
   \hsnewpar\belowdisplayskip
   \ignorespacesafterend}

  {\hsnewpar\abovedisplayskip
   \advance\leftskip\mathindent
   \hscodestyle
   \let\\=\@normalcr
   \(\pboxed}%
  {\endpboxed\)%
   \hsnewpar\belowdisplayskip
   \ignorespacesafterend}

\newcommand{\plainhs}{\sethscode{plainhscode}}

\plainhs

  {\hsnewpar\abovedisplayskip
   \advance\leftskip\mathindent
   \hscodestyle
   \let\\=\@normalcr
   \(\parray}%
  {\endparray\)%
   \hsnewpar\belowdisplayskip
   \ignorespacesafterend}

  {\parray}{\endparray}

  {\(\parray}{\endparray\)}

\def\codeframewidth{\arrayrulewidth}
\RequirePackage{calc}

  {\parskip=\abovedisplayskip\par\noindent
   \hscodestyle
   \arrayrulewidth=\codeframewidth
   \tabular{@{}|p{\linewidth-2\arraycolsep-2\arrayrulewidth-2pt}|@{}}%
   \hline\framedhslinecorrect\\{-1.5ex}%
   \let\endoflinesave=\\
   \let\\=\@normalcr
   \(\pboxed}%
  {\endpboxed\)%
   \framedhslinecorrect\endoflinesave{.5ex}\hline
   \endtabular
   \parskip=\belowdisplayskip\par\noindent
   \ignorespacesafterend}

\newcommand{\framedhslinecorrect}[2]%
  {#1[#2]}

  {\(\def\column##1##2{}%
   \let\>\undefined\let\<\undefined\let\\\undefined
   \newcommand\>[1][]{}\newcommand\<[1][]{}\newcommand\\[1][]{}%
   \def\fromto##1##2##3{##3}%
   }{\) }%

  {\let\orighscode=\hscode
   \let\origendhscode=\endhscode
   \def\endhscode{\def\hscode{\endgroup\def\@currenvir{hscode}\\}\begingroup}
   \orighscode\def\hscode{\endgroup\def\@currenvir{hscode}}}%
  {\origendhscode
   \global\let\hscode=\orighscode
   \global\let\endhscode=\origendhscode}%

\makeatother
\EndFmtInput
\ReadOnlyOnce{agda.fmt}%

\RequirePackage[T1]{fontenc}
\RequirePackage[utf8x]{inputenc}
\RequirePackage{ucs}
\RequirePackage{amsfonts}

\DeclareUnicodeCharacter{737}{\textsuperscript{l}}
\DeclareUnicodeCharacter{8718}{\ensuremath{\blacksquare}}
\DeclareUnicodeCharacter{8759}{::}
\DeclareUnicodeCharacter{9669}{\ensuremath{\triangleleft}}
\DeclareUnicodeCharacter{8799}{\ensuremath{\stackrel{\scriptscriptstyle ?}{=}}}
\DeclareUnicodeCharacter{10214}{\ensuremath{\llbracket}}
\DeclareUnicodeCharacter{10215}{\ensuremath{\rrbracket}}

\renewcommand\Varid[1]{\mathord{\textsf{#1}}}
\let\Conid\Varid
\newcommand\Keyword[1]{\textsf{\textbf{#1}}}
\EndFmtInput

\tikzcdset{scale cd/.style={every label/.append style={scale=#1},
    cells={nodes={scale=#1}}}}

\tikzcdset{scaleedge cd/.style={every label/.append style={scale=#1}}}
\tikzcdset{scalecell cd/.style={cells={nodes={scale=#1}}}}

\makeatletter
\newcommand*\bigcdot{\mathpalette\bigcdot@{3.0}}
\newcommand*\bigcdot@[2]{\mathbin{\vcenter{\hbox{\scalebox{#2}{$\m@th#1\cdot$}}}}}
\makeatother

\begin{document}

\def\titlerunning{Substitution Without Copy and Paste}
\def\authorrunning{T. Altenkirch, N. Burke \& P. Wadler}

\title{Substitution Without Copy and Paste}

\author{Thorsten Altenkirch
\institute{University of Nottingham\\Nottingham, UK}
\email{thorsten.altenkirch@nottingham.ac.uk}
\and
Nathaniel Burke
\institute{Imperial College London\\London, UK}
\email{nathanielrburke3@gmail.com}
\and 
Philip Wadler
\institute{University of Edinburgh and Input Output\\Edinburgh, UK}
\email{wadler@inf.ed.ac.uk}
}

%
%


 


\AtBeginEnvironment{hscode}{\setlength{\parskip}{0pt} \vspace{-1.5ex}}
\AtEndEnvironment{hscode}{\vspace{-1.5ex}}

\maketitle

\begin{abstract}
Defining substitution for a language with binders
like the simply typed $\lambda$-calculus requires repetition,
defining substitution and renaming separately. To verify the
categorical properties of this calculus, we must repeat the same
argument many times. We present a lightweight method
that avoids repetition and that 
gives rise to a simply typed category with families (CwF)
isomorphic to the initial simply typed CwF.
Our paper is a literate Agda script.
\end{abstract}

\section{Introduction}
\label{sec:introduction}

\begin{quote}
Some half dozen persons have written technically on combinatory logic,
and most of these, including ourselves, have published something
erroneous. \cite{curry1958combinatory}
\end{quote}

The first author was writing an introduction to
category theory for functional programmers. One example
was the category of simply-typed $\lambda$-terms and
substitutions, and proving the expected category laws
seemed a suitable exercise.
We attempted to mechanise the
solution in Agda \cite{agda}, and hit a setback: multiple proofs had to be
repeated multiple times. A guideline of
good software engineering is to \textbf{not write code
by copy and paste}, and this applies doubly to
formal proofs.

This paper is the result of our effort to refactor the proof.
The method used also applies to other problems; in
particular, we see the current construction as a warmup for the
definition of substitution for dependent type theory, which
may have interesting applications for interpreting dependent types in
higher categories (coherence).

\subsection{In a nutshell}
\label{sec:nutshell}

When working with substitution for a calculus with binders,
we have to differentiate between renamings (\ensuremath{\Delta\;\Vdash\!{\Varid{v}}\;\Gamma}),
where variables are substituted only for variables (\ensuremath{\Gamma\;\ni\;\Conid{A}}),
and proper substitutions (\ensuremath{\Delta\;\Vdash\;\Gamma}),
where variables are replaced with terms (\ensuremath{\Gamma\;\vdash\;\Conid{A}}).
This results in several similar operations:
\begin{center}
\begin{minipage}{0.5\textwidth}
\begin{hscode}\SaveRestoreHook
\column{B}{@{}>{\hspre}l<{\hspost}@{}}%
\column{3}{@{}>{\hspre}l<{\hspost}@{}}%
\column{11}{@{}>{\hspre}l<{\hspost}@{}}%
\column{20}{@{}>{\hspre}l<{\hspost}@{}}%
\column{30}{@{}>{\hspre}l<{\hspost}@{}}%
\column{E}{@{}>{\hspre}l<{\hspost}@{}}%
\>[3]{}\_\Varid{v}[\_]\Varid{v}\;{}\<[11]%
\>[11]{}\mathbin{:}\;\Gamma\;\ni\;\Conid{A}\;{}\<[20]%
\>[20]{}\rightarrow\;\Delta\;\Vdash\!{\Varid{v}}\;\Gamma\;{}\<[30]%
\>[30]{}\rightarrow\;\Delta\;\ni\;\Conid{A}{}\<[E]%
\\
\>[3]{}\_\Varid{v}[\_]\;{}\<[11]%
\>[11]{}\mathbin{:}\;\Gamma\;\ni\;\Conid{A}\;{}\<[20]%
\>[20]{}\rightarrow\;\Delta\;\Vdash\;\Gamma\;{}\<[30]%
\>[30]{}\rightarrow\;\Delta\;\vdash\;\Conid{A}{}\<[E]%
\ColumnHook
\end{hscode}\resethooks
\end{minipage}
\begin{minipage}{0.45\textwidth}
\begin{hscode}\SaveRestoreHook
\column{B}{@{}>{\hspre}l<{\hspost}@{}}%
\column{3}{@{}>{\hspre}l<{\hspost}@{}}%
\column{11}{@{}>{\hspre}l<{\hspost}@{}}%
\column{20}{@{}>{\hspre}l<{\hspost}@{}}%
\column{30}{@{}>{\hspre}l<{\hspost}@{}}%
\column{E}{@{}>{\hspre}l<{\hspost}@{}}%
\>[3]{}\_[\_]\Varid{v}\;{}\<[11]%
\>[11]{}\mathbin{:}\;\Gamma\;\vdash\;\Conid{A}\;{}\<[20]%
\>[20]{}\rightarrow\;\Delta\;\Vdash\!{\Varid{v}}\;\Gamma\;{}\<[30]%
\>[30]{}\rightarrow\;\Delta\;\vdash\;\Conid{A}{}\<[E]%
\\
\>[3]{}\_[\_]\;{}\<[11]%
\>[11]{}\mathbin{:}\;\Gamma\;\vdash\;\Conid{A}\;{}\<[20]%
\>[20]{}\rightarrow\;\Delta\;\Vdash\;\Gamma\;{}\<[30]%
\>[30]{}\rightarrow\;\Delta\;\vdash\;\Conid{A}{}\<[E]%
\ColumnHook
\end{hscode}\resethooks
\end{minipage}
\end{center}
\noindent The duplication gets worse when we prove properties
of substitution, such as the functor law
\begin{hscode}\SaveRestoreHook
\column{B}{@{}>{\hspre}l<{\hspost}@{}}%
\column{E}{@{}>{\hspre}l<{\hspost}@{}}%
\>[B]{}\Varid{x}\;[\mskip1.5mu \;\Varid{xs}\;\ensuremath{\mbox{$\circ$}}\;\Varid{ys}\;\mskip1.5mu]\;\equiv\;\Varid{x}\;[\mskip1.5mu \;\Varid{xs}\;\mskip1.5mu]\;[\mskip1.5mu \;\Varid{ys}\;\mskip1.5mu]{}\<[E]%
\ColumnHook
\end{hscode}\resethooks
All components \ensuremath{\Varid{x}}, \ensuremath{\Varid{xs}}, \ensuremath{\Varid{ys}} can be either variables/renamings
or terms/substitutions, so we must prove eight combinations.
The repetition extends to the intermediary lemmas. 

Our solution is to introduce a type of sorts with \ensuremath{\Conid{V}\;\mathbin{:}\;\Conid{Sort}} for
variables/renamings and \ensuremath{\Conid{T}\;\mathbin{:}\;\Conid{Sort}} for terms/substitutions, leading
to a single substitution operation 
\begin{hscode}\SaveRestoreHook
\column{B}{@{}>{\hspre}l<{\hspost}@{}}%
\column{E}{@{}>{\hspre}l<{\hspost}@{}}%
\>[B]{}\_[\_]\;\mathbin{:}\;\Gamma\;\vdash\![\;\Varid{q}\;\mskip1.5mu]\;\Conid{A}\;\rightarrow\;\Delta\;\Vdash\![\;\Varid{r}\;\mskip1.5mu]\;\Gamma\;\rightarrow\;\Delta\;\vdash\![\;\Varid{q}\;\sqcup\;\Varid{r}\;\mskip1.5mu]\;\Conid{A}{}\<[E]%
\ColumnHook
\end{hscode}\resethooks
where \ensuremath{\Varid{q,}\;\Varid{r}\;\mathbin{:}\;\Conid{Sort}} and \ensuremath{\Varid{q}\;\sqcup\;\Varid{r}} is the least upper bound in the
lattice of sorts with \ensuremath{\Conid{V}\;\sqsubseteq\;\Conid{T}}.
Now we need only prove one variant of the
functor law, relying on the fact that \ensuremath{\_\sqcup\_} is associative.
Our mutually recursive definitions are accepted by Agda,
as we can convince its termination checker that \ensuremath{\Conid{V}} is
structurally smaller than \ensuremath{\Conid{T}} (see Section~\ref{sec:fact-with-sorts}).

As a specification, we formulate an explicit substitution calculus as
a quotient-inductive type, or QIT (a mutual 
inductive type with equations). Here, substitution itself becomes a term former.
In our specification,
the substitution laws correspond to the equations of a simply-typed
category with families (CwF)---a variant of a CwF
where the types do not depend on a context.
Our recursive substitution operations lead to a simply typed CwF
isomorphic to the initial one, yielding
a normalisation result where $\lambda$-terms without explicit
substitutions are \emph{substitution normal forms}.

\subsection{Related work}
\label{sec:related-work}

De Bruijn introduced his eponymous indices and
simultaneous substitution in \cite{bruijn1972lambda}. We use typed
typed de Bruijn indices as in \cite{alti:csl99}.

In \cite{alti:csl99}, termination of substitution was shown using
well-founded recursion. Our approach is
simpler and scales better.
Andreas Abel used a similar technique to ours to mechanise \cite{alti:csl99},
without manual well-founded recursion,
in an unpublished Agda proof \cite{abel:subst11}.

The duplication between renaming and substitution operations is factored into 
\emph{kits} in \cite{mcbride2006type}. In \cite{allais2017type}, it was
further shown
how to extend this factoring to the proofs (by developing a 
``fusion framework'').
In languages supporting lexicographic recursion, 
our technique is simpler.

All the works listed so far also embrace the monadic perspective. 
That is, encoding 
substitutions as functions from variables to terms (indeed, this is
one of the motivations for relative monads \cite{altenkirch2015monads}). 
However, it is not clear how to extend this approach to dependently typed
languages without
``very dependent'' \cite{hickey1996formal, altenkirch2023munchhausen}
function types.


There have been a number of other publications on mechanising substitution.
Sch{\"{a}}fer and Stark~\emph{et al}~ 
\cite{schafer2015autosubst, stark2019autosubst} develop a Rocq library which 
automatically derives
substitution lemmas, but the proofs are repeated for renamings and
substitutions (as in
Section \ref{sec:naive-approach}). 
Their equational theory is also similar to 
the simply
typed CwFs in Section \ref{sec:initiality}.
Saffrich~\cite{saffrich2024abstractions} uses Agda with an \emph{extrinsic}
formulation (with preterms and typing separate), and applies
\cite{allais2017type} to factor the construction using kits.
In contrast, Saffrich~\cite{saffrich2024intrinsically} uses Agda with
an \emph{intrinsic} formulation (as here, indexing terms by types),
but defines renaming and substitution separately, and the relevant 
substitution lemmas are repeated for all required combinations.

%
%
%



\section{The naive approach}
\label{sec:naive-approach}

First, we review the copy-and-paste approach. 
We define types (\ensuremath{\Conid{A,}\;\Conid{B,}\;\Conid{C}}) and contexts (\ensuremath{\Gamma\Varid{,}\;\Delta\Varid{,}\;\Theta}):

\noindent
\begin{minipage}{0.45\textwidth}
\begin{hscode}\SaveRestoreHook
\column{B}{@{}>{\hspre}l<{\hspost}@{}}%
\column{3}{@{}>{\hspre}l<{\hspost}@{}}%
\column{8}{@{}>{\hspre}l<{\hspost}@{}}%
\column{E}{@{}>{\hspre}l<{\hspost}@{}}%
\>[B]{}\Keyword{data}\;\Conid{Ty}\;\mathbin{:}\;\Conid{Set}\;\Keyword{where}{}\<[E]%
\\
\>[B]{}\hsindent{3}{}\<[3]%
\>[3]{}\Varid{o}\;{}\<[8]%
\>[8]{}\mathbin{:}\;\Conid{Ty}{}\<[E]%
\\
\>[B]{}\hsindent{3}{}\<[3]%
\>[3]{}\_\Rightarrow\_\;{}\<[8]%
\>[8]{}\mathbin{:}\;\Conid{Ty}\;\rightarrow\;\Conid{Ty}\;\rightarrow\;\Conid{Ty}{}\<[E]%
\ColumnHook
\end{hscode}\resethooks
\end{minipage}
\begin{minipage}{0.5\textwidth}
\begin{hscode}\SaveRestoreHook
\column{B}{@{}>{\hspre}l<{\hspost}@{}}%
\column{3}{@{}>{\hspre}l<{\hspost}@{}}%
\column{8}{@{}>{\hspre}l<{\hspost}@{}}%
\column{E}{@{}>{\hspre}l<{\hspost}@{}}%
\>[B]{}\Keyword{data}\;\Conid{Con}\;\mathbin{:}\;\Conid{Set}\;\Keyword{where}{}\<[E]%
\\
\>[B]{}\hsindent{3}{}\<[3]%
\>[3]{}\Varid{•}\;{}\<[8]%
\>[8]{}\mathbin{:}\;\Conid{Con}{}\<[E]%
\\
\>[B]{}\hsindent{3}{}\<[3]%
\>[3]{}\_\rhd\_\;{}\<[8]%
\>[8]{}\mathbin{:}\;\Conid{Con}\;\rightarrow\;\Conid{Ty}\;\rightarrow\;\Conid{Con}{}\<[E]%
\ColumnHook
\end{hscode}\resethooks
\end{minipage}

\noindent
Next, we introduce intrinsically typed de Bruijn variables (\ensuremath{\Varid{i,}\;\Varid{j,}\;\Varid{k}}) and
$\lambda$-terms (\ensuremath{\Varid{t,}\;\Varid{u,}\;\Varid{v}}):

\noindent
\begin{minipage}{0.45\textwidth}
\begin{hscode}\SaveRestoreHook
\column{B}{@{}>{\hspre}l<{\hspost}@{}}%
\column{3}{@{}>{\hspre}l<{\hspost}@{}}%
\column{9}{@{}>{\hspre}l<{\hspost}@{}}%
\column{12}{@{}>{\hspre}l<{\hspost}@{}}%
\column{E}{@{}>{\hspre}l<{\hspost}@{}}%
\>[B]{}\Keyword{data}\;\_\ni\_\;\mathbin{:}\;\Conid{Con}\;\rightarrow\;\Conid{Ty}\;\rightarrow\;\Conid{Set}\;\Keyword{where}{}\<[E]%
\\
\>[B]{}\hsindent{3}{}\<[3]%
\>[3]{}\Varid{zero}\;{}\<[9]%
\>[9]{}\mathbin{:}\;{}\<[12]%
\>[12]{}\Gamma\;\rhd\;\Conid{A}\;\ni\;\Conid{A}{}\<[E]%
\\
\>[B]{}\hsindent{3}{}\<[3]%
\>[3]{}\Varid{suc}\;{}\<[9]%
\>[9]{}\mathbin{:}\;{}\<[12]%
\>[12]{}\Gamma\;\ni\;\Conid{A}\;\rightarrow\;(\Conid{B}\;\mathbin{:}\;\Conid{Ty})\;{}\<[E]%
\\
\>[9]{}\rightarrow\;{}\<[12]%
\>[12]{}\Gamma\;\rhd\;\Conid{B}\;\ni\;\Conid{A}{}\<[E]%
\ColumnHook
\end{hscode}\resethooks
\end{minipage}
\begin{minipage}{0.5\textwidth}
\begin{hscode}\SaveRestoreHook
\column{B}{@{}>{\hspre}l<{\hspost}@{}}%
\column{3}{@{}>{\hspre}l<{\hspost}@{}}%
\column{8}{@{}>{\hspre}l<{\hspost}@{}}%
\column{11}{@{}>{\hspre}l<{\hspost}@{}}%
\column{32}{@{}>{\hspre}l<{\hspost}@{}}%
\column{E}{@{}>{\hspre}l<{\hspost}@{}}%
\>[B]{}\Keyword{data}\;\_\vdash\_\;\mathbin{:}\;\Conid{Con}\;\rightarrow\;\Conid{Ty}\;\rightarrow\;\Conid{Set}\;\Keyword{where}{}\<[E]%
\\
\>[B]{}\hsindent{3}{}\<[3]%
\>[3]{}\texttt{\textasciigrave}\_\;{}\<[8]%
\>[8]{}\mathbin{:}\;{}\<[11]%
\>[11]{}\Gamma\;\ni\;\Conid{A}\;\rightarrow\;\Gamma\;\vdash\;\Conid{A}{}\<[E]%
\\
\>[B]{}\hsindent{3}{}\<[3]%
\>[3]{}\_\cdot\_\;{}\<[8]%
\>[8]{}\mathbin{:}\;{}\<[11]%
\>[11]{}\Gamma\;\vdash\;\Conid{A}\;\Rightarrow\;\Conid{B}\;\rightarrow\;\Gamma\;\vdash\;\Conid{A}\;\rightarrow\;{}\<[32]%
\>[32]{}\Gamma\;\vdash\;\Conid{B}{}\<[E]%
\\
\>[B]{}\hsindent{3}{}\<[3]%
\>[3]{}\lambda\_\;{}\<[8]%
\>[8]{}\mathbin{:}\;{}\<[11]%
\>[11]{}\Gamma\;\rhd\;\Conid{A}\;\vdash\;\Conid{B}\;\rightarrow\;\Gamma\;\vdash\;\Conid{A}\;\Rightarrow\;\Conid{B}{}\<[E]%
\ColumnHook
\end{hscode}\resethooks
\end{minipage}

\noindent
The constructor \ensuremath{\texttt{\textasciigrave}\_} embeds variables in
$\lambda$-terms and we write applications as \ensuremath{\Varid{t}\;\cdot\;\Varid{u}}.
Following de~Bruijn, lambda abstraction \ensuremath{\lambda\_} doesn't bind a name explicitly.
Instead, variables count the number of binders between them and their binding site. 
Substitutions (\ensuremath{\Varid{ts,}\;\Varid{us,}\;\Varid{vs}}) are sequences of terms:
\begin{hscode}\SaveRestoreHook
\column{B}{@{}>{\hspre}l<{\hspost}@{}}%
\column{3}{@{}>{\hspre}l<{\hspost}@{}}%
\column{8}{@{}>{\hspre}l<{\hspost}@{}}%
\column{E}{@{}>{\hspre}l<{\hspost}@{}}%
\>[B]{}\Keyword{data}\;\_\Vdash\_\;\mathbin{:}\;\Conid{Con}\;\rightarrow\;\Conid{Con}\;\rightarrow\;\Conid{Set}\;\Keyword{where}{}\<[E]%
\\
\>[B]{}\hsindent{3}{}\<[3]%
\>[3]{}\varepsilon\;{}\<[8]%
\>[8]{}\mathbin{:}\;\Gamma\;\Vdash\;\Varid{•}{}\<[E]%
\\
\>[B]{}\hsindent{3}{}\<[3]%
\>[3]{}\_\Varid{,}\_\;{}\<[8]%
\>[8]{}\mathbin{:}\;\Gamma\;\Vdash\;\Delta\;\rightarrow\;\Gamma\;\vdash\;\Conid{A}\;\rightarrow\;\Gamma\;\Vdash\;\Delta\;\rhd\;\Conid{A}{}\<[E]%
\ColumnHook
\end{hscode}\resethooks
Now we attempt to define the action of substitution for terms and variables:

\noindent
\begin{minipage}{0.45\textwidth}
\begin{hscode}\SaveRestoreHook
\column{B}{@{}>{\hspre}l<{\hspost}@{}}%
\column{12}{@{}>{\hspre}l<{\hspost}@{}}%
\column{E}{@{}>{\hspre}l<{\hspost}@{}}%
\>[B]{}\_\Varid{v}[\_]\;\mathbin{:}\;\Gamma\;\ni\;\Conid{A}\;\rightarrow\;\Delta\;\Vdash\;\Gamma\;\rightarrow\;\Delta\;\vdash\;\Conid{A}{}\<[E]%
\\
\>[B]{}\Varid{zero}\;{}\<[12]%
\>[12]{}\Varid{v}[\;\Varid{ts}\;\Varid{,}\;\Varid{t}\;\mskip1.5mu]\;\mathrel{=}\;\Varid{t}{}\<[E]%
\\
\>[B]{}(\Varid{suc}\;\Varid{i}\;\anonymous )\;{}\<[12]%
\>[12]{}\Varid{v}[\;\Varid{ts}\;\Varid{,}\;\Varid{t}\;\mskip1.5mu]\;\mathrel{=}\;\Varid{i}\;\Varid{v}[\;\Varid{ts}\;\mskip1.5mu]{}\<[E]%
\ColumnHook
\end{hscode}\resethooks
\end{minipage}
\begin{minipage}{0.5\textwidth}
\begin{hscode}\SaveRestoreHook
\column{B}{@{}>{\hspre}l<{\hspost}@{}}%
\column{10}{@{}>{\hspre}l<{\hspost}@{}}%
\column{E}{@{}>{\hspre}l<{\hspost}@{}}%
\>[B]{}\_[\_]\;\mathbin{:}\;\Gamma\;\vdash\;\Conid{A}\;\rightarrow\;\Delta\;\Vdash\;\Gamma\;\rightarrow\;\Delta\;\vdash\;\Conid{A}{}\<[E]%
\\
\>[B]{}(\texttt{\textasciigrave}\;\Varid{i})\;{}\<[10]%
\>[10]{}[\mskip1.5mu \;\Varid{ts}\;\mskip1.5mu]\;\mathrel{=}\;\Varid{i}\;\Varid{v}[\;\Varid{ts}\;\mskip1.5mu]{}\<[E]%
\\
\>[B]{}(\Varid{t}\;\cdot\;\Varid{u})\;{}\<[10]%
\>[10]{}[\mskip1.5mu \;\Varid{ts}\;\mskip1.5mu]\;\mathrel{=}\;(\Varid{t}\;[\mskip1.5mu \;\Varid{ts}\;\mskip1.5mu])\;\cdot\;(\Varid{u}\;[\mskip1.5mu \;\Varid{ts}\;\mskip1.5mu]){}\<[E]%
\\
\>[B]{}(\lambda\;\Varid{t})\;{}\<[10]%
\>[10]{}[\mskip1.5mu \;\Varid{ts}\;\mskip1.5mu]\;\mathrel{=}\;\lambda\;\Varid{?}{}\<[E]%
\ColumnHook
\end{hscode}\resethooks
\end{minipage}

\noindent
We encounter a problem with the case for binders \ensuremath{\lambda\_}. We are given a
substitution \ensuremath{\Varid{ts}\;\mathbin{:}\;\Delta\;\Vdash\;\Gamma} but the body lives in the extended context
\ensuremath{\Varid{t}\;\mathbin{:}\;\Gamma\;\Varid{,}\;\Conid{A}\;\vdash\;\Conid{B}}. We exploit functoriality of context extension (\ensuremath{\_\rhd\_}),
\ensuremath{\_\uparrow\_\;\mathbin{:}\;\Gamma\;\Vdash\;\Delta\;\rightarrow\;(\Conid{A}\;\mathbin{:}\;\Conid{Ty})\;\rightarrow\;\Gamma\;\rhd\;\Conid{A}\;\Vdash\;\Delta\;\rhd\;\Conid{A}}, and write 
\ensuremath{(\lambda\;\Varid{t})\;[\mskip1.5mu \;\Varid{ts}\;\mskip1.5mu]\;\mathrel{=}\;\lambda\;(\Varid{t}\;[\mskip1.5mu \;\Varid{ts}\;\uparrow\;\anonymous \;\mskip1.5mu])}.

Now, we must define \ensuremath{\_\uparrow\_}. This is easy (isn't it?), but we
need weakening of substitutions (\ensuremath{\_^{+}\_}):

\noindent
\begin{minipage}{0.45\textwidth}
\begin{hscode}\SaveRestoreHook
\column{B}{@{}>{\hspre}l<{\hspost}@{}}%
\column{E}{@{}>{\hspre}l<{\hspost}@{}}%
\>[B]{}\Varid{ts}\;\uparrow\;\Conid{A}\;\mathrel{=}\;\Varid{ts}\;^{+}\;\Conid{A}\;\Varid{,}\;\texttt{\textasciigrave}\;\Varid{zero}{}\<[E]%
\ColumnHook
\end{hscode}\resethooks
\end{minipage}
\begin{minipage}{0.5\textwidth}
\begin{hscode}\SaveRestoreHook
\column{B}{@{}>{\hspre}l<{\hspost}@{}}%
\column{E}{@{}>{\hspre}l<{\hspost}@{}}%
\>[B]{}\_^{+}\_\;\mathbin{:}\;\Gamma\;\Vdash\;\Delta\;\rightarrow\;(\Conid{A}\;\mathbin{:}\;\Conid{Ty})\;\rightarrow\;\Gamma\;\rhd\;\Conid{A}\;\Vdash\;\Delta{}\<[E]%
\ColumnHook
\end{hscode}\resethooks
\end{minipage}

\noindent
Which, in turn, is just a fold of term-weakening (\ensuremath{\Varid{suc-tm}}) over substitutions:

\noindent
\begin{minipage}{0.45\textwidth}
\begin{hscode}\SaveRestoreHook
\column{B}{@{}>{\hspre}l<{\hspost}@{}}%
\column{11}{@{}>{\hspre}l<{\hspost}@{}}%
\column{16}{@{}>{\hspre}l<{\hspost}@{}}%
\column{E}{@{}>{\hspre}l<{\hspost}@{}}%
\>[B]{}\varepsilon\;{}\<[11]%
\>[11]{}^{+}\;\Conid{A}\;{}\<[16]%
\>[16]{}\mathrel{=}\;\varepsilon{}\<[E]%
\\
\>[B]{}(\Varid{ts}\;\Varid{,}\;\Varid{t})\;{}\<[11]%
\>[11]{}^{+}\;\Conid{A}\;{}\<[16]%
\>[16]{}\mathrel{=}\;\Varid{ts}\;^{+}\;\Conid{A}\;\Varid{,}\;\Varid{suc-tm}\;\Varid{t}\;\Conid{A}{}\<[E]%
\ColumnHook
\end{hscode}\resethooks
\end{minipage}
\begin{minipage}{0.5\textwidth}
\begin{hscode}\SaveRestoreHook
\column{B}{@{}>{\hspre}l<{\hspost}@{}}%
\column{E}{@{}>{\hspre}l<{\hspost}@{}}%
\>[B]{}\Varid{suc-tm}\;\mathbin{:}\;\Gamma\;\vdash\;\Conid{B}\;\rightarrow\;(\Conid{A}\;\mathbin{:}\;\Conid{Ty})\;\rightarrow\;\Gamma\;\rhd\;\Conid{A}\;\vdash\;\Conid{B}{}\<[E]%
\ColumnHook
\end{hscode}\resethooks
\end{minipage}

\noindent
But how can we define \ensuremath{\Varid{suc-tm}} when we only have weakening for variables (\ensuremath{\Varid{vs}})? 
If we
already had identity \ensuremath{\Varid{id}\;\mathbin{:}\;\Gamma\;\Vdash\;\Gamma} and substitution we could write:
\ensuremath{\Varid{suc-tm}\;\Varid{t}\;\Conid{A}\;\mathrel{=}\;\Varid{t}\;[\mskip1.5mu \;\Varid{id}\;^{+}\;\Conid{A}\;\mskip1.5mu]}, 
but this is not structurally recursive (and is rejected
by Agda's termination checker).

To fix this, we use that \ensuremath{\Varid{id}} is a renaming, i.e.
a substitution only containing variables,
and defining \ensuremath{\_^{+}\!{\Varid{v}}\_} for renamings is easy.
This leads to a structurally recursive definition,
though with some repetition.
\vspace{-3ex}
\noindent
\begin{hscode}\SaveRestoreHook
\column{B}{@{}>{\hspre}l<{\hspost}@{}}%
\column{3}{@{}>{\hspre}l<{\hspost}@{}}%
\column{7}{@{}>{\hspre}l<{\hspost}@{}}%
\column{E}{@{}>{\hspre}l<{\hspost}@{}}%
\>[B]{}\Keyword{data}\;\_\Vdash\!{\Varid{v}}\_\;\mathbin{:}\;\Conid{Con}\;\rightarrow\;\Conid{Con}\;\rightarrow\;\Conid{Set}\;\Keyword{where}{}\<[E]%
\\
\>[B]{}\hsindent{3}{}\<[3]%
\>[3]{}\varepsilon\;{}\<[7]%
\>[7]{}\mathbin{:}\;\Gamma\;\Vdash\!{\Varid{v}}\;\Varid{•}{}\<[E]%
\\
\>[B]{}\hsindent{3}{}\<[3]%
\>[3]{}\_\Varid{,}\_\;\mathbin{:}\;\Gamma\;\Vdash\!{\Varid{v}}\;\Delta\;\rightarrow\;\Gamma\;\ni\;\Conid{A}\;\rightarrow\;\Gamma\;\Vdash\!{\Varid{v}}\;\Delta\;\rhd\;\Conid{A}{}\<[E]%
\ColumnHook
\end{hscode}\resethooks
\noindent
\begin{minipage}{0.5\textwidth}
\begin{hscode}\SaveRestoreHook
\column{B}{@{}>{\hspre}l<{\hspost}@{}}%
\column{11}{@{}>{\hspre}l<{\hspost}@{}}%
\column{12}{@{}>{\hspre}l<{\hspost}@{}}%
\column{26}{@{}>{\hspre}l<{\hspost}@{}}%
\column{E}{@{}>{\hspre}l<{\hspost}@{}}%
\>[B]{}\_\Varid{v}[\_]\Varid{v}\;\mathbin{:}\;\Gamma\;\ni\;\Conid{A}\;\rightarrow\;\Delta\;\Vdash\!{\Varid{v}}\;\Gamma\;\rightarrow\;\Delta\;\ni\;\Conid{A}{}\<[E]%
\\
\>[B]{}\Varid{zero}\;{}\<[12]%
\>[12]{}\Varid{v}[\;\Varid{is}\;\Varid{,}\;\Varid{i}\;]\Varid{v}\;{}\<[26]%
\>[26]{}\mathrel{=}\;\Varid{i}{}\<[E]%
\\
\>[B]{}(\Varid{suc}\;\Varid{i}\;\anonymous )\;{}\<[12]%
\>[12]{}\Varid{v}[\;\Varid{is}\;\Varid{,}\;\Varid{j}\;]\Varid{v}\;{}\<[26]%
\>[26]{}\mathrel{=}\;\Varid{i}\;\Varid{v}[\;\Varid{is}\;]\Varid{v}{}\<[E]%
\\[\blanklineskip]%
\>[B]{}\_^{+}\!{\Varid{v}}\_\;\mathbin{:}\;\Gamma\;\Vdash\!{\Varid{v}}\;\Delta\;\rightarrow\;\forall{}\;\Conid{A}\;\rightarrow\;\Gamma\;\rhd\;\Conid{A}\;\Vdash\!{\Varid{v}}\;\Delta{}\<[E]%
\\
\>[B]{}\varepsilon\;{}\<[11]%
\>[11]{}^{+}\!{\Varid{v}}\;\Conid{A}\;{}\<[26]%
\>[26]{}\mathrel{=}\;\varepsilon{}\<[E]%
\\
\>[B]{}(\Varid{is}\;\Varid{,}\;\Varid{i})\;{}\<[11]%
\>[11]{}^{+}\!{\Varid{v}}\;\Conid{A}\;{}\<[26]%
\>[26]{}\mathrel{=}\;\Varid{is}\;^{+}\!{\Varid{v}}\;\Conid{A}\;\Varid{,}\;\Varid{suc}\;\Varid{i}\;\Conid{A}{}\<[E]%
\\[\blanklineskip]%
\>[B]{}\_\uparrow\!{\Varid{v}}\_\;\mathbin{:}\;\Gamma\;\Vdash\!{\Varid{v}}\;\Delta\;\rightarrow\;\forall{}\;\Conid{A}\;\rightarrow\;\Gamma\;\rhd\;\Conid{A}\;\Vdash\!{\Varid{v}}\;\Delta\;\rhd\;\Conid{A}{}\<[E]%
\\
\>[B]{}\Varid{is}\;\uparrow\!{\Varid{v}}\;\Conid{A}\;{}\<[26]%
\>[26]{}\mathrel{=}\;\Varid{is}\;^{+}\!{\Varid{v}}\;\Conid{A}\;\Varid{,}\;\Varid{zero}{}\<[E]%
\ColumnHook
\end{hscode}\resethooks
\end{minipage}
\hfill
\begin{minipage}{0.45\textwidth}
\begin{hscode}\SaveRestoreHook
\column{B}{@{}>{\hspre}l<{\hspost}@{}}%
\column{10}{@{}>{\hspre}l<{\hspost}@{}}%
\column{18}{@{}>{\hspre}l<{\hspost}@{}}%
\column{19}{@{}>{\hspre}l<{\hspost}@{}}%
\column{22}{@{}>{\hspre}l<{\hspost}@{}}%
\column{E}{@{}>{\hspre}l<{\hspost}@{}}%
\>[B]{}\_[\_]\Varid{v}\;\mathbin{:}\;\Gamma\;\vdash\;\Conid{A}\;\rightarrow\;\Delta\;\Vdash\!{\Varid{v}}\;\Gamma\;\rightarrow\;\Delta\;\vdash\;\Conid{A}{}\<[E]%
\\
\>[B]{}(\texttt{\textasciigrave}\;\Varid{i})\;{}\<[10]%
\>[10]{}[\mskip1.5mu \;\Varid{is}\;]\Varid{v}\;{}\<[19]%
\>[19]{}\mathrel{=}\;{}\<[22]%
\>[22]{}\texttt{\textasciigrave}\;(\Varid{i}\;\Varid{v}[\;\Varid{is}\;]\Varid{v}){}\<[E]%
\\
\>[B]{}(\Varid{t}\;\cdot\;\Varid{u})\;{}\<[10]%
\>[10]{}[\mskip1.5mu \;\Varid{is}\;]\Varid{v}\;{}\<[19]%
\>[19]{}\mathrel{=}\;{}\<[22]%
\>[22]{}(\Varid{t}\;[\mskip1.5mu \;\Varid{is}\;]\Varid{v})\;\cdot\;(\Varid{u}\;[\mskip1.5mu \;\Varid{is}\;]\Varid{v}){}\<[E]%
\\
\>[B]{}(\lambda\;\Varid{t})\;{}\<[10]%
\>[10]{}[\mskip1.5mu \;\Varid{is}\;]\Varid{v}\;{}\<[19]%
\>[19]{}\mathrel{=}\;{}\<[22]%
\>[22]{}\lambda\;(\Varid{t}\;[\mskip1.5mu \;\Varid{is}\;\uparrow\!{\Varid{v}}\;\anonymous \;]\Varid{v}){}\<[E]%
\\[\blanklineskip]%
\>[B]{}\Varid{idv}\;\mathbin{:}\;\Gamma\;\Vdash\!{\Varid{v}}\;\Gamma{}\<[E]%
\\
\>[B]{}\Varid{idv}\;\{\mskip1.5mu \Gamma\;\mathrel{=}\;\Varid{•}\mskip1.5mu\}\;{}\<[18]%
\>[18]{}\mathrel{=}\;\varepsilon{}\<[E]%
\\
\>[B]{}\Varid{idv}\;\{\mskip1.5mu \Gamma\;\mathrel{=}\;\Gamma\;\rhd\;\Conid{A}\mskip1.5mu\}\;{}\<[18]%
\>[18]{}\mathrel{=}\;\Varid{idv}\;\uparrow\!{\Varid{v}}\;\Conid{A}{}\<[E]%
\\[\blanklineskip]%
\>[B]{}\Varid{suc-tm}\;\Varid{t}\;\Conid{A}\;{}\<[18]%
\>[18]{}\mathrel{=}\;\Varid{t}\;[\mskip1.5mu \;\Varid{idv}\;^{+}\!{\Varid{v}}\;\Conid{A}\;]\Varid{v}{}\<[E]%
\ColumnHook
\end{hscode}\resethooks
\end{minipage}

\noindent
This may not seem too bad, but it gets worse when proving the laws.
Let \ensuremath{\_\ensuremath{\mbox{$\circ$}}\_} be composition of substitutions.
To prove associativity, we first need functoriality,
\ensuremath{[\circ]\;\mathbin{:}\;\Varid{t}\;[\mskip1.5mu \;\Varid{us}\;\ensuremath{\mbox{$\circ$}}\;\Varid{vs}\;\mskip1.5mu]\;\equiv\;\Varid{t}\;[\mskip1.5mu \;\Varid{us}\;\mskip1.5mu]\;[\mskip1.5mu \;\Varid{vs}\;\mskip1.5mu]} but to prove this, we also need
to cover all variations where \ensuremath{\Varid{t,}\;\Varid{us,}\;\Varid{vs}} are variables/renamings rather than 
terms/substitutions. This leads to eight combinations, with the cases for
each constructor of \ensuremath{\Varid{t}} reading near-identically.
This repetition is emblematic of many prior attempts at mechanising
substitution
\cite{adams2004formal, benton2012strongly, schafer2015autosubst, 
stark2019autosubst, saffrich2024intrinsically}.

The rest of the paper shows how to factor these definitions
and proofs, using only (lexicographic) structural recursion.

\section{Factorising with sorts}
\label{sec:fact-with-sorts}

Our main idea is to turn the distinction between
variables and terms into a parameter. The first approximation of this idea is to
define a type \ensuremath{\Conid{Sort}} (\ensuremath{\Varid{q,}\;\Varid{r,}\;\Varid{s}}):
\begin{hscode}\SaveRestoreHook
\column{B}{@{}>{\hspre}l<{\hspost}@{}}%
\column{4}{@{}>{\hspre}l<{\hspost}@{}}%
\column{E}{@{}>{\hspre}l<{\hspost}@{}}%
\>[B]{}\Keyword{data}\;\Conid{Sort}\;\mathbin{:}\;\Conid{Set}\;\Keyword{where}{}\<[E]%
\\
\>[B]{}\hsindent{4}{}\<[4]%
\>[4]{}\Conid{V}\;\Conid{T}\;\mathbin{:}\;\Conid{Sort}{}\<[E]%
\ColumnHook
\end{hscode}\resethooks
But this is not quite what we want.
Agda's termination checker uses structural orderings.
Following our intuition that variable weakening is trivial but
term weakening requires renaming, we would like the sort
of variables \ensuremath{\Conid{V}} to be structurally smaller than the sort of terms \ensuremath{\Conid{T}}.

With the following definition, we
make \ensuremath{\Conid{V}} structurally smaller than \ensuremath{\Conid{T>V}\;\Conid{V}\;\Varid{isV}},
while maintaining that \ensuremath{\Conid{Sort}} has only two elements.

\noindent
\begin{minipage}{0.55\textwidth}
\begin{hscode}\SaveRestoreHook
\column{B}{@{}>{\hspre}l<{\hspost}@{}}%
\column{3}{@{}>{\hspre}l<{\hspost}@{}}%
\column{5}{@{}>{\hspre}l<{\hspost}@{}}%
\column{10}{@{}>{\hspre}l<{\hspost}@{}}%
\column{E}{@{}>{\hspre}l<{\hspost}@{}}%
\>[3]{}\Keyword{data}\;\Conid{Sort}\;\mathbin{:}\;\Conid{Set}\;\Keyword{where}{}\<[E]%
\\
\>[3]{}\hsindent{2}{}\<[5]%
\>[5]{}\Conid{V}\;{}\<[10]%
\>[10]{}\mathbin{:}\;\Conid{Sort}{}\<[E]%
\\
\>[3]{}\hsindent{2}{}\<[5]%
\>[5]{}\Conid{T>V}\;{}\<[10]%
\>[10]{}\mathbin{:}\;(\Varid{s}\;\mathbin{:}\;\Conid{Sort})\;\rightarrow\;\Conid{IsV}\;\Varid{s}\;\rightarrow\;\Conid{Sort}{}\<[E]%
\ColumnHook
\end{hscode}\resethooks
\end{minipage}
\begin{minipage}{0.4\textwidth}
\begin{hscode}\SaveRestoreHook
\column{B}{@{}>{\hspre}l<{\hspost}@{}}%
\column{3}{@{}>{\hspre}l<{\hspost}@{}}%
\column{5}{@{}>{\hspre}l<{\hspost}@{}}%
\column{E}{@{}>{\hspre}l<{\hspost}@{}}%
\>[3]{}\Keyword{data}\;\Conid{IsV}\;\mathbin{:}\;\Conid{Sort}\;\rightarrow\;\Conid{Set}\;\Keyword{where}{}\<[E]%
\\
\>[3]{}\hsindent{2}{}\<[5]%
\>[5]{}\Varid{isV}\;\mathbin{:}\;\Conid{IsV}\;\Conid{V}{}\<[E]%
\ColumnHook
\end{hscode}\resethooks
\end{minipage}

\noindent
The predicate \ensuremath{\Varid{isV}} only holds for \ensuremath{\Conid{V}}. This encoding makes
use of Agda's support for inductive-inductive datatypes (IITs), but a
pair of a natural number \ensuremath{\Varid{n}} and a proof \ensuremath{\Varid{n}\;\leq\;\Varid{1}} would also work, i.e.
\ensuremath{\Conid{Sort}\;\mathrel{=}\;\Sigma\;\mathbb{N}\;(\_\leq\;\Varid{1})}.
We make \ensuremath{\Conid{T}\;\mathbin{:}\;\Conid{Sort}} an abbreviation for \ensuremath{\Conid{T>V}\;\Conid{V}\;\Varid{isV}} with a pattern declaration.
\begin{hscode}\SaveRestoreHook
\column{B}{@{}>{\hspre}l<{\hspost}@{}}%
\column{E}{@{}>{\hspre}l<{\hspost}@{}}%
\>[B]{}\Varid{pattern}\;\Conid{T}\;\mathrel{=}\;\Conid{T>V}\;\Conid{V}\;\Varid{isV}{}\<[E]%
\ColumnHook
\end{hscode}\resethooks
Now we can pattern match over \ensuremath{\Conid{Sort}} with \ensuremath{\Conid{V}} and \ensuremath{\Conid{T}},
while Agda's termination checker treats \ensuremath{\Conid{V}} as structurally 
smaller than \ensuremath{\Conid{T}}.

It is now possible to define terms and variables (\ensuremath{\Varid{x,}\;\Varid{y,}\;\Varid{z}}) in one go:
\begin{hscode}\SaveRestoreHook
\column{B}{@{}>{\hspre}l<{\hspost}@{}}%
\column{3}{@{}>{\hspre}l<{\hspost}@{}}%
\column{9}{@{}>{\hspre}l<{\hspost}@{}}%
\column{14}{@{}>{\hspre}l<{\hspost}@{}}%
\column{22}{@{}>{\hspre}l<{\hspost}@{}}%
\column{29}{@{}>{\hspre}l<{\hspost}@{}}%
\column{37}{@{}>{\hspre}l<{\hspost}@{}}%
\column{44}{@{}>{\hspre}l<{\hspost}@{}}%
\column{52}{@{}>{\hspre}l<{\hspost}@{}}%
\column{E}{@{}>{\hspre}l<{\hspost}@{}}%
\>[B]{}\Keyword{data}\;\_\vdash\![\_]\_\;\mathbin{:}\;\Conid{Con}\;\rightarrow\;\Conid{Sort}\;\rightarrow\;\Conid{Ty}\;\rightarrow\;\Conid{Set}\;\Keyword{where}{}\<[E]%
\\
\>[B]{}\hsindent{3}{}\<[3]%
\>[3]{}\Varid{zero}\;{}\<[9]%
\>[9]{}\mathbin{:}\;\Gamma\;\rhd\;\Conid{A}\;\vdash\![\;\Conid{V}\;\mskip1.5mu]\;\Conid{A}{}\<[E]%
\\
\>[B]{}\hsindent{3}{}\<[3]%
\>[3]{}\Varid{suc}\;{}\<[9]%
\>[9]{}\mathbin{:}\;\Gamma\;{}\<[14]%
\>[14]{}\vdash\![\;\Conid{V}\;\mskip1.5mu]\;{}\<[22]%
\>[22]{}\Conid{A}\;\rightarrow\;(\Conid{B}\;\mathbin{:}\;\Conid{Ty})\;\rightarrow\;\Gamma\;\rhd\;\Conid{B}\;{}\<[44]%
\>[44]{}\vdash\![\;\Conid{V}\;\mskip1.5mu]\;{}\<[52]%
\>[52]{}\Conid{A}{}\<[E]%
\\
\>[B]{}\hsindent{3}{}\<[3]%
\>[3]{}\texttt{\textasciigrave}\_\;{}\<[9]%
\>[9]{}\mathbin{:}\;\Gamma\;{}\<[14]%
\>[14]{}\vdash\![\;\Conid{V}\;\mskip1.5mu]\;{}\<[22]%
\>[22]{}\Conid{A}\;\rightarrow\;\Gamma\;{}\<[29]%
\>[29]{}\vdash\![\;\Conid{T}\;\mskip1.5mu]\;{}\<[37]%
\>[37]{}\Conid{A}{}\<[E]%
\\
\>[B]{}\hsindent{3}{}\<[3]%
\>[3]{}\_\cdot\_\;{}\<[9]%
\>[9]{}\mathbin{:}\;\Gamma\;\vdash\![\;\Conid{T}\;\mskip1.5mu]\;\Conid{A}\;\Rightarrow\;\Conid{B}\;\rightarrow\;\Gamma\;\vdash\![\;\Conid{T}\;\mskip1.5mu]\;\Conid{A}\;\rightarrow\;\Gamma\;\vdash\![\;\Conid{T}\;\mskip1.5mu]\;\Conid{B}{}\<[E]%
\\
\>[B]{}\hsindent{3}{}\<[3]%
\>[3]{}\lambda\_\;{}\<[9]%
\>[9]{}\mathbin{:}\;\Gamma\;\rhd\;\Conid{A}\;\vdash\![\;\Conid{T}\;\mskip1.5mu]\;\Conid{B}\;\rightarrow\;\Gamma\;\vdash\![\;\Conid{T}\;\mskip1.5mu]\;\Conid{A}\;\Rightarrow\;\Conid{B}{}\<[E]%
\ColumnHook
\end{hscode}\resethooks
This recapitulates our previous definitions 
(in Section \ref{sec:naive-approach}), where
\ensuremath{\Gamma\;\vdash\![\;\Conid{V}\;\mskip1.5mu]\;\Conid{A}} corresponds to \ensuremath{\Gamma\;\ni\;\Conid{A}} and \ensuremath{\Gamma\;\vdash\![\;\Conid{T}\;\mskip1.5mu]\;\Conid{A}} to \ensuremath{\Gamma\;\vdash\;\Conid{A}}.
Now we can parametrise our previous development.
As a first step, we generalise renamings and substitutions (\ensuremath{\Varid{xs,}\;\Varid{ys,}\;\Varid{zs}}):
\begin{hscode}\SaveRestoreHook
\column{B}{@{}>{\hspre}l<{\hspost}@{}}%
\column{3}{@{}>{\hspre}l<{\hspost}@{}}%
\column{8}{@{}>{\hspre}l<{\hspost}@{}}%
\column{E}{@{}>{\hspre}l<{\hspost}@{}}%
\>[B]{}\Keyword{data}\;\_\Vdash\![\_]\_\;\mathbin{:}\;\Conid{Con}\;\rightarrow\;\Conid{Sort}\;\rightarrow\;\Conid{Con}\;\rightarrow\;\Conid{Set}\;\Keyword{where}{}\<[E]%
\\
\>[B]{}\hsindent{3}{}\<[3]%
\>[3]{}\varepsilon\;{}\<[8]%
\>[8]{}\mathbin{:}\;\Gamma\;\Vdash\![\;\Varid{q}\;\mskip1.5mu]\;\Varid{•}{}\<[E]%
\\
\>[B]{}\hsindent{3}{}\<[3]%
\>[3]{}\_\Varid{,}\_\;{}\<[8]%
\>[8]{}\mathbin{:}\;\Gamma\;\Vdash\![\;\Varid{q}\;\mskip1.5mu]\;\Delta\;\rightarrow\;\Gamma\;\vdash\![\;\Varid{q}\;\mskip1.5mu]\;\Conid{A}\;\rightarrow\;\Gamma\;\Vdash\![\;\Varid{q}\;\mskip1.5mu]\;\Delta\;\rhd\;\Conid{A}{}\<[E]%
\ColumnHook
\end{hscode}\resethooks

We model the structural order on sorts as
an explicit relation with a least upper bound.
The latter will help with substitution and composition,
where the result is \ensuremath{\Conid{V}} only if both inputs are \ensuremath{\Conid{V}}.

\noindent
\begin{minipage}{0.55\textwidth}
\begin{hscode}\SaveRestoreHook
\column{B}{@{}>{\hspre}l<{\hspost}@{}}%
\column{3}{@{}>{\hspre}l<{\hspost}@{}}%
\column{8}{@{}>{\hspre}l<{\hspost}@{}}%
\column{E}{@{}>{\hspre}l<{\hspost}@{}}%
\>[B]{}\Keyword{data}\;\_\sqsubseteq\_\;\mathbin{:}\;\Conid{Sort}\;\rightarrow\;\Conid{Sort}\;\rightarrow\;\Conid{Set}\;\Keyword{where}{}\<[E]%
\\
\>[B]{}\hsindent{3}{}\<[3]%
\>[3]{}\Varid{rfl}\;{}\<[8]%
\>[8]{}\mathbin{:}\;\Varid{s}\;\sqsubseteq\;\Varid{s}{}\<[E]%
\\
\>[B]{}\hsindent{3}{}\<[3]%
\>[3]{}\Varid{v}\!\sqsubseteq\!\Varid{t}\;{}\<[8]%
\>[8]{}\mathbin{:}\;\Conid{V}\;\sqsubseteq\;\Conid{T}{}\<[E]%
\ColumnHook
\end{hscode}\resethooks
\end{minipage}
\begin{minipage}{0.4\textwidth}
\begin{hscode}\SaveRestoreHook
\column{B}{@{}>{\hspre}l<{\hspost}@{}}%
\column{8}{@{}>{\hspre}l<{\hspost}@{}}%
\column{11}{@{}>{\hspre}l<{\hspost}@{}}%
\column{E}{@{}>{\hspre}l<{\hspost}@{}}%
\>[B]{}\_\sqcup\_\;\mathbin{:}\;\Conid{Sort}\;\rightarrow\;\Conid{Sort}\;\rightarrow\;\Conid{Sort}{}\<[E]%
\\
\>[B]{}\Conid{V}\;\sqcup\;\Varid{r}\;{}\<[8]%
\>[8]{}\mathrel{=}\;{}\<[11]%
\>[11]{}\Varid{r}{}\<[E]%
\\
\>[B]{}\Conid{T}\;\sqcup\;\Varid{r}\;{}\<[8]%
\>[8]{}\mathrel{=}\;{}\<[11]%
\>[11]{}\Conid{T}{}\<[E]%
\ColumnHook
\end{hscode}\resethooks
\end{minipage}

\noindent
This is just Boolean algebra. We need a number of laws:

\noindent
\begin{minipage}{0.55\textwidth}
\begin{hscode}\SaveRestoreHook
\column{B}{@{}>{\hspre}l<{\hspost}@{}}%
\column{6}{@{}>{\hspre}l<{\hspost}@{}}%
\column{E}{@{}>{\hspre}l<{\hspost}@{}}%
\>[B]{}\sqsubseteq\!\Varid{t}\;{}\<[6]%
\>[6]{}\mathbin{:}\;\Varid{s}\;\sqsubseteq\;\Conid{T}{}\<[E]%
\\
\>[B]{}\Varid{v}\!\sqsubseteq\;{}\<[6]%
\>[6]{}\mathbin{:}\;\Conid{V}\;\sqsubseteq\;\Varid{s}{}\<[E]%
\\
\>[B]{}\sqsubseteq\!\Varid{q}\!\sqcup\;{}\<[6]%
\>[6]{}\mathbin{:}\;\Varid{q}\;\sqsubseteq\;(\Varid{q}\;\sqcup\;\Varid{r}){}\<[E]%
\\
\>[B]{}\sqsubseteq\!\sqcup\Varid{r}\;{}\<[6]%
\>[6]{}\mathbin{:}\;\Varid{r}\;\sqsubseteq\;(\Varid{q}\;\sqcup\;\Varid{r}){}\<[E]%
\ColumnHook
\end{hscode}\resethooks
\end{minipage}
\begin{minipage}{0.4\textwidth}
\begin{hscode}\SaveRestoreHook
\column{B}{@{}>{\hspre}l<{\hspost}@{}}%
\column{5}{@{}>{\hspre}l<{\hspost}@{}}%
\column{E}{@{}>{\hspre}l<{\hspost}@{}}%
\>[B]{}\sqcup\!\sqcup\;{}\<[5]%
\>[5]{}\mathbin{:}\;\Varid{q}\;\sqcup\;(\Varid{r}\;\sqcup\;\Varid{s})\;\equiv\;(\Varid{q}\;\sqcup\;\Varid{r})\;\sqcup\;\Varid{s}{}\<[E]%
\\
\>[B]{}\sqcup\!\Varid{v}\;{}\<[5]%
\>[5]{}\mathbin{:}\;\Varid{q}\;\sqcup\;\Conid{V}\;\equiv\;\Varid{q}{}\<[E]%
\\
\>[B]{}\sqcup\!\Varid{t}\;{}\<[5]%
\>[5]{}\mathbin{:}\;\Varid{q}\;\sqcup\;\Conid{T}\;\equiv\;\Conid{T}{}\<[E]%
\ColumnHook
\end{hscode}\resethooks
\end{minipage}

\noindent
These are easy to prove by case analysis,
e.g. \ensuremath{\sqsubseteq\!\Varid{t}\;\{\mskip1.5mu \Conid{V}\mskip1.5mu\}\;\mathrel{=}\;\Varid{v}\!\sqsubseteq\!\Varid{t}} and \ensuremath{\sqsubseteq\!\Varid{t}\;\{\mskip1.5mu \Conid{T}\mskip1.5mu\}\;\mathrel{=}\;\Varid{rfl}}.

Further, we turn the equations
($\sqcup\sqcup$, $\sqcup\Varid{v}$, $\sqcup\Varid{t}$) into rewrite rules with
\ensuremath{\Varid{\{-\#}\;\Keyword{REWRITE}\;\sqcup\!\sqcup\;\sqcup\!\Varid{v}\;\sqcup\!\Varid{t}\;\Varid{\#-\}}}.
This introduces new definitional equalities, allowing the
type checker to directly exploit e.g. associativity of \ensuremath{\_\sqcup\_} 
(effectively, this feature allows a selective use of 
extensional type theory).

Functoriality of context extension is now parametric
\begin{hscode}\SaveRestoreHook
\column{B}{@{}>{\hspre}l<{\hspost}@{}}%
\column{E}{@{}>{\hspre}l<{\hspost}@{}}%
\>[B]{}\_\uparrow\_\;\mathbin{:}\;\Gamma\;\Vdash\![\;\Varid{q}\;\mskip1.5mu]\;\Delta\;\rightarrow\;\forall{}\;\Conid{A}\;\rightarrow\;\Gamma\;\rhd\;\Conid{A}\;\Vdash\![\;\Varid{q}\;\mskip1.5mu]\;\Delta\;\rhd\;\Conid{A}{}\<[E]%
\ColumnHook
\end{hscode}\resethooks
We'll derive this later. Meanwhile,
the order on sorts gives rise to another functorial action.
\begin{hscode}\SaveRestoreHook
\column{B}{@{}>{\hspre}l<{\hspost}@{}}%
\column{12}{@{}>{\hspre}l<{\hspost}@{}}%
\column{E}{@{}>{\hspre}l<{\hspost}@{}}%
\>[B]{}\Varid{tm}\!\sqsubseteq\;\mathbin{:}\;\Varid{q}\;\sqsubseteq\;\Varid{s}\;\rightarrow\;\Gamma\;\vdash\![\;\Varid{q}\;\mskip1.5mu]\;\Conid{A}\;\rightarrow\;\Gamma\;\vdash\![\;\Varid{s}\;\mskip1.5mu]\;\Conid{A}{}\<[E]%
\\
\>[B]{}\Varid{tm}\!\sqsubseteq\;\Varid{rfl}\;\Varid{x}\;{}\<[12]%
\>[12]{}\mathrel{=}\;\Varid{x}{}\<[E]%
\\
\>[B]{}\Varid{tm}\!\sqsubseteq\;\Varid{v}\!\sqsubseteq\!\Varid{t}\;\Varid{i}\;{}\<[12]%
\>[12]{}\mathrel{=}\;\texttt{\textasciigrave}\;\Varid{i}{}\<[E]%
\ColumnHook
\end{hscode}\resethooks
Now we can define substitution and renaming in one go:

\noindent
\begin{minipage}{0.55\textwidth}
\begin{hscode}\SaveRestoreHook
\column{B}{@{}>{\hspre}l<{\hspost}@{}}%
\column{12}{@{}>{\hspre}l<{\hspost}@{}}%
\column{24}{@{}>{\hspre}l<{\hspost}@{}}%
\column{E}{@{}>{\hspre}l<{\hspost}@{}}%
\>[B]{}\_[\_]\;\mathbin{:}\;\Gamma\;\vdash\![\;\Varid{q}\;\mskip1.5mu]\;\Conid{A}\;\rightarrow\;\Delta\;\Vdash\![\;\Varid{r}\;\mskip1.5mu]\;\Gamma\;\rightarrow\;\Delta\;\vdash\![\;\Varid{q}\;\sqcup\;\Varid{r}\;\mskip1.5mu]\;\Conid{A}{}\<[E]%
\\
\>[B]{}\Varid{zero}\;{}\<[12]%
\>[12]{}[\mskip1.5mu \;\Varid{xs}\;\Varid{,}\;\Varid{x}\;\mskip1.5mu]\;{}\<[24]%
\>[24]{}\mathrel{=}\;\Varid{x}{}\<[E]%
\\
\>[B]{}(\Varid{suc}\;\Varid{i}\;\anonymous )\;{}\<[12]%
\>[12]{}[\mskip1.5mu \;\Varid{xs}\;\Varid{,}\;\Varid{x}\;\mskip1.5mu]\;{}\<[24]%
\>[24]{}\mathrel{=}\;\Varid{i}\;[\mskip1.5mu \;\Varid{xs}\;\mskip1.5mu]{}\<[E]%
\ColumnHook
\end{hscode}\resethooks
\end{minipage}
\begin{minipage}{0.4\textwidth}
\begin{hscode}\SaveRestoreHook
\column{B}{@{}>{\hspre}l<{\hspost}@{}}%
\column{12}{@{}>{\hspre}l<{\hspost}@{}}%
\column{24}{@{}>{\hspre}l<{\hspost}@{}}%
\column{31}{@{}>{\hspre}l<{\hspost}@{}}%
\column{35}{@{}>{\hspre}l<{\hspost}@{}}%
\column{E}{@{}>{\hspre}l<{\hspost}@{}}%
\>[B]{}(\texttt{\textasciigrave}\;\Varid{i})\;{}\<[12]%
\>[12]{}[\mskip1.5mu \;\Varid{xs}\;\mskip1.5mu]\;{}\<[24]%
\>[24]{}\mathrel{=}\;\Varid{tm}\!\sqsubseteq\;{}\<[31]%
\>[31]{}\sqsubseteq\!\Varid{t}\;{}\<[35]%
\>[35]{}(\Varid{i}\;[\mskip1.5mu \;\Varid{xs}\;\mskip1.5mu]){}\<[E]%
\\
\>[B]{}(\Varid{t}\;\cdot\;\Varid{u})\;{}\<[12]%
\>[12]{}[\mskip1.5mu \;\Varid{xs}\;\mskip1.5mu]\;{}\<[24]%
\>[24]{}\mathrel{=}\;(\Varid{t}\;[\mskip1.5mu \;\Varid{xs}\;\mskip1.5mu])\;\cdot\;(\Varid{u}\;[\mskip1.5mu \;\Varid{xs}\;\mskip1.5mu]){}\<[E]%
\\
\>[B]{}(\lambda\;\Varid{t})\;{}\<[12]%
\>[12]{}[\mskip1.5mu \;\Varid{xs}\;\mskip1.5mu]\;{}\<[24]%
\>[24]{}\mathrel{=}\;\lambda\;(\Varid{t}\;[\mskip1.5mu \;\Varid{xs}\;\uparrow\;\anonymous \;\mskip1.5mu]){}\<[E]%
\ColumnHook
\end{hscode}\resethooks
\end{minipage}

\noindent
Here \ensuremath{\_\sqcup\_} ensures substitution returns a variable
only if both inputs are variables/renamings.
We use \ensuremath{\Varid{tm}\!\sqsubseteq} when substituting for variables because \ensuremath{\Varid{i}\;[\mskip1.5mu \;\Varid{xs}\;\mskip1.5mu]} 
will return a variable if
\ensuremath{\Varid{xs}} is a renaming, but \ensuremath{(\texttt{\textasciigrave}\;\Varid{i})\;[\mskip1.5mu \;\Varid{xs}\;\mskip1.5mu]} must return a term.

We define \ensuremath{\Varid{id}} using \ensuremath{\_\uparrow\_}, recursing over contexts.
To define \ensuremath{\_\uparrow\_} itself, we need parametric versions of \ensuremath{\Varid{zero}} and \ensuremath{\Varid{suc}}.
Defining \ensuremath{\Varid{zero}} is easy.

\noindent
\begin{minipage}{0.55\textwidth}
\begin{hscode}\SaveRestoreHook
\column{B}{@{}>{\hspre}l<{\hspost}@{}}%
\column{17}{@{}>{\hspre}l<{\hspost}@{}}%
\column{20}{@{}>{\hspre}l<{\hspost}@{}}%
\column{E}{@{}>{\hspre}l<{\hspost}@{}}%
\>[B]{}\Varid{id}\;\mathbin{:}\;\Gamma\;\Vdash\![\;\Conid{V}\;\mskip1.5mu]\;\Gamma{}\<[E]%
\\
\>[B]{}\Varid{id}\;\{\mskip1.5mu \Gamma\;\mathrel{=}\;\Varid{•}\mskip1.5mu\}\;{}\<[17]%
\>[17]{}\mathrel{=}\;{}\<[20]%
\>[20]{}\varepsilon{}\<[E]%
\\
\>[B]{}\Varid{id}\;\{\mskip1.5mu \Gamma\;\mathrel{=}\;\Gamma\;\rhd\;\Conid{A}\mskip1.5mu\}\;{}\<[17]%
\>[17]{}\mathrel{=}\;{}\<[20]%
\>[20]{}\Varid{id}\;\uparrow\;\Conid{A}{}\<[E]%
\ColumnHook
\end{hscode}\resethooks
\end{minipage}
\begin{minipage}{0.4\textwidth}
\begin{hscode}\SaveRestoreHook
\column{B}{@{}>{\hspre}l<{\hspost}@{}}%
\column{16}{@{}>{\hspre}l<{\hspost}@{}}%
\column{19}{@{}>{\hspre}l<{\hspost}@{}}%
\column{E}{@{}>{\hspre}l<{\hspost}@{}}%
\>[B]{}\Varid{zero}[\_]\;\mathbin{:}\;\forall{}\;\Varid{q}\;\rightarrow\;\Gamma\;\rhd\;\Conid{A}\;\vdash\![\;\Varid{q}\;\mskip1.5mu]\;\Conid{A}{}\<[E]%
\\
\>[B]{}\Varid{zero}[\;\Conid{V}\;\mskip1.5mu]\;{}\<[16]%
\>[16]{}\mathrel{=}\;{}\<[19]%
\>[19]{}\Varid{zero}{}\<[E]%
\\
\>[B]{}\Varid{zero}[\;\Conid{T}\;\mskip1.5mu]\;{}\<[16]%
\>[16]{}\mathrel{=}\;{}\<[19]%
\>[19]{}\texttt{\textasciigrave}\;\Varid{zero}{}\<[E]%
\ColumnHook
\end{hscode}\resethooks
\end{minipage}

However, \ensuremath{\Varid{suc}} is more subtle since the case for \ensuremath{\Conid{T}} depends on
weakening over substitutions:

\noindent
\begin{minipage}{0.55\textwidth}
\begin{hscode}\SaveRestoreHook
\column{B}{@{}>{\hspre}l<{\hspost}@{}}%
\column{9}{@{}>{\hspre}l<{\hspost}@{}}%
\column{12}{@{}>{\hspre}l<{\hspost}@{}}%
\column{13}{@{}>{\hspre}l<{\hspost}@{}}%
\column{16}{@{}>{\hspre}l<{\hspost}@{}}%
\column{28}{@{}>{\hspre}l<{\hspost}@{}}%
\column{E}{@{}>{\hspre}l<{\hspost}@{}}%
\>[B]{}\Varid{suc}[\_]\;{}\<[9]%
\>[9]{}\mathbin{:}\;{}\<[12]%
\>[12]{}\forall{}\;\Varid{q}\;\rightarrow\;\Gamma\;\vdash\![\;\Varid{q}\;\mskip1.5mu]\;\Conid{B}\;\rightarrow\;\forall{}\;\Conid{A}\;{}\<[E]%
\\
\>[9]{}\rightarrow\;{}\<[12]%
\>[12]{}\Gamma\;\rhd\;\Conid{A}\;\vdash\![\;\Varid{q}\;\mskip1.5mu]\;\Conid{B}{}\<[E]%
\\
\>[B]{}\Varid{suc}[\;\Conid{V}\;\mskip1.5mu]\;\Varid{i}\;{}\<[13]%
\>[13]{}\Conid{A}\;{}\<[16]%
\>[16]{}\mathrel{=}\;\Varid{suc}\;\Varid{i}\;\Conid{A}{}\<[E]%
\\
\>[B]{}\Varid{suc}[\;\Conid{T}\;\mskip1.5mu]\;\Varid{t}\;{}\<[13]%
\>[13]{}\Conid{A}\;{}\<[16]%
\>[16]{}\mathrel{=}\;\Varid{t}\;[\mskip1.5mu \;\Varid{id}\;^{+}\;{}\<[28]%
\>[28]{}\Conid{A}\;\mskip1.5mu]{}\<[E]%
\ColumnHook
\end{hscode}\resethooks
\end{minipage}
\begin{minipage}{0.4\textwidth}
\begin{hscode}\SaveRestoreHook
\column{B}{@{}>{\hspre}l<{\hspost}@{}}%
\column{6}{@{}>{\hspre}l<{\hspost}@{}}%
\column{9}{@{}>{\hspre}l<{\hspost}@{}}%
\column{11}{@{}>{\hspre}l<{\hspost}@{}}%
\column{E}{@{}>{\hspre}l<{\hspost}@{}}%
\>[B]{}\_^{+}\_\;{}\<[6]%
\>[6]{}\mathbin{:}\;{}\<[9]%
\>[9]{}\Gamma\;\Vdash\![\;\Varid{q}\;\mskip1.5mu]\;\Delta\;\rightarrow\;\forall{}\;\Conid{A}\;{}\<[E]%
\\
\>[6]{}\rightarrow\;{}\<[9]%
\>[9]{}\Gamma\;\rhd\;\Conid{A}\;\Vdash\![\;\Varid{q}\;\mskip1.5mu]\;\Delta{}\<[E]%
\\
\>[B]{}\varepsilon\;{}\<[11]%
\>[11]{}^{+}\;\Conid{A}\;\mathrel{=}\;\varepsilon{}\<[E]%
\\
\>[B]{}(\Varid{xs}\;\Varid{,}\;\Varid{x})\;{}\<[11]%
\>[11]{}^{+}\;\Conid{A}\;\mathrel{=}\;\Varid{xs}\;^{+}\;\Conid{A}\;\Varid{,}\;\Varid{suc}[\;\anonymous \;\mskip1.5mu]\;\Varid{x}\;\Conid{A}{}\<[E]%
\ColumnHook
\end{hscode}\resethooks
\end{minipage}\\
And finally we can define \ensuremath{\_\uparrow\_} and \ensuremath{\_\ensuremath{\mbox{$\circ$}}\_}.

\noindent
\begin{minipage}{0.4\textwidth}
\begin{hscode}\SaveRestoreHook
\column{B}{@{}>{\hspre}l<{\hspost}@{}}%
\column{11}{@{}>{\hspre}l<{\hspost}@{}}%
\column{E}{@{}>{\hspre}l<{\hspost}@{}}%
\>[B]{}\Varid{xs}\;\uparrow\;\Conid{A}\;\mathrel{=}\;{}\<[11]%
\>[11]{}\Varid{xs}\;^{+}\;\Conid{A}\;\Varid{,}\;\Varid{zero}[\;\anonymous \;\mskip1.5mu]{}\<[E]%
\ColumnHook
\end{hscode}\resethooks
\end{minipage}
\begin{minipage}{0.55\textwidth}
\begin{hscode}\SaveRestoreHook
\column{B}{@{}>{\hspre}l<{\hspost}@{}}%
\column{16}{@{}>{\hspre}l<{\hspost}@{}}%
\column{E}{@{}>{\hspre}l<{\hspost}@{}}%
\>[B]{}\_\ensuremath{\mbox{$\circ$}}\_\;\mathbin{:}\;\Gamma\;\Vdash\![\;\Varid{q}\;\mskip1.5mu]\;\Theta\;\rightarrow\;\Delta\;\Vdash\![\;\Varid{r}\;\mskip1.5mu]\;\Gamma\;\rightarrow\;\Delta\;\Vdash\![\;\Varid{q}\;\sqcup\;\Varid{r}\;\mskip1.5mu]\;\Theta{}\<[E]%
\\
\>[B]{}\varepsilon\;\ensuremath{\mbox{$\circ$}}\;\Varid{ys}\;{}\<[16]%
\>[16]{}\mathrel{=}\;\varepsilon{}\<[E]%
\\
\>[B]{}(\Varid{xs}\;\Varid{,}\;\Varid{x})\;\ensuremath{\mbox{$\circ$}}\;\Varid{ys}\;{}\<[16]%
\>[16]{}\mathrel{=}\;(\Varid{xs}\;\ensuremath{\mbox{$\circ$}}\;\Varid{ys})\;\Varid{,}\;\Varid{x}\;[\mskip1.5mu \;\Varid{ys}\;\mskip1.5mu]{}\<[E]%
\ColumnHook
\end{hscode}\resethooks
\end{minipage}

\subsection{Termination}
\label{sec:termination}

Unfortunately (as of Agda 2.7.0.1) we now hit a termination error.
\begin{hscode}\SaveRestoreHook
\column{B}{@{}>{\hspre}l<{\hspost}@{}}%
\column{3}{@{}>{\hspre}l<{\hspost}@{}}%
\column{E}{@{}>{\hspre}l<{\hspost}@{}}%
\>[B]{}\Conid{Termination}\;\Varid{checking}\;\Varid{failed}\;\Varid{for}\;\Varid{the}\;\Varid{following}\;\Varid{functions:}\;{}\<[E]%
\\
\>[B]{}\hsindent{3}{}\<[3]%
\>[3]{}\Varid{\char95 \char94 \char95 ,}\;\Varid{\char95 [\char95 ],}\;\Varid{id,}\;\_^{+}\_,\;\Varid{suc}[\_]{}\<[E]%
\ColumnHook
\end{hscode}\resethooks
The cause turns out to be \ensuremath{\Varid{id}}.
Termination here hinges on weakening for terms
(\ensuremath{\Varid{suc}[\;\Conid{T}\;\mskip1.5mu]\;\Varid{t}\;\Conid{A}}) applying a renaming rather than a full substitution.
Note that if instead we had \ensuremath{\Varid{id}\;\mathbin{:}\;\Gamma\;\Vdash\![\;\Conid{T}\;\mskip1.5mu]\;\Gamma}, or if
weakening for variables (\ensuremath{\Varid{suc}[\;\Conid{V}\;\mskip1.5mu]\;\Varid{i}\;\Conid{A}}) was implemented by \ensuremath{\Varid{i}\;[\mskip1.5mu \;\Varid{id}\;^{+}\;\Conid{A}\;\mskip1.5mu]},
our operations would still be type-correct but would genuinely loop,
so perhaps Agda is right to be careful.

We have appropriately specialised weakening for variables though, so 
why doesn't Agda accept our program? The limitation is ultimately
technical: Agda only looks at direct arguments to function calls when 
building the call graph from which it identifies termination order 
\cite{alti:jfp02}. Because \ensuremath{\Varid{id}} is not passed a sort, the sort cannot be 
considered as decreasing in the case of term weakening (\ensuremath{\Varid{suc}[\;\Conid{T}\;\mskip1.5mu]\;\Varid{t}\;\Conid{A}}).

\noindent
\begin{minipage}{0.675\textwidth}
\begin{tikzcd}[scaleedge cd=1.1, sep=large]
& \ensuremath{\Varid{suc}[\;\Varid{q}_4\;\mskip1.5mu]\;\Varid{t}_4\vphantom{a}_{\Gamma_4}^{\Varid{q}_4}}
\arrow[dd, bend left, "\substack{\ensuremath{\Varid{r}_3\;\ensuremath{<}\;\Varid{q}_4}}"]
\arrow[ldd, bend right, swap, "\substack{\ensuremath{\Varid{r}_2\;\ensuremath{<}\;\Varid{q}_4}}"]
\arrow[rdd, bend left, "\ensuremath{\Varid{r}_1\;\ensuremath{<}\;\Varid{q}_4}"]
\\
\\
\ensuremath{\Varid{id}_{\Gamma_2}^{\Varid{r}_2}} 
\arrow[r, swap, "\substack{\ensuremath{\Varid{r}_3\;\mathrel{=}\;\Varid{r}_2}}"]
\arrow[in=300, out=240, loop, swap, "\substack{\ensuremath{\Varid{r}_2^{\prime}\;\mathrel{=}\;\Varid{r}_2} \\ \ensuremath{\Gamma_2^{\prime}\;\ensuremath{<}\;\Gamma_2}}"]
& \ensuremath{^{\Varid{r}_3}\sigma_3\vphantom{a}_{\Gamma_3}^{\Delta_3}\;^{+}\;\Conid{A}} 
\arrow[uu, bend left, "\substack{\ensuremath{\Varid{q}_4\;\mathrel{=}\;\Varid{r}_3}}"]
\arrow[in=300, out=240, loop, swap, "\substack{\ensuremath{\Varid{r}_3^{\prime}\;\mathrel{=}\;\Varid{r}_3} \\ \ensuremath{\sigma_3^{\prime}\;\ensuremath{<}\;\sigma_3}}"]
& \ensuremath{\Varid{t}_1\vphantom{a}_{\Gamma_1}^{\Varid{q}_1}\;[\mskip1.5mu \;^{\Varid{r}_1}\sigma_1\vphantom{a}_{\Gamma_1}^{\Delta_1}\;\mskip1.5mu]}
\arrow[l, "\ensuremath{\Varid{r}_3\;\mathrel{=}\;\Varid{r}_1}"]
\arrow[in=300, out=240, loop, swap, "\substack{\ensuremath{\Varid{r}_1^{\prime}\;\mathrel{=}\;\Varid{r}_1} \\ \ensuremath{\Varid{t}_1^{\prime}\;\ensuremath{<}\;\Varid{t}_1}}"]
\end{tikzcd}

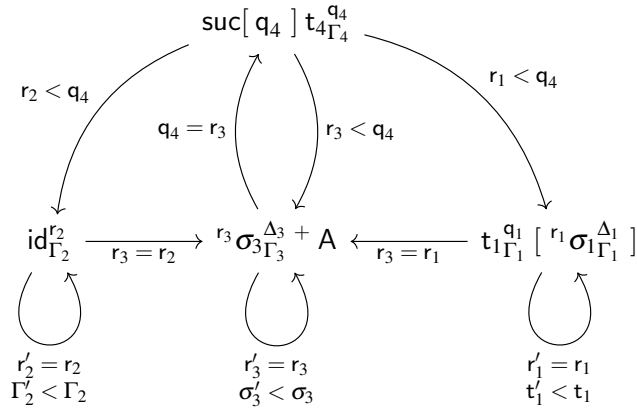
\captionof{figure}{Call graph of substitution operations}
\label{figure:termination}
\end{minipage}
\begin{minipage}{0.3\textwidth}
\renewcommand{\arraystretch}{1.2}
\begin{center}
\begin{tabular}{ |c|c|c| }
\hline
Function & Measure \\
\hline \hline
\ensuremath{\Varid{t}_1\vphantom{a}_{\Gamma_1}^{\Varid{q}_1}\;[\mskip1.5mu \;^{\Varid{r}_1}\sigma_1\vphantom{a}_{\Gamma_1}^{\Delta_1}\;\mskip1.5mu]} & \ensuremath{(\Varid{r}_1\;\Varid{,}\;\Varid{t}_1)} \\ [0.2ex]
\hline
\ensuremath{\Varid{id}_{\Gamma_2}^{\Varid{r}_2}} & \ensuremath{(\Varid{r}_2\;\Varid{,}\;\Gamma_2)} \\ [0.2ex]
\hline
\ensuremath{^{\Varid{r}_3}\sigma_3\vphantom{a}_{\Gamma_3}^{\Delta_3}\;^{+}\;\Conid{A}} & \ensuremath{(\Varid{r}_3\;\Varid{,}\;\sigma_3)} \\ [0.2ex]
\hline
\ensuremath{\Varid{suc}[\;\Varid{q}_4\;\mskip1.5mu]\;\Varid{t}_4\vphantom{a}_{\Gamma_4}^{\Varid{q}_4}} & \ensuremath{(\Varid{q}_4)} \\ [0.2ex]
\hline
\end{tabular}
\end{center}
\captionof{table}{Per-function termination measures}
\label{table:termination}
\end{minipage}
\\[2.0ex]

Luckily, there is an easy solution: making \ensuremath{\Varid{id}} polymorphic in its \ensuremath{\Conid{Sort}}
and instantiating with \ensuremath{\Conid{V}} at the call-sites
enables the decrease
to be tracked and termination to be correctly inferred by Agda.
We present the call graph diagrammatically (inlining \ensuremath{\_\uparrow\_}), 
in the style of \cite{keller2010hereditary} (Figure~\ref{figure:termination}).

To justify termination, we note that along all cycles in the graph,
either the \ensuremath{\Conid{Sort}} strictly decreases,
or the \ensuremath{\Conid{Sort}} is preserved and some other argument
(the context, substitution, or term) decreases. Following this, we can
assign lexicographically-decreasing measures to each of the functions
(\textsf{Table \ref{table:termination}}).

In practice, we will generally require identity renamings, rather than
substitutions. We define \ensuremath{\Conid{Sort}}-polymorphic \ensuremath{\Varid{id-poly}}, and then recover
our original \ensuremath{\Varid{id}} by instantiating it at \ensuremath{\Conid{V}} (and using an \ensuremath{\Keyword{INLINE}} pragma
so Agda's termination checker cannot tell the difference).

\begin{minipage}{0.4\textwidth}
\begin{hscode}\SaveRestoreHook
\column{B}{@{}>{\hspre}l<{\hspost}@{}}%
\column{10}{@{}>{\hspre}l<{\hspost}@{}}%
\column{E}{@{}>{\hspre}l<{\hspost}@{}}%
\>[B]{}\Varid{id-poly}\;{}\<[10]%
\>[10]{}\mathbin{:}\;\Gamma\;\Vdash\![\;\Varid{q}\;\mskip1.5mu]\;\Gamma{}\<[E]%
\\
\>[B]{}\Varid{id}\;{}\<[10]%
\>[10]{}\mathbin{:}\;\Gamma\;\Vdash\![\;\Conid{V}\;\mskip1.5mu]\;\Gamma{}\<[E]%
\ColumnHook
\end{hscode}\resethooks
\end{minipage}
\begin{minipage}{0.5\textwidth}
\begin{hscode}\SaveRestoreHook
\column{B}{@{}>{\hspre}l<{\hspost}@{}}%
\column{E}{@{}>{\hspre}l<{\hspost}@{}}%
\>[B]{}\Varid{id}\;\mathrel{=}\;\Varid{id-poly}\;\{\mskip1.5mu \Varid{q}\;\mathrel{=}\;\Conid{V}\mskip1.5mu\}{}\<[E]%
\\
\>[B]{}\Varid{\{-\#}\;\Keyword{INLINE}\;\Varid{id}\;\Varid{\#-\}}{}\<[E]%
\ColumnHook
\end{hscode}\resethooks
\end{minipage}

(All this fuss with \ensuremath{\Conid{Sort}}-polymorphic \ensuremath{\Varid{id}} may be unnecessary.
At a cost in performance, it is possible to extend Agda's 
termination checker so that our original definitions are accepted directly.
See \href{https://github.com/agda/agda/pull/7695}{\#7695}.)

\section{Proving the laws}
\label{sec:proving-laws}

We now present a formal proof of the categorical laws, proving each
lemma only once while only using structural induction. Indeed 
termination isn't completely trivial but is still inferred by the termination
checker.

\subsection{The right identity law}
\label{sec:right-identity-law}

Let's get the easy case out of the way: the right-identity law 
(\ensuremath{\Varid{xs}\;\ensuremath{\mbox{$\circ$}}\;\Varid{id}\;\equiv\;\Varid{xs}}). It is easy because it doesn't depend on any other
categorical equations.

The main lemma is the identity law for the substitution functor 
\ensuremath{\Varid{[id]}\;\mathbin{:}\;\Varid{x}\;[\mskip1.5mu \;\Varid{id}\;\mskip1.5mu]\;\equiv\;\Varid{x}}.
To prove the successor case, we need naturality of \ensuremath{\Varid{suc}[\;\Varid{q}\;\mskip1.5mu]} applied to a 
variable, which can be shown by simple induction over said variable:
\begin{hscode}\SaveRestoreHook
\column{B}{@{}>{\hspre}l<{\hspost}@{}}%
\column{25}{@{}>{\hspre}l<{\hspost}@{}}%
\column{E}{@{}>{\hspre}l<{\hspost}@{}}%
\>[B]{}^{+}\Varid{-nat[]v}\;\mathbin{:}\;\Varid{i}\;[\mskip1.5mu \;\Varid{xs}\;^{+}\;\Conid{A}\;\mskip1.5mu]\;\equiv\;\Varid{suc}[\;\Varid{q}\;\mskip1.5mu]\;(\Varid{i}\;[\mskip1.5mu \;\Varid{xs}\;\mskip1.5mu])\;\Conid{A}{}\<[E]%
\\
\>[B]{}^{+}\Varid{-nat[]v}\;\{\mskip1.5mu \Varid{i}\;\mathrel{=}\;\Varid{zero}\mskip1.5mu\}\;{}\<[25]%
\>[25]{}\{\mskip1.5mu \Varid{xs}\;\mathrel{=}\;\Varid{xs}\;\Varid{,}\;\Varid{x}\mskip1.5mu\}\;\mathrel{=}\;\Varid{refl}{}\<[E]%
\\
\>[B]{}^{+}\Varid{-nat[]v}\;\{\mskip1.5mu \Varid{i}\;\mathrel{=}\;\Varid{suc}\;\Varid{j}\;\Conid{A}\mskip1.5mu\}\;{}\<[25]%
\>[25]{}\{\mskip1.5mu \Varid{xs}\;\mathrel{=}\;\Varid{xs}\;\Varid{,}\;\Varid{x}\mskip1.5mu\}\;\mathrel{=}\;^{+}\Varid{-nat[]v}\;\{\mskip1.5mu \Varid{i}\;\mathrel{=}\;\Varid{j}\mskip1.5mu\}{}\<[E]%
\ColumnHook
\end{hscode}\resethooks
The identity law is now easily provable by structural induction:

\noindent
\begin{minipage}{0.5\textwidth}
\begin{hscode}\SaveRestoreHook
\column{B}{@{}>{\hspre}l<{\hspost}@{}}%
\column{4}{@{}>{\hspre}l<{\hspost}@{}}%
\column{18}{@{}>{\hspre}l<{\hspost}@{}}%
\column{21}{@{}>{\hspre}l<{\hspost}@{}}%
\column{E}{@{}>{\hspre}l<{\hspost}@{}}%
\>[B]{}\Varid{[id]}\;\{\mskip1.5mu \Varid{x}\;\mathrel{=}\;\Varid{zero}\mskip1.5mu\}\;{}\<[21]%
\>[21]{}\mathrel{=}\;\Varid{refl}{}\<[E]%
\\
\>[B]{}\Varid{[id]}\;\{\mskip1.5mu \Varid{x}\;\mathrel{=}\;\Varid{suc}\;\Varid{i}\;\Conid{A}\mskip1.5mu\}\;{}\<[21]%
\>[21]{}\mathrel{=}\;{}\<[E]%
\\
\>[B]{}\hsindent{4}{}\<[4]%
\>[4]{}\Varid{i}\;[\mskip1.5mu \;\Varid{id}\;^{+}\;\Conid{A}\;\mskip1.5mu]\;{}\<[18]%
\>[18]{}\equiv\!\langle\;^{+}\Varid{-nat[]v}\;\{\mskip1.5mu \Varid{i}\;\mathrel{=}\;\Varid{i}\mskip1.5mu\}\;\Varid{⟩}\;{}\<[E]%
\\
\>[B]{}\hsindent{4}{}\<[4]%
\>[4]{}\Varid{suc}\;(\Varid{i}\;[\mskip1.5mu \;\Varid{id}\;\mskip1.5mu])\;\Conid{A}\;{}\<[E]%
\\
\>[B]{}\hsindent{4}{}\<[4]%
\>[4]{}\equiv\!\langle\;\Varid{cong}\;(\lambda\;\Varid{j}\;\rightarrow\;\Varid{suc}\;\Varid{j}\;\Conid{A})\;(\Varid{[id]}\;\{\mskip1.5mu \Varid{x}\;\mathrel{=}\;\Varid{i}\mskip1.5mu\})\;\Varid{⟩}\;{}\<[E]%
\\
\>[B]{}\hsindent{4}{}\<[4]%
\>[4]{}\Varid{suc}\;\Varid{i}\;\Conid{A}\;{}\<[18]%
\>[18]{}\blacksquare{}\<[E]%
\ColumnHook
\end{hscode}\resethooks
\end{minipage}
\begin{minipage}{0.45\textwidth}
\begin{hscode}\SaveRestoreHook
\column{B}{@{}>{\hspre}l<{\hspost}@{}}%
\column{4}{@{}>{\hspre}l<{\hspost}@{}}%
\column{19}{@{}>{\hspre}l<{\hspost}@{}}%
\column{E}{@{}>{\hspre}l<{\hspost}@{}}%
\>[B]{}\Varid{[id]}\;\{\mskip1.5mu \Varid{x}\;\mathrel{=}\;\texttt{\textasciigrave}\;\Varid{i}\mskip1.5mu\}\;{}\<[19]%
\>[19]{}\mathrel{=}\;{}\<[E]%
\\
\>[B]{}\hsindent{4}{}\<[4]%
\>[4]{}\Varid{cong}\;\texttt{\textasciigrave}\_\;(\Varid{[id]}\;\{\mskip1.5mu \Varid{x}\;\mathrel{=}\;\Varid{i}\mskip1.5mu\}){}\<[E]%
\\
\>[B]{}\Varid{[id]}\;\{\mskip1.5mu \Varid{x}\;\mathrel{=}\;\Varid{t}\;\cdot\;\Varid{u}\mskip1.5mu\}\;{}\<[19]%
\>[19]{}\mathrel{=}\;{}\<[E]%
\\
\>[B]{}\hsindent{4}{}\<[4]%
\>[4]{}\Varid{cong}_{2}\;\_\cdot\_\;(\Varid{[id]}\;\{\mskip1.5mu \Varid{x}\;\mathrel{=}\;\Varid{t}\mskip1.5mu\})\;(\Varid{[id]}\;\{\mskip1.5mu \Varid{x}\;\mathrel{=}\;\Varid{u}\mskip1.5mu\}){}\<[E]%
\\
\>[B]{}\Varid{[id]}\;\{\mskip1.5mu \Varid{x}\;\mathrel{=}\;\lambda\;\Varid{t}\mskip1.5mu\}\;{}\<[19]%
\>[19]{}\mathrel{=}\;{}\<[E]%
\\
\>[B]{}\hsindent{4}{}\<[4]%
\>[4]{}\Varid{cong}\;\lambda\_\;(\Varid{[id]}\;\{\mskip1.5mu \Varid{x}\;\mathrel{=}\;\Varid{t}\mskip1.5mu\}){}\<[E]%
\ColumnHook
\end{hscode}\resethooks
\end{minipage}

Note that the \ensuremath{\lambda\_} case is easy here: we need the law to hold for
\ensuremath{\Varid{t}\;\mathbin{:}\;\Gamma\;\Varid{,}\;\Conid{A}\;\vdash\![\;\Conid{T}\;\mskip1.5mu]\;\Conid{B}}, but this is still covered by the inductive hypothesis 
because \ensuremath{\Varid{id}\;\{\mskip1.5mu \Gamma\;\mathrel{=}\;\Gamma\;\Varid{,}\;\Conid{A}\mskip1.5mu\}\;\mathrel{=}\;\Varid{id}\;\uparrow\;\Conid{A}}.

Note also that is the first time we use Agda's syntax for equational derivations.
This is just syntactic sugar for constructing an equational
derivation using transitivity, exploiting Agda's
flexible syntax. Here \ensuremath{\Varid{e}\;\equiv\!\langle\;\Varid{p}\;\Varid{⟩}\;\Varid{e'}} means that \ensuremath{\Varid{p}} is a proof of
\ensuremath{\Varid{e}\;\equiv\;\Varid{e'}}. Later we will also use the special case \ensuremath{\Varid{e}\;\equiv\!\langle\rangle\;\Varid{e'}} which
means that \ensuremath{\Varid{e}} and \ensuremath{\Varid{e'}} are definitionally equal (this corresponds to
\ensuremath{\Varid{e}\;\equiv\!\langle\;\Varid{refl}\;\Varid{⟩}\;\Varid{e'}} and is just used to make the proof more
readable).  The proof is terminated with \ensuremath{\blacksquare} which inserts \ensuremath{\Varid{refl}}.
We also make heavy use of congruence \ensuremath{\Varid{cong}\;\Varid{f}\;\mathbin{:}\;\Varid{a}\;\equiv\;\Varid{b}\;\rightarrow\;\Varid{f}\;\Varid{a}\;\equiv\;\Varid{f}\;\Varid{b}}
and a version for binary functions
\ensuremath{\Varid{cong}_{2}\;\Varid{g}\;\mathbin{:}\;\Varid{a}\;\equiv\;\Varid{b}\;\rightarrow\;\Varid{c}\;\equiv\;\Varid{d}\;\rightarrow\;\Varid{g}\;\Varid{a}\;\Varid{c}\;\equiv\;\Varid{g}\;\Varid{b}\;\Varid{d}}.

The category law \ensuremath{\circ\Varid{id}\;\mathbin{:}\;\Varid{xs}\;\ensuremath{\mbox{$\circ$}}\;\Varid{id}\;\equiv\;\Varid{xs}} is now simply a fold of the functor law
(\ensuremath{\Varid{[id]}}).

\subsection{The left identity law}
\label{sec:right-ident}

We need to prove the left identity law mutually with the second
functor law for substitution. This is the main lemma for
associativity. 

Let's state the functor law but postpone the proof until the next section:
\ensuremath{[\circ]\;\mathbin{:}\;\Varid{x}\;[\mskip1.5mu \;\Varid{xs}\;\ensuremath{\mbox{$\circ$}}\;\Varid{ys}\;\mskip1.5mu]\;\equiv\;\Varid{x}\;[\mskip1.5mu \;\Varid{xs}\;\mskip1.5mu]\;[\mskip1.5mu \;\Varid{ys}\;\mskip1.5mu]}. 
Even stating this signature requires (definitional) associativity of \ensuremath{\_\sqcup\_},
since the left hand side has type \ensuremath{\Delta\;\vdash\![\;\Varid{q}\;\sqcup\;(\Varid{r}\;\sqcup\;\Varid{s})\;\mskip1.5mu]\;\Conid{A}}
while the right hand side has type \ensuremath{\Delta\;\vdash\![\;(\Varid{q}\;\sqcup\;\Varid{r})\;\sqcup\;\Varid{s}\;\mskip1.5mu]\;\Conid{A}}.
Fortunately, we obtain this via the $\sqcup\sqcup$ rewrite rule, but
alternatively we would have to insert a transport using \ensuremath{\Varid{subst}}.

Of course, we must also state the left-identity law \ensuremath{\Varid{id}\circ\;\mathbin{:}\;\Varid{id}\;\ensuremath{\mbox{$\circ$}}\;\Varid{xs}\;\equiv\;\Varid{xs}}. 
Similarly to \ensuremath{\Varid{id}}, Agda will not accept a direct implementation of \ensuremath{\Varid{id}\circ} as 
structurally recursive. Unfortunately, adapting the law to deal with a
\ensuremath{\Conid{Sort}}-polymorphic \ensuremath{\Varid{id}} complicates matters: when \ensuremath{\Varid{xs}} is a renaming 
(i.e. at sort \ensuremath{\Conid{V}})
composed with an identity substitution (i.e. at sort \ensuremath{\Conid{T}}), its sort must be lifted
on
the RHS (e.g. by extending the \ensuremath{\Varid{tm}\!\sqsubseteq} functor to lists of terms) to
obey \ensuremath{\_\sqcup\_}. 

Accounting for this lifting is certainly do-able, but in keeping with the
single-responsibility principle of software design, we argue it is neater
to consider only \ensuremath{\Conid{V}}-sorted \ensuremath{\Varid{id}} here and worry about equations involving
\ensuremath{\Conid{Sort}}-coercions later (in \ref{sec:cwf-recurs-subst}).
Therefore, we instead add a ``dummy'' \ensuremath{\Conid{Sort}} argument
(i.e. \ensuremath{\Varid{id}\circ^{\prime}\;\mathbin{:}\;\Conid{Sort}\;\rightarrow\;\Varid{id}\;\ensuremath{\mbox{$\circ$}}\;\Varid{xs}\;\equiv\;\Varid{xs}}) to track the size
decrease (such that we can eventually just use \ensuremath{\Varid{id}\circ\;\mathrel{=}\;\Varid{id}\circ^{\prime}\;\Conid{V}}).

(Perhaps surprisingly, this ``dummy'' argument does not even need to
be of type \ensuremath{\Conid{Sort}} to satisfy Agda here. More discussion on this trick 
can be found at Agda issue
\href{https://github.com/agda/agda/issues/7693}{\#7693}, but in summary:
(i) Agda considers all base constructors (constructors with no parameters) 
   to be of minimal size structurally, so their presence can track size
   preservation of other base-constructor arguments across function calls.
(ii) It turns out that
   a strict decrease in \ensuremath{\Conid{Sort}} is not necessary everywhere for termination: 
   note that the context also gets structurally smaller in the call to \ensuremath{\_^{+}\_} 
   from \ensuremath{\Varid{id}}.)

To prove \ensuremath{\Varid{id}\circ^{\prime}}, we need the $\beta$-law for \ensuremath{\_^{+}\_},
\ensuremath{\Varid{xs}\;^{+}\;\Conid{A}\;\ensuremath{\mbox{$\circ$}}\;(\Varid{ys}\;\Varid{,}\;\Varid{x})\;\equiv\;\Varid{xs}\;\ensuremath{\mbox{$\circ$}}\;\Varid{ys}}, which can be shown with a fold over a
corresponding property for \ensuremath{\Varid{suc}[\_]}.

\noindent
\begin{minipage}{0.5\textwidth}
\begin{hscode}\SaveRestoreHook
\column{B}{@{}>{\hspre}l<{\hspost}@{}}%
\column{3}{@{}>{\hspre}l<{\hspost}@{}}%
\column{13}{@{}>{\hspre}l<{\hspost}@{}}%
\column{16}{@{}>{\hspre}l<{\hspost}@{}}%
\column{30}{@{}>{\hspre}l<{\hspost}@{}}%
\column{31}{@{}>{\hspre}l<{\hspost}@{}}%
\column{E}{@{}>{\hspre}l<{\hspost}@{}}%
\>[B]{}\Varid{suc[]}\;\mathbin{:}\;(\Varid{suc}[\;\Varid{q}\;\mskip1.5mu]\;\Varid{x}\;\anonymous )\;[\mskip1.5mu \;\Varid{ys}\;\Varid{,}\;\Varid{y}\;\mskip1.5mu]\;\equiv\;\Varid{x}\;[\mskip1.5mu \;\Varid{ys}\;\mskip1.5mu]{}\<[E]%
\\
\>[B]{}\Varid{suc[]}\;\{\mskip1.5mu \Varid{q}\;\mathrel{=}\;\Conid{V}\mskip1.5mu\}\;\mathrel{=}\;\Varid{refl}{}\<[E]%
\\
\>[B]{}\Varid{suc[]}\;\{\mskip1.5mu \Varid{q}\;\mathrel{=}\;\Conid{T}\mskip1.5mu\}\;\{\mskip1.5mu \Varid{x}\;\mathrel{=}\;\Varid{x}\mskip1.5mu\}\;\{\mskip1.5mu \Varid{ys}\;\mathrel{=}\;\Varid{ys}\mskip1.5mu\}\;\{\mskip1.5mu \Varid{y}\;\mathrel{=}\;\Varid{y}\mskip1.5mu\}\;\mathrel{=}\;{}\<[E]%
\\
\>[B]{}\hsindent{3}{}\<[3]%
\>[3]{}(\Varid{suc}[\;\Conid{T}\;\mskip1.5mu]\;\Varid{x}\;\anonymous )\;[\mskip1.5mu \;\Varid{ys}\;\Varid{,}\;\Varid{y}\;\mskip1.5mu]\;{}\<[30]%
\>[30]{}\equiv\!\langle\rangle\;{}\<[E]%
\\
\>[B]{}\hsindent{3}{}\<[3]%
\>[3]{}\Varid{x}\;[\mskip1.5mu \;\Varid{id}\;^{+}\;\anonymous \;\mskip1.5mu]\;[\mskip1.5mu \;\Varid{ys}\;\Varid{,}\;\Varid{y}\;\mskip1.5mu]\;{}\<[E]%
\\
\>[B]{}\hsindent{3}{}\<[3]%
\>[3]{}\equiv\!\langle\;\Varid{sym}\;([\circ]\;\{\mskip1.5mu \Varid{x}\;\mathrel{=}\;\Varid{x}\mskip1.5mu\})\;\Varid{⟩}\;{}\<[E]%
\\
\>[B]{}\hsindent{3}{}\<[3]%
\>[3]{}\Varid{x}\;[\mskip1.5mu \;(\Varid{id}\;^{+}\;\anonymous )\;\ensuremath{\mbox{$\circ$}}\;(\Varid{ys}\;\Varid{,}\;\Varid{y})\;\mskip1.5mu]\;{}\<[E]%
\\
\>[B]{}\hsindent{3}{}\<[3]%
\>[3]{}\equiv\!\langle\;\Varid{cong}\;(\lambda\;\rho\;\rightarrow\;\Varid{x}\;[\mskip1.5mu \;\rho\;\mskip1.5mu])\;{}^+\!\circ\;{}\<[31]%
\>[31]{}\Varid{⟩}\;{}\<[E]%
\\
\>[B]{}\hsindent{3}{}\<[3]%
\>[3]{}\Varid{x}\;[\mskip1.5mu \;\Varid{id}\;\ensuremath{\mbox{$\circ$}}\;\Varid{ys}\;{}\<[16]%
\>[16]{}\mskip1.5mu]\;{}\<[E]%
\\
\>[B]{}\hsindent{3}{}\<[3]%
\>[3]{}\equiv\!\langle\;\Varid{cong}\;(\lambda\;\rho\;\rightarrow\;\Varid{x}\;[\mskip1.5mu \;\rho\;\mskip1.5mu])\;\Varid{id}\circ\;\Varid{⟩}\;{}\<[E]%
\\
\>[B]{}\hsindent{3}{}\<[3]%
\>[3]{}\Varid{x}\;[\mskip1.5mu \;\Varid{ys}\;\mskip1.5mu]\;{}\<[13]%
\>[13]{}\blacksquare{}\<[E]%
\ColumnHook
\end{hscode}\resethooks
\end{minipage}
\hfill
\begin{minipage}{0.45\textwidth}
\begin{hscode}\SaveRestoreHook
\column{B}{@{}>{\hspre}l<{\hspost}@{}}%
\column{4}{@{}>{\hspre}l<{\hspost}@{}}%
\column{15}{@{}>{\hspre}l<{\hspost}@{}}%
\column{19}{@{}>{\hspre}l<{\hspost}@{}}%
\column{21}{@{}>{\hspre}l<{\hspost}@{}}%
\column{24}{@{}>{\hspre}l<{\hspost}@{}}%
\column{E}{@{}>{\hspre}l<{\hspost}@{}}%
\>[B]{}{}^+\!\circ\;\mathbin{:}\;\Varid{xs}\;^{+}\;\Conid{A}\;\ensuremath{\mbox{$\circ$}}\;(\Varid{ys}\;\Varid{,}\;\Varid{x})\;\equiv\;\Varid{xs}\;\ensuremath{\mbox{$\circ$}}\;\Varid{ys}{}\<[E]%
\\
\>[B]{}{}^+\!\circ\;\{\mskip1.5mu \Varid{xs}\;\mathrel{=}\;\varepsilon\mskip1.5mu\}\;{}\<[19]%
\>[19]{}\mathrel{=}\;\Varid{refl}{}\<[E]%
\\
\>[B]{}{}^+\!\circ\;\{\mskip1.5mu \Varid{xs}\;\mathrel{=}\;\Varid{xs}\;\Varid{,}\;\Varid{x}\mskip1.5mu\}\;{}\<[19]%
\>[19]{}\mathrel{=}\;{}\<[E]%
\\
\>[B]{}\hsindent{4}{}\<[4]%
\>[4]{}\Varid{cong}_{2}\;\_\Varid{,}\_\;{}\<[15]%
\>[15]{}({}^+\!\circ\;\{\mskip1.5mu \Varid{xs}\;\mathrel{=}\;\Varid{xs}\mskip1.5mu\})\;{}\<[E]%
\\
\>[15]{}(\Varid{suc[]}\;\{\mskip1.5mu \Varid{x}\;\mathrel{=}\;\Varid{x}\mskip1.5mu\}){}\<[E]%
\\[\blanklineskip]%
\>[B]{}\Varid{id}\circ^{\prime}\;\{\mskip1.5mu \Varid{xs}\;\mathrel{=}\;\varepsilon\mskip1.5mu\}\;{}\<[21]%
\>[21]{}\anonymous \;\mathrel{=}\;\Varid{refl}{}\<[E]%
\\
\>[B]{}\Varid{id}\circ^{\prime}\;\{\mskip1.5mu \Varid{xs}\;\mathrel{=}\;\Varid{xs}\;\Varid{,}\;\Varid{x}\mskip1.5mu\}\;{}\<[21]%
\>[21]{}\anonymous \;\mathrel{=}\;\Varid{cong}_{2}\;\_\Varid{,}\_\;{}\<[E]%
\\
\>[B]{}\hsindent{4}{}\<[4]%
\>[4]{}(\Varid{id}\;^{+}\;\anonymous \;\ensuremath{\mbox{$\circ$}}\;(\Varid{xs}\;\Varid{,}\;\Varid{x})\;{}\<[24]%
\>[24]{}\equiv\!\langle\;{}^+\!\circ\;\{\mskip1.5mu \Varid{xs}\;\mathrel{=}\;\Varid{id}\mskip1.5mu\}\;\Varid{⟩}{}\<[E]%
\\
\>[B]{}\hsindent{4}{}\<[4]%
\>[4]{}\Varid{id}\;\ensuremath{\mbox{$\circ$}}\;\Varid{xs}\;{}\<[24]%
\>[24]{}\equiv\!\langle\;\Varid{id}\circ\;\Varid{⟩}{}\<[E]%
\\
\>[B]{}\hsindent{4}{}\<[4]%
\>[4]{}\Varid{xs}\;{}\<[24]%
\>[24]{}\blacksquare)\;{}\<[E]%
\\
\>[B]{}\hsindent{4}{}\<[4]%
\>[4]{}\Varid{refl}{}\<[E]%
\ColumnHook
\end{hscode}\resethooks
\end{minipage}

One may note that
\ensuremath{{}^+\!\circ} relies on itself indirectly via \ensuremath{\Varid{suc[]}}. Like with the substitution
operations, termination is justified here by 
the \ensuremath{\Conid{Sort}} decreasing.

\subsection{Associativity}
\label{sec:associativity}
We finally get to the proof of the second functor law
(\ensuremath{[\circ]\;\mathbin{:}\;\Varid{x}\;[\mskip1.5mu \;\Varid{xs}\;\ensuremath{\mbox{$\circ$}}\;\Varid{ys}\;\mskip1.5mu]\;\equiv\;\Varid{x}\;[\mskip1.5mu \;\Varid{xs}\;\mskip1.5mu]\;[\mskip1.5mu \;\Varid{ys}\;\mskip1.5mu]}), the main lemma for
associativity. The main obstacle is that for the \ensuremath{\lambda\_} case; we need the
second functor law for context extension:
\ensuremath{\uparrow\!\circ\;\mathbin{:}\;(\Varid{xs}\;\ensuremath{\mbox{$\circ$}}\;\Varid{ys})\;\uparrow\;\Conid{A}\;\equiv\;(\Varid{xs}\;\uparrow\;\Conid{A})\;\ensuremath{\mbox{$\circ$}}\;(\Varid{ys}\;\uparrow\;\Conid{A})}.

To verify the variable case we also need that \ensuremath{\Varid{tm}\!\sqsubseteq} commutes with substitution,
\ensuremath{\Varid{tm[]}\;\mathbin{:}\;\Varid{tm}\!\sqsubseteq\;\sqsubseteq\!\Varid{t}\;(\Varid{x}\;[\mskip1.5mu \;\Varid{xs}\;\mskip1.5mu])\;\equiv\;(\Varid{tm}\!\sqsubseteq\;\sqsubseteq\!\Varid{t}\;\Varid{x})\;[\mskip1.5mu \;\Varid{xs}\;\mskip1.5mu]},
which is easy to prove by case analysis.

We are now ready to prove \ensuremath{[\circ]} by structural induction:

\begin{hscode}\SaveRestoreHook
\column{B}{@{}>{\hspre}l<{\hspost}@{}}%
\column{4}{@{}>{\hspre}l<{\hspost}@{}}%
\column{20}{@{}>{\hspre}l<{\hspost}@{}}%
\column{29}{@{}>{\hspre}l<{\hspost}@{}}%
\column{31}{@{}>{\hspre}l<{\hspost}@{}}%
\column{32}{@{}>{\hspre}l<{\hspost}@{}}%
\column{42}{@{}>{\hspre}l<{\hspost}@{}}%
\column{43}{@{}>{\hspre}l<{\hspost}@{}}%
\column{62}{@{}>{\hspre}l<{\hspost}@{}}%
\column{E}{@{}>{\hspre}l<{\hspost}@{}}%
\>[B]{}[\circ]\;\{\mskip1.5mu \Varid{x}\;\mathrel{=}\;\Varid{zero}\mskip1.5mu\}\;{}\<[20]%
\>[20]{}\{\mskip1.5mu \Varid{xs}\;\mathrel{=}\;\Varid{xs}\;\Varid{,}\;\Varid{x}\mskip1.5mu\}\;{}\<[42]%
\>[42]{}\mathrel{=}\;\Varid{refl}{}\<[E]%
\\
\>[B]{}[\circ]\;\{\mskip1.5mu \Varid{x}\;\mathrel{=}\;\Varid{suc}\;\Varid{i}\;\anonymous \mskip1.5mu\}\;{}\<[20]%
\>[20]{}\{\mskip1.5mu \Varid{xs}\;\mathrel{=}\;\Varid{xs}\;\Varid{,}\;\Varid{x}\mskip1.5mu\}\;{}\<[42]%
\>[42]{}\mathrel{=}\;[\circ]\;\{\mskip1.5mu \Varid{x}\;\mathrel{=}\;\Varid{i}\mskip1.5mu\}{}\<[E]%
\\
\>[B]{}[\circ]\;\{\mskip1.5mu \Varid{x}\;\mathrel{=}\;\texttt{\textasciigrave}\;\Varid{x}\mskip1.5mu\}\;{}\<[20]%
\>[20]{}\{\mskip1.5mu \Varid{xs}\;\mathrel{=}\;\Varid{xs}\mskip1.5mu\}\;{}\<[31]%
\>[31]{}\{\mskip1.5mu \Varid{ys}\;\mathrel{=}\;\Varid{ys}\mskip1.5mu\}\;{}\<[42]%
\>[42]{}\mathrel{=}\;{}\<[E]%
\\
\>[B]{}\hsindent{4}{}\<[4]%
\>[4]{}\Varid{tm}\!\sqsubseteq\;\sqsubseteq\!\Varid{t}\;(\Varid{x}\;[\mskip1.5mu \;\Varid{xs}\;\ensuremath{\mbox{$\circ$}}\;\Varid{ys}\;\mskip1.5mu])\;{}\<[32]%
\>[32]{}\equiv\!\langle\;\Varid{cong}\;(\Varid{tm}\!\sqsubseteq\;\sqsubseteq\!\Varid{t})\;([\circ]\;\{\mskip1.5mu \Varid{x}\;\mathrel{=}\;\Varid{x}\mskip1.5mu\})\;\Varid{⟩}\;{}\<[E]%
\\
\>[B]{}\hsindent{4}{}\<[4]%
\>[4]{}\Varid{tm}\!\sqsubseteq\;\sqsubseteq\!\Varid{t}\;(\Varid{x}\;[\mskip1.5mu \;\Varid{xs}\;\mskip1.5mu]\;[\mskip1.5mu \;\Varid{ys}\;\mskip1.5mu])\;{}\<[32]%
\>[32]{}\equiv\!\langle\;\Varid{tm[]}\;\{\mskip1.5mu \Varid{x}\;\mathrel{=}\;\Varid{x}\;[\mskip1.5mu \;\Varid{xs}\;\mskip1.5mu]\mskip1.5mu\}\;\Varid{⟩}\;{}\<[E]%
\\
\>[B]{}\hsindent{4}{}\<[4]%
\>[4]{}(\Varid{tm}\!\sqsubseteq\;\sqsubseteq\!\Varid{t}\;(\Varid{x}\;[\mskip1.5mu \;\Varid{xs}\;\mskip1.5mu]))\;[\mskip1.5mu \;\Varid{ys}\;\mskip1.5mu]\;{}\<[32]%
\>[32]{}\blacksquare{}\<[E]%
\\
\>[B]{}[\circ]\;\{\mskip1.5mu \Varid{x}\;\mathrel{=}\;\Varid{t}\;\cdot\;\Varid{u}\mskip1.5mu\}\;{}\<[43]%
\>[43]{}\mathrel{=}\;\Varid{cong}_{2}\;\_\cdot\_\;([\circ]\;\{\mskip1.5mu \Varid{x}\;\mathrel{=}\;\Varid{t}\mskip1.5mu\})\;([\circ]\;\{\mskip1.5mu \Varid{x}\;\mathrel{=}\;\Varid{u}\mskip1.5mu\}){}\<[E]%
\\
\>[B]{}[\circ]\;\{\mskip1.5mu \Varid{x}\;\mathrel{=}\;\lambda\;\Varid{t}\mskip1.5mu\}\;{}\<[20]%
\>[20]{}\{\mskip1.5mu \Varid{xs}\;\mathrel{=}\;\Varid{xs}\mskip1.5mu\}\;{}\<[31]%
\>[31]{}\{\mskip1.5mu \Varid{ys}\;\mathrel{=}\;\Varid{ys}\mskip1.5mu\}\;{}\<[43]%
\>[43]{}\mathrel{=}\;\Varid{cong}\;\lambda\_\;({}\<[E]%
\\
\>[B]{}\hsindent{4}{}\<[4]%
\>[4]{}\Varid{t}\;[\mskip1.5mu \;(\Varid{xs}\;\ensuremath{\mbox{$\circ$}}\;\Varid{ys})\;\uparrow\;\anonymous \;\mskip1.5mu]\;{}\<[32]%
\>[32]{}\equiv\!\langle\;\Varid{cong}\;(\lambda\;\Varid{zs}\;\rightarrow\;\Varid{t}\;[\mskip1.5mu \;\Varid{zs}\;\mskip1.5mu])\;\uparrow\!\circ\;{}\<[62]%
\>[62]{}\Varid{⟩}{}\<[E]%
\\
\>[B]{}\hsindent{4}{}\<[4]%
\>[4]{}\Varid{t}\;[\mskip1.5mu \;(\Varid{xs}\;\uparrow\;\anonymous )\;\ensuremath{\mbox{$\circ$}}\;(\Varid{ys}\;\uparrow\;\anonymous )\;{}\<[29]%
\>[29]{}\mskip1.5mu]\;{}\<[32]%
\>[32]{}\equiv\!\langle\;[\circ]\;\{\mskip1.5mu \Varid{x}\;\mathrel{=}\;\Varid{t}\mskip1.5mu\}\;\Varid{⟩}{}\<[E]%
\\
\>[B]{}\hsindent{4}{}\<[4]%
\>[4]{}(\Varid{t}\;[\mskip1.5mu \;\Varid{xs}\;\uparrow\;\anonymous \;\mskip1.5mu])\;[\mskip1.5mu \;\Varid{ys}\;\uparrow\;\anonymous \;\mskip1.5mu]\;{}\<[32]%
\>[32]{}\blacksquare){}\<[E]%
\ColumnHook
\end{hscode}\resethooks


\noindent
Associativity \ensuremath{\circ\!\circ\;\mathbin{:}\;\Varid{xs}\;\ensuremath{\mbox{$\circ$}}\;(\Varid{ys}\;\ensuremath{\mbox{$\circ$}}\;\Varid{zs})\;\equiv\;(\Varid{xs}\;\ensuremath{\mbox{$\circ$}}\;\Varid{ys})\;\ensuremath{\mbox{$\circ$}}\;\Varid{zs}} can be proven by folding 
\ensuremath{[\circ]} over substitutions.

However, we are not done yet. We still need to prove
the second functor law for \ensuremath{\_\uparrow\_} (\ensuremath{\uparrow\!\circ}). It turns out that this depends on
the naturality of weakening \ensuremath{{}^+\!-\Varid{nat}\circ\;\mathbin{:}\;\Varid{xs}\;\ensuremath{\mbox{$\circ$}}\;(\Varid{ys}\;^{+}\;\Conid{A})\;\equiv\;(\Varid{xs}\;\ensuremath{\mbox{$\circ$}}\;\Varid{ys})\;^{+}\;\Conid{A}},
which unsurprisingly must be shown by establishing a corresponding
property for substitutions: \\\ensuremath{^{+}\Varid{-nat[]}\;\mathbin{:}\;\Varid{x}\;[\mskip1.5mu \;\Varid{xs}\;^{+}\;\Conid{A}\;\mskip1.5mu]\;\equiv\;\Varid{suc}[\;\anonymous \;\mskip1.5mu]\;(\Varid{x}\;[\mskip1.5mu \;\Varid{xs}\;\mskip1.5mu])\;\Conid{A}}.
The case \ensuremath{\Varid{q}\;\mathrel{=}\;\Conid{V}} is just the naturality for variables which we have
already proven (\ensuremath{^{+}\Varid{-nat[]v}}).
The case for \ensuremath{\Varid{q}\;\mathrel{=}\;\Conid{T}} is more interesting and relies again on \ensuremath{[\circ]} and
\ensuremath{\circ\Varid{id}}:
\begin{hscode}\SaveRestoreHook
\column{B}{@{}>{\hspre}l<{\hspost}@{}}%
\column{4}{@{}>{\hspre}l<{\hspost}@{}}%
\column{25}{@{}>{\hspre}l<{\hspost}@{}}%
\column{E}{@{}>{\hspre}l<{\hspost}@{}}%
\>[B]{}^{+}\Varid{-nat[]}\;\{\mskip1.5mu \Varid{q}\;\mathrel{=}\;\Conid{T}\mskip1.5mu\}\;\{\mskip1.5mu \Conid{A}\;\mathrel{=}\;\Conid{A}\mskip1.5mu\}\;\{\mskip1.5mu \Varid{x}\;\mathrel{=}\;\Varid{x}\mskip1.5mu\}\;\{\mskip1.5mu \Varid{xs}\;\mathrel{=}\;\Varid{xs}\mskip1.5mu\}\;\mathrel{=}\;{}\<[E]%
\\
\>[B]{}\hsindent{4}{}\<[4]%
\>[4]{}\Varid{x}\;[\mskip1.5mu \;\Varid{xs}\;^{+}\;\Conid{A}\;\mskip1.5mu]\;{}\<[25]%
\>[25]{}\equiv\!\langle\;\Varid{cong}\;(\lambda\;\Varid{zs}\;\rightarrow\;\Varid{x}\;[\mskip1.5mu \;\Varid{zs}\;^{+}\;\Conid{A}\;\mskip1.5mu])\;(\Varid{sym}\;\circ\Varid{id})\;\Varid{⟩}\;{}\<[E]%
\\
\>[B]{}\hsindent{4}{}\<[4]%
\>[4]{}\Varid{x}\;[\mskip1.5mu \;(\Varid{xs}\;\ensuremath{\mbox{$\circ$}}\;\Varid{id})\;^{+}\;\Conid{A}\;\mskip1.5mu]\;{}\<[25]%
\>[25]{}\equiv\!\langle\;\Varid{cong}\;(\lambda\;\Varid{zs}\;\rightarrow\;\Varid{x}\;[\mskip1.5mu \;\Varid{zs}\;\mskip1.5mu])\;(\Varid{sym}\;({}^+\!-\Varid{nat}\circ\;\{\mskip1.5mu \Varid{xs}\;\mathrel{=}\;\Varid{xs}\mskip1.5mu\}))\;\Varid{⟩}\;{}\<[E]%
\\
\>[B]{}\hsindent{4}{}\<[4]%
\>[4]{}\Varid{x}\;[\mskip1.5mu \;\Varid{xs}\;\ensuremath{\mbox{$\circ$}}\;(\Varid{id}\;^{+}\;\Conid{A})\;\mskip1.5mu]\;{}\<[25]%
\>[25]{}\equiv\!\langle\;[\circ]\;\{\mskip1.5mu \Varid{x}\;\mathrel{=}\;\Varid{x}\mskip1.5mu\}\;\Varid{⟩}\;{}\<[E]%
\\
\>[B]{}\hsindent{4}{}\<[4]%
\>[4]{}\Varid{x}\;[\mskip1.5mu \;\Varid{xs}\;\mskip1.5mu]\;[\mskip1.5mu \;\Varid{id}\;^{+}\;\Conid{A}\;\mskip1.5mu]\;{}\<[25]%
\>[25]{}\blacksquare{}\<[E]%
\ColumnHook
\end{hscode}\resethooks

It also turns out we need 
\ensuremath{\Varid{zero[]}\;\mathbin{:}\;\Varid{zero}[\;\Varid{q}\;\mskip1.5mu]\;[\mskip1.5mu \;\Varid{xs}\;\Varid{,}\;\Varid{x}\;\mskip1.5mu]\;\equiv\;\Varid{tm}\!\sqsubseteq\;(\sqsubseteq\!\sqcup\Varid{r}\;\{\mskip1.5mu \Varid{q}\;\mathrel{=}\;\Varid{q}\mskip1.5mu\})\;\Varid{x}}, the $\beta$-law for 
\ensuremath{\Varid{zero}[\_]}, which holds
definitionally in the case for either \ensuremath{\Conid{Sort}}.
And we need that zero commutes with \ensuremath{\Varid{tm}\!\sqsubseteq}, that is, for any
\ensuremath{\Varid{q}\!\sqsubseteq\!\Varid{r}\;\mathbin{:}\;\Varid{q}\;\sqsubseteq\;\Varid{r}} we have that \ensuremath{\Varid{tm}\!\sqsubseteq\!\Varid{zero}\;\Varid{q}\!\sqsubseteq\!\Varid{r}\;\mathbin{:}\;\Varid{zero}[\;\Varid{r}\;\mskip1.5mu]\;\equiv\;\Varid{tm}\!\sqsubseteq\;\Varid{q}\!\sqsubseteq\!\Varid{r}\;\Varid{zero}[\;\Varid{q}\;\mskip1.5mu]}.

Finally, we have all the ingredients to prove the second functor law \ensuremath{\uparrow\!\circ}:
\begin{hscode}\SaveRestoreHook
\column{B}{@{}>{\hspre}l<{\hspost}@{}}%
\column{5}{@{}>{\hspre}l<{\hspost}@{}}%
\column{7}{@{}>{\hspre}l<{\hspost}@{}}%
\column{17}{@{}>{\hspre}l<{\hspost}@{}}%
\column{36}{@{}>{\hspre}l<{\hspost}@{}}%
\column{50}{@{}>{\hspre}l<{\hspost}@{}}%
\column{80}{@{}>{\hspre}l<{\hspost}@{}}%
\column{E}{@{}>{\hspre}l<{\hspost}@{}}%
\>[B]{}\uparrow\!\circ\;\{\mskip1.5mu \Varid{r}\;\mathrel{=}\;\Varid{r}\mskip1.5mu\}\;\{\mskip1.5mu \Varid{s}\;\mathrel{=}\;\Varid{s}\mskip1.5mu\}\;\{\mskip1.5mu \Varid{xs}\;\mathrel{=}\;\Varid{xs}\mskip1.5mu\}\;\{\mskip1.5mu \Varid{ys}\;\mathrel{=}\;\Varid{ys}\mskip1.5mu\}\;\{\mskip1.5mu \Conid{A}\;\mathrel{=}\;\Conid{A}\mskip1.5mu\}\;\mathrel{=}\;{}\<[E]%
\\
\>[B]{}\hsindent{5}{}\<[5]%
\>[5]{}(\Varid{xs}\;\ensuremath{\mbox{$\circ$}}\;\Varid{ys})\;\uparrow\;\Conid{A}\;{}\<[36]%
\>[36]{}\equiv\!\langle\rangle\;{}\<[E]%
\\
\>[B]{}\hsindent{5}{}\<[5]%
\>[5]{}(\Varid{xs}\;\ensuremath{\mbox{$\circ$}}\;\Varid{ys})\;^{+}\;\Conid{A}\;\Varid{,}\;\Varid{zero}[\;\Varid{r}\;\sqcup\;\Varid{s}\;\mskip1.5mu]\;{}\<[36]%
\>[36]{}\equiv\!\langle\;\Varid{cong}_{2}\;\_\Varid{,}\_\;(\Varid{sym}\;({}^+\!-\Varid{nat}\circ\;\{\mskip1.5mu \Varid{xs}\;\mathrel{=}\;\Varid{xs}\mskip1.5mu\}))\;\Varid{refl}\;\Varid{⟩}\;{}\<[E]%
\\
\>[B]{}\hsindent{5}{}\<[5]%
\>[5]{}\Varid{xs}\;\ensuremath{\mbox{$\circ$}}\;(\Varid{ys}\;^{+}\;\Conid{A})\;\Varid{,}\;\Varid{zero}[\;\Varid{r}\;\sqcup\;\Varid{s}\;\mskip1.5mu]\;{}\<[36]%
\>[36]{}\equiv\!\langle\;\Varid{cong}_{2}\;\_\Varid{,}\_\;\Varid{refl}\;(\Varid{tm}\!\sqsubseteq\!\Varid{zero}\;(\sqsubseteq\!\sqcup\Varid{r}\;\{\mskip1.5mu \Varid{r}\;\mathrel{=}\;\Varid{s}\mskip1.5mu\}\;\{\mskip1.5mu \Varid{q}\;\mathrel{=}\;\Varid{r}\mskip1.5mu\}))\;\Varid{⟩}\;{}\<[E]%
\\
\>[B]{}\hsindent{5}{}\<[5]%
\>[5]{}\Varid{xs}\;\ensuremath{\mbox{$\circ$}}\;(\Varid{ys}\;^{+}\;\Conid{A})\;\Varid{,}\;\Varid{tm}\!\sqsubseteq\;(\sqsubseteq\!\sqcup\Varid{r}\;\{\mskip1.5mu \Varid{q}\;\mathrel{=}\;\Varid{r}\mskip1.5mu\})\;\Varid{zero}[\;\Varid{s}\;\mskip1.5mu]\;{}\<[E]%
\\
\>[5]{}\hsindent{2}{}\<[7]%
\>[7]{}\equiv\!\langle\;\Varid{cong}_{2}\;\_\Varid{,}\_\;(\Varid{sym}\;({}^+\!\circ\;\{\mskip1.5mu \Varid{xs}\;\mathrel{=}\;\Varid{xs}\mskip1.5mu\}))\;(\Varid{sym}\;(\Varid{zero[]}\;\{\mskip1.5mu \Varid{q}\;\mathrel{=}\;\Varid{r}\mskip1.5mu\}\;\{\mskip1.5mu \Varid{x}\;\mathrel{=}\;\Varid{zero}[\;\Varid{s}\;\mskip1.5mu]\mskip1.5mu\}))\;{}\<[80]%
\>[80]{}\Varid{⟩}\;{}\<[E]%
\\
\>[B]{}\hsindent{5}{}\<[5]%
\>[5]{}(\Varid{xs}\;^{+}\;\Conid{A})\;\ensuremath{\mbox{$\circ$}}\;{}\<[17]%
\>[17]{}(\Varid{ys}\;\uparrow\;\Conid{A})\;\Varid{,}\;\Varid{zero}[\;\Varid{r}\;\mskip1.5mu]\;[\mskip1.5mu \;\Varid{ys}\;\uparrow\;\Conid{A}\;\mskip1.5mu]\;{}\<[50]%
\>[50]{}\equiv\!\langle\rangle\;{}\<[E]%
\\
\>[B]{}\hsindent{5}{}\<[5]%
\>[5]{}(\Varid{xs}\;\uparrow\;\Conid{A})\;\ensuremath{\mbox{$\circ$}}\;(\Varid{ys}\;\uparrow\;\Conid{A})\;{}\<[50]%
\>[50]{}\blacksquare{}\<[E]%
\ColumnHook
\end{hscode}\resethooks

\section{Initiality}
\label{sec:initiality}

We can do more than just prove that we have a category. Indeed we
can verify the laws of a simply typed category with families
(CwF). CwFs are mostly known as models of dependent type theory, but
they can be specialised to simple types \cite{castellan2021categories}. We 
summarise the definition of a simply typed CwF as follows:

\vspace{-1ex}
\begin{itemize}
\item A category of contexts (\ensuremath{\Conid{Con}}) and substitutions (\ensuremath{\_\Vdash\_}),
\item A set of types \ensuremath{\Conid{Ty}},
\item For every type \ensuremath{\Conid{A}} a presheaf of terms \ensuremath{\anonymous \;\vdash\;\Conid{A}} over the category of contexts (i.e. a
  contravariant functor into the category of sets),
\item A terminal object (the empty context) and a context extension
  operation \ensuremath{\_\rhd\_} such that \ensuremath{\Gamma\;\Vdash\;\Delta\;\rhd\;\Conid{A}} is naturally isomorphic to
  \ensuremath{(\Gamma\;\Vdash\;\Delta)\;\times\;(\Gamma\;\vdash\;\Conid{A}}).
\end{itemize}

\vspace{-1ex}
That is, a simply typed CwF is just a CwF where the presheaf of types is constant.
We will give the precise definition in the next section, hence it
isn't necessary to be familiar with the categorical terminology to follow the 
rest of the paper. 

We can add further constructors like function types \ensuremath{\_\Rightarrow\_}. These usually
come with a natural isomorphisms, giving rise to $\beta$ and $\eta$ laws,
but since we are only interested in substitutions, we don't assume these. 
Instead we add the term formers for application
(\ensuremath{\_\cdot\_}) and lambda-abstraction \ensuremath{\lambda} as natural transformations.



We start with a precise definition of a simply typed CwF with the
additional structure to model simply typed $\lambda$-calculus (Section
\ref{sec:simply-typed-cwfs}) and then we show that the recursive
definition of substitution gives rise to a simply typed CwF (Section
\ref{sec:cwf-recurs-subst}). We can define the initial CwF as a
quotient inductive-inductive type (QIIT).
We postulate the existence of this QIIT in Agda, with
the associated $\beta$-laws implemented with rewrite rules
(alternatively, we could use a truncated Cubical Agda HIT,
but Cubical Agda still lacks essential automation,
e.g. it does not integrate no-confusion properties into pattern matching).
By initiality, there is an evaluation
functor from the initial CwF to the recursively defined CwF (defined
in Section \ref{sec:cwf-recurs-subst}). On the
other hand, we can embed the recursive CwF into the initial CwF;
this corresponds to the embedding of normal forms into
$\lambda$-terms, only that here we talk about \emph{substitution normal
forms}. We then show that these two structure maps are inverse to each
other and
hence that the recursively defined CwF is indeed initial (Section
\ref{sec:proving-initiality}). The two identities correspond to
completeness and stability in the language of normalisation functions.  

\subsection{Simply Typed CwFs}
\label{sec:simply-typed-cwfs}

We define a record to capture simply typed CWFs,
\ensuremath{\Keyword{record}\;\Conid{CwF-simple}\;\mathbin{:}\;\Varid{Set}_1}.

For the contents, we begin with the category of contexts, using the 
same naming conventions as introduced previously:

\begin{minipage}{0.5\textwidth}
\begin{hscode}\SaveRestoreHook
\column{B}{@{}>{\hspre}l<{\hspost}@{}}%
\column{3}{@{}>{\hspre}l<{\hspost}@{}}%
\column{8}{@{}>{\hspre}l<{\hspost}@{}}%
\column{E}{@{}>{\hspre}l<{\hspost}@{}}%
\>[3]{}\Conid{Con}\;{}\<[8]%
\>[8]{}\mathbin{:}\;\Conid{Set}{}\<[E]%
\\
\>[3]{}\_\Vdash\_\;{}\<[8]%
\>[8]{}\mathbin{:}\;\Conid{Con}\;\rightarrow\;\Conid{Con}\;\rightarrow\;\Conid{Set}{}\<[E]%
\\[\blanklineskip]%
\>[3]{}\Varid{id}\;{}\<[8]%
\>[8]{}\mathbin{:}\;\Gamma\;\Vdash\;\Gamma{}\<[E]%
\\
\>[3]{}\_\ensuremath{\mbox{$\circ$}}\_\;{}\<[8]%
\>[8]{}\mathbin{:}\;\Delta\;\Vdash\;\Theta\;\rightarrow\;\Gamma\;\Vdash\;\Delta\;\rightarrow\;\Gamma\;\Vdash\;\Theta{}\<[E]%
\ColumnHook
\end{hscode}\resethooks
\end{minipage}
\begin{minipage}{0.45\textwidth}
\begin{hscode}\SaveRestoreHook
\column{B}{@{}>{\hspre}l<{\hspost}@{}}%
\column{3}{@{}>{\hspre}l<{\hspost}@{}}%
\column{8}{@{}>{\hspre}l<{\hspost}@{}}%
\column{E}{@{}>{\hspre}l<{\hspost}@{}}%
\>[3]{}\Varid{id}\circ\;{}\<[8]%
\>[8]{}\mathbin{:}\;\Varid{id}\;\ensuremath{\mbox{$\circ$}}\;\delta\;\equiv\;\delta{}\<[E]%
\\
\>[3]{}\circ\Varid{id}\;{}\<[8]%
\>[8]{}\mathbin{:}\;\delta\;\ensuremath{\mbox{$\circ$}}\;\Varid{id}\;\equiv\;\delta{}\<[E]%
\\
\>[3]{}\circ\!\circ\;{}\<[8]%
\>[8]{}\mathbin{:}\;(\xi\;\ensuremath{\mbox{$\circ$}}\;\theta)\;\ensuremath{\mbox{$\circ$}}\;\delta\;\equiv\;\xi\;\ensuremath{\mbox{$\circ$}}\;(\theta\;\ensuremath{\mbox{$\circ$}}\;\delta){}\<[E]%
\ColumnHook
\end{hscode}\resethooks
\end{minipage}\\
We introduce the set of types and associate a presheaf with each type:

\begin{minipage}{0.45\textwidth}
\begin{hscode}\SaveRestoreHook
\column{B}{@{}>{\hspre}l<{\hspost}@{}}%
\column{3}{@{}>{\hspre}l<{\hspost}@{}}%
\column{9}{@{}>{\hspre}l<{\hspost}@{}}%
\column{E}{@{}>{\hspre}l<{\hspost}@{}}%
\>[3]{}\Conid{Ty}\;{}\<[9]%
\>[9]{}\mathbin{:}\;\Conid{Set}{}\<[E]%
\\
\>[3]{}\_\vdash\_\;{}\<[9]%
\>[9]{}\mathbin{:}\;\Conid{Con}\;\rightarrow\;\Conid{Ty}\;\rightarrow\;\Conid{Set}{}\<[E]%
\\[\blanklineskip]%
\>[3]{}\_[\_]\;{}\<[9]%
\>[9]{}\mathbin{:}\;\Gamma\;\vdash\;\Conid{A}\;\rightarrow\;\Delta\;\Vdash\;\Gamma\;\rightarrow\;\Delta\;\vdash\;\Conid{A}{}\<[E]%
\ColumnHook
\end{hscode}\resethooks
\end{minipage}
\begin{minipage}{0.45\textwidth}
\begin{hscode}\SaveRestoreHook
\column{B}{@{}>{\hspre}l<{\hspost}@{}}%
\column{3}{@{}>{\hspre}l<{\hspost}@{}}%
\column{9}{@{}>{\hspre}l<{\hspost}@{}}%
\column{E}{@{}>{\hspre}l<{\hspost}@{}}%
\>[3]{}\Varid{[id]}\;{}\<[9]%
\>[9]{}\mathbin{:}\;(\Varid{t}\;[\mskip1.5mu \;\Varid{id}\;\mskip1.5mu])\;\equiv\;\Varid{t}{}\<[E]%
\\
\>[3]{}[\circ]\;{}\<[9]%
\>[9]{}\mathbin{:}\;\Varid{t}\;[\mskip1.5mu \;\theta\;\mskip1.5mu]\;[\mskip1.5mu \;\delta\;\mskip1.5mu]\;\equiv\;\Varid{t}\;[\mskip1.5mu \;\theta\;\ensuremath{\mbox{$\circ$}}\;\delta\;\mskip1.5mu]{}\<[E]%
\ColumnHook
\end{hscode}\resethooks
\end{minipage}\\
The category of contexts has a terminal object (the empty context), and
context extension resembles categorical products but mixing contexts
and types:

\noindent
\begin{minipage}{0.55\textwidth}
\begin{hscode}\SaveRestoreHook
\column{B}{@{}>{\hspre}l<{\hspost}@{}}%
\column{3}{@{}>{\hspre}l<{\hspost}@{}}%
\column{7}{@{}>{\hspre}l<{\hspost}@{}}%
\column{8}{@{}>{\hspre}l<{\hspost}@{}}%
\column{E}{@{}>{\hspre}l<{\hspost}@{}}%
\>[3]{}\Varid{•}\;{}\<[7]%
\>[7]{}\mathbin{:}\;\Conid{Con}{}\<[E]%
\\
\>[3]{}\varepsilon\;{}\<[7]%
\>[7]{}\mathbin{:}\;\Gamma\;\Vdash\;\Varid{•}{}\<[E]%
\\[\blanklineskip]%
\>[3]{}\_\rhd\_\;{}\<[8]%
\>[8]{}\mathbin{:}\;\Conid{Con}\;\rightarrow\;\Conid{Ty}\;\rightarrow\;\Conid{Con}{}\<[E]%
\\
\>[3]{}\_\Varid{,}\_\;{}\<[8]%
\>[8]{}\mathbin{:}\;\Gamma\;\Vdash\;\Delta\;\rightarrow\;\Gamma\;\vdash\;\Conid{A}\;\rightarrow\;\Gamma\;\Vdash\;(\Delta\;\rhd\;\Conid{A}){}\<[E]%
\\
\>[3]{}\pi_0\;{}\<[8]%
\>[8]{}\mathbin{:}\;\Gamma\;\Vdash\;(\Delta\;\rhd\;\Conid{A})\;\rightarrow\;\Gamma\;\Vdash\;\Delta{}\<[E]%
\\
\>[3]{}\pi_1\;{}\<[8]%
\>[8]{}\mathbin{:}\;\Gamma\;\Vdash\;(\Delta\;\rhd\;\Conid{A})\;\rightarrow\;\Gamma\;\vdash\;\Conid{A}{}\<[E]%
\ColumnHook
\end{hscode}\resethooks
\end{minipage}
\begin{minipage}{0.4\textwidth}
\begin{hscode}\SaveRestoreHook
\column{B}{@{}>{\hspre}l<{\hspost}@{}}%
\column{3}{@{}>{\hspre}l<{\hspost}@{}}%
\column{9}{@{}>{\hspre}l<{\hspost}@{}}%
\column{E}{@{}>{\hspre}l<{\hspost}@{}}%
\>[3]{}\Varid{•}\!-\!\eta\;\mathbin{:}\;\delta\;\equiv\;\varepsilon{}\<[E]%
\\[\blanklineskip]%
\>[3]{}\rhd\!-\!\beta_0\;{}\<[9]%
\>[9]{}\mathbin{:}\;\pi_0\;(\delta\;\Varid{,}\;\Varid{t})\;\equiv\;\delta{}\<[E]%
\\
\>[3]{}\rhd\!-\!\beta_1\;{}\<[9]%
\>[9]{}\mathbin{:}\;\pi_1\;(\delta\;\Varid{,}\;\Varid{t})\;\equiv\;\Varid{t}{}\<[E]%
\\
\>[3]{}\rhd\!-\!\eta\;{}\<[9]%
\>[9]{}\mathbin{:}\;(\pi_0\;\delta\;\Varid{,}\;\pi_1\;\delta)\;\equiv\;\delta{}\<[E]%
\\
\>[3]{}\pi_0\!\circ\;{}\<[9]%
\>[9]{}\mathbin{:}\;\pi_0\;(\theta\;\ensuremath{\mbox{$\circ$}}\;\delta)\;\equiv\;\pi_0\;\theta\;\ensuremath{\mbox{$\circ$}}\;\delta{}\<[E]%
\\
\>[3]{}\pi_1\!\circ\;{}\<[9]%
\>[9]{}\mathbin{:}\;\pi_1\;(\theta\;\ensuremath{\mbox{$\circ$}}\;\delta)\;\equiv\;(\pi_1\;\theta)\;[\mskip1.5mu \;\delta\;\mskip1.5mu]{}\<[E]%
\ColumnHook
\end{hscode}\resethooks
\end{minipage}\\
We can define the morphism part of the context extension functor as
before, \ensuremath{\delta\;\uparrow\;\Conid{A}\;\mathrel{=}\;(\delta\;\ensuremath{\mbox{$\circ$}}\;(\pi_0\;\Varid{id}))\;\Varid{,}\;\pi_1\;\Varid{id}}.
We need to add the specific components for simply typed
$\lambda$-calculus; we add the type constructors, the term
constructors and the corresponding naturality laws:

\noindent
\begin{minipage}{0.525\textwidth}
\begin{hscode}\SaveRestoreHook
\column{B}{@{}>{\hspre}l<{\hspost}@{}}%
\column{3}{@{}>{\hspre}l<{\hspost}@{}}%
\column{8}{@{}>{\hspre}l<{\hspost}@{}}%
\column{E}{@{}>{\hspre}l<{\hspost}@{}}%
\>[3]{}\Varid{o}\;{}\<[8]%
\>[8]{}\mathbin{:}\;\Conid{Ty}{}\<[E]%
\\
\>[3]{}\_\Rightarrow\_\;{}\<[8]%
\>[8]{}\mathbin{:}\;\Conid{Ty}\;\rightarrow\;\Conid{Ty}\;\rightarrow\;\Conid{Ty}{}\<[E]%
\\
\>[3]{}\_\cdot\_\;{}\<[8]%
\>[8]{}\mathbin{:}\;\Gamma\;\vdash\;\Conid{A}\;\Rightarrow\;\Conid{B}\;\rightarrow\;\Gamma\;\vdash\;\Conid{A}\;\rightarrow\;\Gamma\;\vdash\;\Conid{B}{}\<[E]%
\ColumnHook
\end{hscode}\resethooks
\end{minipage}
\begin{minipage}{0.4\textwidth}
\begin{hscode}\SaveRestoreHook
\column{B}{@{}>{\hspre}l<{\hspost}@{}}%
\column{3}{@{}>{\hspre}l<{\hspost}@{}}%
\column{8}{@{}>{\hspre}l<{\hspost}@{}}%
\column{E}{@{}>{\hspre}l<{\hspost}@{}}%
\>[3]{}\lambda\_\;{}\<[8]%
\>[8]{}\mathbin{:}\;\Gamma\;\rhd\;\Conid{A}\;\vdash\;\Conid{B}\;\rightarrow\;\Gamma\;\vdash\;\Conid{A}\;\Rightarrow\;\Conid{B}{}\<[E]%
\\
\>[3]{}\cdot[]\;{}\<[8]%
\>[8]{}\mathbin{:}\;(\Varid{t}\;\cdot\;\Varid{u})\;[\mskip1.5mu \;\delta\;\mskip1.5mu]\;\equiv\;(\Varid{t}\;[\mskip1.5mu \;\delta\;\mskip1.5mu])\;\cdot\;(\Varid{u}\;[\mskip1.5mu \;\delta\;\mskip1.5mu]){}\<[E]%
\\
\>[3]{}\lambda[]\;{}\<[8]%
\>[8]{}\mathbin{:}\;(\lambda\;\Varid{t})\;[\mskip1.5mu \;\delta\;\mskip1.5mu]\;\equiv\;\lambda\;(\Varid{t}\;[\mskip1.5mu \;\delta\;\uparrow\;\anonymous \;\mskip1.5mu]){}\<[E]%
\ColumnHook
\end{hscode}\resethooks
\end{minipage}

\subsection{The CwF of recursive substitutions}
\label{sec:cwf-recurs-subst}

We are building towards a proof of initiality for our recursive substitution
syntax, but shall start by showing that our recursive substitution syntax obeys 
the specified CwF laws, specifically that \ensuremath{\Conid{CwF-simple}} can be instantiated with 
\ensuremath{\_\vdash\![\_]\_}/\ensuremath{\_\Vdash\![\_]\_}. This will be more-or-less enough to implement the 
``normalisation'' direction of our initial CwF \ensuremath{\simeq} recursive substitution syntax 
isomorphism.

Most of the work to prove these laws was already done in Section
\ref{sec:proving-laws} but there are a couple tricky details with fitting
into the exact structure the \ensuremath{\Conid{CwF-simple}} record requires.

Our first non-trivial decision is which type family to interpret substitutions 
into. 
In our first attempt, we tried to pair renamings/substitutions with their sorts 
to stay polymorphic, \ensuremath{\Varid{is-cwf}\;\Varid{.CwF.}\_\Vdash\_\;\mathrel{=}\;\Sigma\;\Conid{Sort}\;(\Delta\;\Vdash\![\_]\;\Gamma)}.
Unfortunately, this approach quickly breaks. The \ensuremath{\Varid{•}\!-\!\eta} CwF law forces us to 
provide a 
unique morphism to the terminal context (i.e. a unique weakening from the empty 
context); \ensuremath{\Sigma\;\Conid{Sort}\;(\Delta\;\Vdash\![\_]\;\Gamma)} is simply too flexible here, allowing
both \ensuremath{\Conid{V}\;\Varid{,}\;\varepsilon} and \ensuremath{\Conid{T}\;\Varid{,}\;\varepsilon}.

\noindent
Therefore, we instead fix the sort to \ensuremath{\Conid{T}}.

\noindent
\begin{minipage}{0.45\textwidth}
\begin{hscode}\SaveRestoreHook
\column{B}{@{}>{\hspre}l<{\hspost}@{}}%
\column{3}{@{}>{\hspre}l<{\hspost}@{}}%
\column{20}{@{}>{\hspre}l<{\hspost}@{}}%
\column{E}{@{}>{\hspre}l<{\hspost}@{}}%
\>[3]{}\Varid{is-cwf}\;\Varid{.CwF.}\_\Vdash\_\;{}\<[20]%
\>[20]{}\mathrel{=}\;\_\Vdash\![\;\Conid{T}\;]\_{}\<[E]%
\\
\>[3]{}\Varid{is-cwf}\;\Varid{.CwF.•}\;{}\<[20]%
\>[20]{}\mathrel{=}\;\Varid{•}{}\<[E]%
\\
\>[3]{}\Varid{is-cwf}\;\Varid{.CwF.}\varepsilon\;{}\<[20]%
\>[20]{}\mathrel{=}\;\varepsilon{}\<[E]%
\ColumnHook
\end{hscode}\resethooks
\end{minipage}
\begin{minipage}{0.45\textwidth}
\begin{hscode}\SaveRestoreHook
\column{B}{@{}>{\hspre}l<{\hspost}@{}}%
\column{3}{@{}>{\hspre}l<{\hspost}@{}}%
\column{28}{@{}>{\hspre}l<{\hspost}@{}}%
\column{E}{@{}>{\hspre}l<{\hspost}@{}}%
\>[3]{}\Varid{is-cwf}\;\Varid{.CwF.}\Varid{•}\!-\!\eta\;\{\mskip1.5mu \delta\;\mathrel{=}\;\varepsilon\mskip1.5mu\}\;{}\<[28]%
\>[28]{}\mathrel{=}\;\Varid{refl}{}\<[E]%
\\
\>[3]{}\Varid{is-cwf}\;\Varid{.CwF.}\_\ensuremath{\mbox{$\circ$}}\_\;{}\<[28]%
\>[28]{}\mathrel{=}\;\_\ensuremath{\mbox{$\circ$}}\_{}\<[E]%
\\
\>[3]{}\Varid{is-cwf}\;\Varid{.CwF.}\circ\!\circ\;{}\<[28]%
\>[28]{}\mathrel{=}\;\Varid{sym}\;\circ\!\circ{}\<[E]%
\ColumnHook
\end{hscode}\resethooks
\end{minipage}

The lack of flexibility over sorts when constructing substitutions does, 
however, make identity a little trickier. 
\ensuremath{\Varid{id}} doesn't fit \ensuremath{\Conid{CwF.id}} directly as it produces a renaming \ensuremath{\Gamma\;\Vdash\![\;\Conid{V}\;\mskip1.5mu]\;\Gamma}. 
We need the equivalent substitution \ensuremath{\Gamma\;\Vdash\![\;\Conid{T}\;\mskip1.5mu]\;\Gamma}.

We first extend \ensuremath{\Varid{tm}\!\sqsubseteq} to renamings/substitutions: 
\ensuremath{\Varid{tm}\Varid{*}\!\sqsubseteq\;\mathbin{:}\;\Varid{q}\;\sqsubseteq\;\Varid{s}\;\rightarrow\;\Gamma\;\Vdash\![\;\Varid{q}\;\mskip1.5mu]\;\Delta\;\rightarrow\;\Gamma\;\Vdash\![\;\Varid{s}\;\mskip1.5mu]\;\Delta} with a fold, and prove various lemmas 
about how \ensuremath{\Varid{tm}\Varid{*}\!\sqsubseteq} coercions can be lifted outside of our substitution operators:

\noindent
\begin{minipage}{0.4\textwidth}
\begin{hscode}\SaveRestoreHook
\column{B}{@{}>{\hspre}l<{\hspost}@{}}%
\column{3}{@{}>{\hspre}l<{\hspost}@{}}%
\column{8}{@{}>{\hspre}l<{\hspost}@{}}%
\column{29}{@{}>{\hspre}l<{\hspost}@{}}%
\column{E}{@{}>{\hspre}l<{\hspost}@{}}%
\>[3]{}\sqsubseteq\!\circ\;{}\<[8]%
\>[8]{}\mathbin{:}\;\Varid{tm}\Varid{*}\!\sqsubseteq\;\Varid{v}\!\sqsubseteq\!\Varid{t}\;\Varid{xs}\;\ensuremath{\mbox{$\circ$}}\;\Varid{ys}\;{}\<[29]%
\>[29]{}\equiv\;\Varid{xs}\;\ensuremath{\mbox{$\circ$}}\;\Varid{ys}{}\<[E]%
\\
\>[3]{}\circ\!\sqsubseteq\;{}\<[8]%
\>[8]{}\mathbin{:}\;\Varid{xs}\;\ensuremath{\mbox{$\circ$}}\;\Varid{tm}\Varid{*}\!\sqsubseteq\;\Varid{v}\!\sqsubseteq\!\Varid{t}\;\Varid{ys}\;{}\<[29]%
\>[29]{}\equiv\;\Varid{xs}\;\ensuremath{\mbox{$\circ$}}\;\Varid{ys}{}\<[E]%
\\
\>[3]{}\Varid{t}[\sqsubseteq]\;\mathbin{:}\;\Varid{t}\;[\mskip1.5mu \;\Varid{tm}\Varid{*}\!\sqsubseteq\;\Varid{v}\!\sqsubseteq\!\Varid{t}\;\Varid{ys}\;\mskip1.5mu]\;{}\<[29]%
\>[29]{}\equiv\;\Varid{t}\;[\mskip1.5mu \;\Varid{ys}\;\mskip1.5mu]{}\<[E]%
\ColumnHook
\end{hscode}\resethooks
\end{minipage}
\begin{minipage}{0.55\textwidth}
\begin{hscode}\SaveRestoreHook
\column{B}{@{}>{\hspre}l<{\hspost}@{}}%
\column{3}{@{}>{\hspre}l<{\hspost}@{}}%
\column{9}{@{}>{\hspre}l<{\hspost}@{}}%
\column{30}{@{}>{\hspre}l<{\hspost}@{}}%
\column{E}{@{}>{\hspre}l<{\hspost}@{}}%
\>[3]{}\sqsubseteq^{+}\;{}\<[9]%
\>[9]{}\mathbin{:}\;\Varid{tm}\Varid{*}\!\sqsubseteq\;\sqsubseteq\!\Varid{t}\;\Varid{xs}\;^{+}\;\Conid{A}\;{}\<[30]%
\>[30]{}\equiv\;\Varid{tm}\Varid{*}\!\sqsubseteq\;\Varid{v}\!\sqsubseteq\!\Varid{t}\;(\Varid{xs}\;^{+}\;\Conid{A}){}\<[E]%
\\
\>[3]{}\sqsubseteq\uparrow\;{}\<[9]%
\>[9]{}\mathbin{:}\;\Varid{tm}\Varid{*}\!\sqsubseteq\;\Varid{v}\!\sqsubseteq\!\Varid{t}\;\Varid{xs}\;\uparrow\;\Conid{A}\;{}\<[30]%
\>[30]{}\equiv\;\Varid{tm}\Varid{*}\!\sqsubseteq\;\Varid{v}\!\sqsubseteq\!\Varid{t}\;(\Varid{xs}\;\uparrow\;\Conid{A}){}\<[E]%
\\
\>[3]{}\Varid{v}[\sqsubseteq]\;{}\<[9]%
\>[9]{}\mathbin{:}\;\Varid{i}\;[\mskip1.5mu \;\Varid{tm}\Varid{*}\!\sqsubseteq\;\Varid{v}\!\sqsubseteq\!\Varid{t}\;\Varid{ys}\;\mskip1.5mu]\;{}\<[30]%
\>[30]{}\equiv\;\Varid{tm}\!\sqsubseteq\;\Varid{v}\!\sqsubseteq\!\Varid{t}\;\Varid{i}\;[\mskip1.5mu \;\Varid{ys}\;\mskip1.5mu]{}\<[E]%
\ColumnHook
\end{hscode}\resethooks
\end{minipage}

Most of these are proofs come out easily by induction on terms and 
substitutions so we skip over them.
Perhaps worth noting though is that \ensuremath{\sqsubseteq^{+}} requires folding over substitutions
using one new law, relating our two
ways of weakening variables.

\begin{hscode}\SaveRestoreHook
\column{B}{@{}>{\hspre}l<{\hspost}@{}}%
\column{3}{@{}>{\hspre}l<{\hspost}@{}}%
\column{31}{@{}>{\hspre}l<{\hspost}@{}}%
\column{49}{@{}>{\hspre}l<{\hspost}@{}}%
\column{E}{@{}>{\hspre}l<{\hspost}@{}}%
\>[3]{}\Varid{suc}[\Varid{id}^{+}]\;\mathbin{:}\;\Varid{i}\;[\mskip1.5mu \;\Varid{id}\;^{+}\;\Conid{A}\;\mskip1.5mu]\;\equiv\;\Varid{suc}\;\Varid{i}\;\Conid{A}{}\<[E]%
\\
\>[3]{}\Varid{suc}[\Varid{id}^{+}]\;\{\mskip1.5mu \Varid{i}\;\mathrel{=}\;\Varid{i}\mskip1.5mu\}\;\{\mskip1.5mu \Conid{A}\;\mathrel{=}\;\Conid{A}\mskip1.5mu\}\;\mathrel{=}\;{}\<[31]%
\>[31]{}\Varid{i}\;[\mskip1.5mu \;\Varid{id}\;^{+}\;\Conid{A}\;\mskip1.5mu]\;{}\<[49]%
\>[49]{}\equiv\!\langle\;^{+}\Varid{-nat[]v}\;\{\mskip1.5mu \Varid{i}\;\mathrel{=}\;\Varid{i}\mskip1.5mu\}\;\Varid{⟩}\;{}\<[E]%
\\
\>[31]{}\Varid{suc}\;(\Varid{i}\;[\mskip1.5mu \;\Varid{id}\;\mskip1.5mu])\;\Conid{A}\;{}\<[49]%
\>[49]{}\equiv\!\langle\;\Varid{cong}\;(\lambda\;\Varid{j}\;\rightarrow\;\Varid{suc}\;\Varid{j}\;\Conid{A})\;\Varid{[id]}\;\Varid{⟩}\;{}\<[E]%
\\
\>[31]{}\Varid{suc}\;\Varid{i}\;\Conid{A}\;\blacksquare{}\<[E]%
\ColumnHook
\end{hscode}\resethooks

We can now build an identity substitution by coercing the
identity renaming: \ensuremath{\Varid{is-cwf}\;\Varid{.CwF.id}\;\mathrel{=}\;\Varid{tm}\Varid{*}\!\sqsubseteq\;\Varid{v}\!\sqsubseteq\!\Varid{t}\;\Varid{id}}.
The left and right identity CwF laws take the form \ensuremath{\Varid{tm}\Varid{*}\!\sqsubseteq\;\Varid{v}\!\sqsubseteq\!\Varid{t}\;\Varid{id}\;\ensuremath{\mbox{$\circ$}}\;\delta\;\equiv\;\delta}
and \ensuremath{\delta\;\ensuremath{\mbox{$\circ$}}\;\Varid{tm}\Varid{*}\!\sqsubseteq\;\Varid{v}\!\sqsubseteq\!\Varid{t}\;\Varid{id}\;\equiv\;\delta}. This is where we can take full advantage of the 
\ensuremath{\Varid{tm}\Varid{*}\!\sqsubseteq} machinery; these lemmas let us reuse our existing \ensuremath{\Varid{id}\circ}/\ensuremath{\circ\Varid{id}} proofs!

\begin{minipage}{0.45\textwidth}
\begin{hscode}\SaveRestoreHook
\column{B}{@{}>{\hspre}l<{\hspost}@{}}%
\column{3}{@{}>{\hspre}l<{\hspost}@{}}%
\column{5}{@{}>{\hspre}l<{\hspost}@{}}%
\column{22}{@{}>{\hspre}l<{\hspost}@{}}%
\column{E}{@{}>{\hspre}l<{\hspost}@{}}%
\>[3]{}\Varid{is-cwf}\;\Varid{.CwF.}\Varid{id}\circ\;\{\mskip1.5mu \delta\;\mathrel{=}\;\delta\mskip1.5mu\}\;\mathrel{=}\;{}\<[E]%
\\
\>[3]{}\hsindent{2}{}\<[5]%
\>[5]{}\Varid{tm}\Varid{*}\!\sqsubseteq\;\Varid{v}\!\sqsubseteq\!\Varid{t}\;\Varid{id}\;\ensuremath{\mbox{$\circ$}}\;\delta\;{}\<[22]%
\>[22]{}\equiv\!\langle\;\sqsubseteq\!\circ\;\Varid{⟩}\;{}\<[E]%
\\
\>[3]{}\hsindent{2}{}\<[5]%
\>[5]{}\Varid{id}\;\ensuremath{\mbox{$\circ$}}\;\delta\;{}\<[22]%
\>[22]{}\equiv\!\langle\;\Varid{id}\circ\;\Varid{⟩}\;{}\<[E]%
\\
\>[3]{}\hsindent{2}{}\<[5]%
\>[5]{}\delta\;\blacksquare{}\<[E]%
\ColumnHook
\end{hscode}\resethooks
\end{minipage}
\begin{minipage}{0.45\textwidth}
\begin{hscode}\SaveRestoreHook
\column{B}{@{}>{\hspre}l<{\hspost}@{}}%
\column{3}{@{}>{\hspre}l<{\hspost}@{}}%
\column{5}{@{}>{\hspre}l<{\hspost}@{}}%
\column{22}{@{}>{\hspre}l<{\hspost}@{}}%
\column{E}{@{}>{\hspre}l<{\hspost}@{}}%
\>[3]{}\Varid{is-cwf}\;\Varid{.CwF.}\circ\Varid{id}\;\{\mskip1.5mu \delta\;\mathrel{=}\;\delta\mskip1.5mu\}\;\mathrel{=}\;{}\<[E]%
\\
\>[3]{}\hsindent{2}{}\<[5]%
\>[5]{}\delta\;\ensuremath{\mbox{$\circ$}}\;\Varid{tm}\Varid{*}\!\sqsubseteq\;\Varid{v}\!\sqsubseteq\!\Varid{t}\;\Varid{id}\;{}\<[22]%
\>[22]{}\equiv\!\langle\;\circ\!\sqsubseteq\;\Varid{⟩}\;{}\<[E]%
\\
\>[3]{}\hsindent{2}{}\<[5]%
\>[5]{}\delta\;\ensuremath{\mbox{$\circ$}}\;\Varid{id}\;{}\<[22]%
\>[22]{}\equiv\!\langle\;\circ\Varid{id}\;\Varid{⟩}\;{}\<[E]%
\\
\>[3]{}\hsindent{2}{}\<[5]%
\>[5]{}\delta\;\blacksquare{}\<[E]%
\ColumnHook
\end{hscode}\resethooks
\end{minipage}

Similarly to substitutions, we must fix the sort of our terms to \ensuremath{\Conid{T}} 
(in this case, so we can prove the identity law - note that applying the 
identity substitution to a variable \ensuremath{\Varid{i}} produces the distinct term \ensuremath{\texttt{\textasciigrave}\;\Varid{i}}).

\begin{minipage}{0.5\textwidth}
\begin{hscode}\SaveRestoreHook
\column{B}{@{}>{\hspre}l<{\hspost}@{}}%
\column{3}{@{}>{\hspre}l<{\hspost}@{}}%
\column{5}{@{}>{\hspre}l<{\hspost}@{}}%
\column{24}{@{}>{\hspre}l<{\hspost}@{}}%
\column{29}{@{}>{\hspre}l<{\hspost}@{}}%
\column{E}{@{}>{\hspre}l<{\hspost}@{}}%
\>[3]{}\Varid{is-cwf}\;\Varid{.CwF.[id]}\;\{\mskip1.5mu \Varid{t}\;\mathrel{=}\;\Varid{t}\mskip1.5mu\}\;{}\<[29]%
\>[29]{}\mathrel{=}\;{}\<[E]%
\\
\>[3]{}\hsindent{2}{}\<[5]%
\>[5]{}\Varid{t}\;[\mskip1.5mu \;\Varid{tm}\Varid{*}\!\sqsubseteq\;\Varid{v}\!\sqsubseteq\!\Varid{t}\;\Varid{id}\;\mskip1.5mu]\;{}\<[24]%
\>[24]{}\equiv\!\langle\;\Varid{t}[\sqsubseteq]\;\{\mskip1.5mu \Varid{t}\;\mathrel{=}\;\Varid{t}\mskip1.5mu\}\;\Varid{⟩}\;{}\<[E]%
\\
\>[3]{}\hsindent{2}{}\<[5]%
\>[5]{}\Varid{t}\;[\mskip1.5mu \;\Varid{id}\;\mskip1.5mu]\;{}\<[24]%
\>[24]{}\equiv\!\langle\;\Varid{[id]}\;\Varid{⟩}\;{}\<[E]%
\\
\>[3]{}\hsindent{2}{}\<[5]%
\>[5]{}\Varid{t}\;{}\<[24]%
\>[24]{}\blacksquare{}\<[E]%
\ColumnHook
\end{hscode}\resethooks
\end{minipage}
\begin{minipage}{0.4\textwidth}
\begin{hscode}\SaveRestoreHook
\column{B}{@{}>{\hspre}l<{\hspost}@{}}%
\column{3}{@{}>{\hspre}l<{\hspost}@{}}%
\column{21}{@{}>{\hspre}l<{\hspost}@{}}%
\column{E}{@{}>{\hspre}l<{\hspost}@{}}%
\>[3]{}\Varid{is-cwf}\;\Varid{.CwF.}\_\vdash\_\;{}\<[21]%
\>[21]{}\mathrel{=}\;\_\vdash\![\;\Conid{T}\;]\_{}\<[E]%
\\[\blanklineskip]%
\>[3]{}\Varid{is-cwf}\;\Varid{.CwF.\char95 [\char95 ]}\;{}\<[21]%
\>[21]{}\mathrel{=}\;\_[\_]{}\<[E]%
\ColumnHook
\end{hscode}\resethooks
\end{minipage}

We now define projections \ensuremath{\pi_0\;(\delta\;\Varid{,}\;\Varid{t})\;\mathrel{=}\;\delta} and \ensuremath{\pi_1\;(\delta\;\Varid{,}\;\Varid{t})\;\mathrel{=}\;\Varid{t}} and
\ensuremath{\rhd\!-\!\beta_0}, \ensuremath{\rhd\!-\!\beta_1}, \ensuremath{\rhd\!-\!\eta}, \ensuremath{\pi_0\!\circ} and \ensuremath{\pi_1\!\circ} all hold by definition (at least, after
matching on the guaranteed-non-empty substitution).

Finally, we can deal with the cases specific to simply typed $\lambda$-calculus.
\ensuremath{\cdot[]} also holds by definition, but the $\beta$-rule for substitutions applied 
to lambdas requires a bit of equational reasoning due to 
differing implementations of \ensuremath{\_\uparrow\_}.

\begin{hscode}\SaveRestoreHook
\column{B}{@{}>{\hspre}l<{\hspost}@{}}%
\column{3}{@{}>{\hspre}l<{\hspost}@{}}%
\column{5}{@{}>{\hspre}l<{\hspost}@{}}%
\column{35}{@{}>{\hspre}l<{\hspost}@{}}%
\column{E}{@{}>{\hspre}l<{\hspost}@{}}%
\>[3]{}\Varid{is-cwf}\;\Varid{.CwF.}\lambda[]\;\{\mskip1.5mu \Conid{A}\;\mathrel{=}\;\Conid{A}\mskip1.5mu\}\;\{\mskip1.5mu \Varid{t}\;\mathrel{=}\;\Varid{x}\mskip1.5mu\}\;\{\mskip1.5mu \delta\;\mathrel{=}\;\Varid{ys}\mskip1.5mu\}\;\mathrel{=}\;{}\<[E]%
\\
\>[3]{}\hsindent{2}{}\<[5]%
\>[5]{}\lambda\;\Varid{x}\;[\mskip1.5mu \;\Varid{ys}\;\uparrow\;\Conid{A}\;\mskip1.5mu]\;{}\<[35]%
\>[35]{}\equiv\!\langle\;\Varid{cong}\;(\lambda\;\rho\;\rightarrow\;\lambda\;\Varid{x}\;[\mskip1.5mu \;\rho\;\uparrow\;\Conid{A}\;\mskip1.5mu])\;(\Varid{sym}\;\circ\Varid{id})\;\Varid{⟩}\;{}\<[E]%
\\
\>[3]{}\hsindent{2}{}\<[5]%
\>[5]{}\lambda\;\Varid{x}\;[\mskip1.5mu \;(\Varid{ys}\;\ensuremath{\mbox{$\circ$}}\;\Varid{id})\;\uparrow\;\Conid{A}\;\mskip1.5mu]\;{}\<[35]%
\>[35]{}\equiv\!\langle\;\Varid{cong}\;(\lambda\;\rho\;\rightarrow\;\lambda\;\Varid{x}\;[\mskip1.5mu \;\rho\;\Varid{,}\;\texttt{\textasciigrave}\;\Varid{zero}\;\mskip1.5mu])\;(\Varid{sym}\;{}^+\!-\Varid{nat}\circ)\;\Varid{⟩}\;{}\<[E]%
\\
\>[3]{}\hsindent{2}{}\<[5]%
\>[5]{}\lambda\;\Varid{x}\;[\mskip1.5mu \;\Varid{ys}\;\ensuremath{\mbox{$\circ$}}\;\Varid{id}\;^{+}\;\Conid{A}\;\Varid{,}\;\texttt{\textasciigrave}\;\Varid{zero}\;\mskip1.5mu]\;{}\<[35]%
\>[35]{}\equiv\!\langle\;\Varid{cong}\;(\lambda\;\rho\;\rightarrow\;\lambda\;\Varid{x}\;[\mskip1.5mu \;\rho\;\Varid{,}\;\texttt{\textasciigrave}\;\Varid{zero}\;\mskip1.5mu])\;(\Varid{sym}\;(\circ\!\sqsubseteq\;\{\mskip1.5mu \Varid{ys}\;\mathrel{=}\;\Varid{id}\;^{+}\;\anonymous \mskip1.5mu\}))\;\Varid{⟩}\;{}\<[E]%
\\
\>[3]{}\hsindent{2}{}\<[5]%
\>[5]{}\lambda\;\Varid{x}\;[\mskip1.5mu \;\Varid{ys}\;\ensuremath{\mbox{$\circ$}}\;\Varid{tm}\Varid{*}\!\sqsubseteq\;\Varid{v}\!\sqsubseteq\!\Varid{t}\;(\Varid{id}\;^{+}\;\Conid{A})\;\Varid{,}\;\texttt{\textasciigrave}\;\Varid{zero}\;\mskip1.5mu]\;\blacksquare{}\<[E]%
\ColumnHook
\end{hscode}\resethooks

We have shown our recursive substitution syntax satisfies the CwF laws, but we
want to go a step further and show initiality: that our syntax is
isomorphic to
the initial CwF.

An important first step is to actually define the initial CwF (and its
eliminator). We use postulates and rewrite rules instead of a Cubical 
Agda higher inductive type (HIT) because of technical limitations mentioned 
previously.
We can reuse our existing datatypes for contexts and types because in STLC 
there are no non-trivial equations on these components.

To avoid name clashes between our existing syntax and the initial CwF 
constructors, we annotate every \ensuremath{\Conid{ICwF}} constructor with \ensuremath{^{\mathrm{I}}}. e.g.
\ensuremath{\_\vdash^{\mathrm{I}}\_\;\mathbin{:}\;\Conid{Con}\;\rightarrow\;\Conid{Ty}\;\rightarrow\;\Conid{Set}}, \ensuremath{\Varid{id}^{\mathrm{I}}\;\mathbin{:}\;\Gamma\;\Vdash^{\mathrm{I}}\;\Gamma} etc.

We state the eliminator for the initial CwF assuming appropriate \ensuremath{\Conid{Motive}\;\mathbin{:}\;\Varid{Set}_1} and 
\ensuremath{\Conid{Methods}\;\mathbin{:}\;\Conid{Motive}\;\rightarrow\;\Varid{Set}_1} records as in \cite{altenkirch2016tt}.
Again to avoid name clashes, we annotate fields of these records (corresponding
to how each type/constructor is eliminated) with \ensuremath{^{\mathrm{M}}}.

\noindent
\begin{minipage}{0.35\textwidth}
\begin{hscode}\SaveRestoreHook
\column{B}{@{}>{\hspre}l<{\hspost}@{}}%
\column{3}{@{}>{\hspre}l<{\hspost}@{}}%
\column{13}{@{}>{\hspre}l<{\hspost}@{}}%
\column{26}{@{}>{\hspre}l<{\hspost}@{}}%
\column{E}{@{}>{\hspre}l<{\hspost}@{}}%
\>[3]{}\Varid{elim-con}\;{}\<[13]%
\>[13]{}\mathbin{:}\;\forall{}\;\Gamma\;\rightarrow\;\Conid{Con}^{\mathrm{M}}\;\Gamma{}\<[E]%
\\
\>[3]{}\Varid{elim-ty}\;{}\<[13]%
\>[13]{}\mathbin{:}\;\forall{}\;\Conid{A}\;\rightarrow\;\Conid{Ty}^{\mathrm{M}}\;{}\<[26]%
\>[26]{}\Conid{A}{}\<[E]%
\ColumnHook
\end{hscode}\resethooks
\end{minipage}
\begin{minipage}{0.45\textwidth}
\begin{hscode}\SaveRestoreHook
\column{B}{@{}>{\hspre}l<{\hspost}@{}}%
\column{3}{@{}>{\hspre}l<{\hspost}@{}}%
\column{14}{@{}>{\hspre}l<{\hspost}@{}}%
\column{E}{@{}>{\hspre}l<{\hspost}@{}}%
\>[3]{}\Varid{elim-cwf}\;{}\<[14]%
\>[14]{}\mathbin{:}\;\forall{}\;\Varid{t}^{\mathrm{I}}\;\rightarrow\;\Conid{Tm}^{\mathrm{M}}\;(\Varid{elim-con}\;\Gamma)\;(\Varid{elim-ty}\;\Conid{A})\;\Varid{t}^{\mathrm{I}}{}\<[E]%
\\
\>[3]{}\Varid{elim-cwf}\Varid{*}\;{}\<[14]%
\>[14]{}\mathbin{:}\;\forall{}\;\delta^{\mathrm{I}}\;\rightarrow\;\Conid{Tms}^{\mathrm{M}}\;(\Varid{elim-con}\;\Delta)\;(\Varid{elim-con}\;\Gamma)\;\delta^{\mathrm{I}}{}\<[E]%
\ColumnHook
\end{hscode}\resethooks
\end{minipage}

To state the dependent equations in \ensuremath{\Conid{Methods}} between outputs of the eliminator,
enforcing congruence of equality (e.g. the functor law, which asks for 
\ensuremath{\Varid{t}^{\mathrm{M}}\;[\mskip1.5mu \;\sigma^{\mathrm{M}}\;\mskip1.5mu]^{\mathrm{M}}\;[\mskip1.5mu \;\delta^{\mathrm{M}}\;\mskip1.5mu]^{\mathrm{M}}} and \ensuremath{\Varid{t}^{\mathrm{M}}\;[\mskip1.5mu \;\sigma^{\mathrm{M}}\;\ensuremath{\mbox{$\circ$}}^{\mathrm{M}}\;\delta^{\mathrm{M}}\;\mskip1.5mu]^{\mathrm{M}}} to be equated)
we need
dependent identity types
\ensuremath{\_\equiv\!\![\_]\!\!\equiv\_\;\mathbin{:}\;\Conid{A}\;\rightarrow\;\Conid{A}\;\equiv\;\Conid{B}\;\rightarrow\;\Conid{B}\;\rightarrow\;\Conid{Set}\;\ell}. 
We can define these simply by matching on the identity
between \ensuremath{\Conid{A}} and \ensuremath{\Conid{B}}, \ensuremath{\Varid{x}\;\equiv\!\![\;\Varid{refl}\;]\!\!\equiv\;\Varid{y}\;\mathrel{=}\;\Varid{x}\;\equiv\;\Varid{y}}.

Normalisation from the initial CwF into substitution normal forms now only
needs a way to connect our notion of ``being a CwF'' with our initial CwF's 
eliminator: specifically, that any set of type families satisfying the CwF laws
gives rise to a \ensuremath{\Conid{Motive}} and associated set of \ensuremath{\Conid{Methods}}. To achieve this,
we define \ensuremath{\Varid{cwf-to-motive}\;\mathbin{:}\;\Conid{CwF-simple}\;\rightarrow\;\Conid{Motive}} and 
\ensuremath{\Varid{cwf-to-methods}\;\mathbin{:}\;\Conid{CwF-simple}\;\rightarrow\;\Conid{Methods}}, which simply project out the relevant 
fields,
and then implement e.g. \ensuremath{\Varid{rec-cwf}\;\mathrel{=}\;\Varid{elim-cwf}\;\Varid{cwf-to-methods}}.

The one extra ingredient we need to make this work out neatly is to introduce
a new reduction for \ensuremath{\Varid{cong}}, \ensuremath{\Varid{cong}\;(\lambda\;\anonymous \;\rightarrow\;\Varid{x})\;\Varid{p}\;\equiv\;\Varid{refl}}, via an Agda rewrite
rule (this identity also holds natively in Cubical).
This enables the no-longer-dependent \ensuremath{\_\equiv\!\![\_]\!\!\equiv\_}s to collapse to \ensuremath{\_\equiv\_}s 
automatically.

\noindent
Normalisation into our substitution normal forms can now be achieved by with:

\begin{hscode}\SaveRestoreHook
\column{B}{@{}>{\hspre}l<{\hspost}@{}}%
\column{E}{@{}>{\hspre}l<{\hspost}@{}}%
\>[B]{}\Varid{norm}\;\mathbin{:}\;\Gamma\;\vdash^{\mathrm{I}}\;\Conid{A}\;\rightarrow\;\Varid{rec-con}\;\Varid{is-cwf}\;\Gamma\;\vdash\![\;\Conid{T}\;\mskip1.5mu]\;\Varid{rec-ty}\;\Varid{is-cwf}\;\Conid{A}{}\<[E]%
\\
\>[B]{}\Varid{norm}\;\mathrel{=}\;\Varid{rec-cwf}\;\Varid{is-cwf}{}\<[E]%
\ColumnHook
\end{hscode}\resethooks

Of course, normalisation shouldn't change the type of a term, or the context it
is in, so we might hope for a simpler signature \ensuremath{\Gamma\;\vdash^{\mathrm{I}}\;\Conid{A}\;\rightarrow\;\Gamma\;\vdash\![\;\Conid{T}\;\mskip1.5mu]\;\Conid{A}} and, 
conveniently, rewrite rules (\ensuremath{\Varid{rec-con}\;\Varid{is-cwf}\;\Gamma\;\equiv\;\Gamma} and \ensuremath{\Varid{rec-ty}\;\Varid{is-cwf}\;\Conid{A}\;\equiv\;\Conid{A}}) 
can get us there!

\begin{minipage}{0.45\textwidth}
\begin{hscode}\SaveRestoreHook
\column{B}{@{}>{\hspre}l<{\hspost}@{}}%
\column{E}{@{}>{\hspre}l<{\hspost}@{}}%
\>[B]{}\Varid{norm}\;\mathbin{:}\;\Gamma\;\vdash^{\mathrm{I}}\;\Conid{A}\;\rightarrow\;\Gamma\;\vdash\![\;\Conid{T}\;\mskip1.5mu]\;\Conid{A}{}\<[E]%
\\
\>[B]{}\Varid{norm}\;\mathrel{=}\;\Varid{rec-cwf}\;\Varid{is-cwf}{}\<[E]%
\ColumnHook
\end{hscode}\resethooks
\end{minipage}
\begin{minipage}{0.45\textwidth}
\begin{hscode}\SaveRestoreHook
\column{B}{@{}>{\hspre}l<{\hspost}@{}}%
\column{E}{@{}>{\hspre}l<{\hspost}@{}}%
\>[B]{}\Varid{norm}\Varid{*}\;\mathbin{:}\;\Delta\;\Vdash^{\mathrm{I}}\;\Gamma\;\rightarrow\;\Delta\;\Vdash\![\;\Conid{T}\;\mskip1.5mu]\;\Gamma{}\<[E]%
\\
\>[B]{}\Varid{norm}\Varid{*}\;\mathrel{=}\;\Varid{rec-cwf}\Varid{*}\;\Varid{is-cwf}{}\<[E]%
\ColumnHook
\end{hscode}\resethooks
\end{minipage}

The inverse operation to inject our syntax back into the initial CwF is easily
implemented by recursion on substitution normal forms.

\begin{minipage}{0.45\textwidth}
\begin{hscode}\SaveRestoreHook
\column{B}{@{}>{\hspre}l<{\hspost}@{}}%
\column{7}{@{}>{\hspre}l<{\hspost}@{}}%
\column{14}{@{}>{\hspre}l<{\hspost}@{}}%
\column{E}{@{}>{\hspre}l<{\hspost}@{}}%
\>[B]{}\ulcorner\_\urcorner\;{}\<[7]%
\>[7]{}\mathbin{:}\;\Gamma\;\vdash\![\;\Varid{q}\;\mskip1.5mu]\;\Conid{A}\;\rightarrow\;\Gamma\;\vdash^{\mathrm{I}}\;\Conid{A}{}\<[E]%
\\
\>[B]{}\ulcorner\_\urcorner\Varid{*}\;{}\<[7]%
\>[7]{}\mathbin{:}\;\Delta\;\Vdash\![\;\Varid{q}\;\mskip1.5mu]\;\Gamma\;\rightarrow\;\Delta\;\Vdash^{\mathrm{I}}\;\Gamma{}\<[E]%
\\[\blanklineskip]%
\>[B]{}\ulcorner\;\Varid{zero}\;\urcorner\;{}\<[14]%
\>[14]{}\mathrel{=}\;\Varid{zero}^{\mathrm{I}}{}\<[E]%
\\
\>[B]{}\ulcorner\;\Varid{suc}\;\Varid{i}\;\Conid{B}\;\urcorner\;{}\<[14]%
\>[14]{}\mathrel{=}\;\Varid{suc}^{\mathrm{I}}\;\ulcorner\;\Varid{i}\;\urcorner\;\Conid{B}{}\<[E]%
\\
\>[B]{}\ulcorner\;\texttt{\textasciigrave}\;\Varid{i}\;\urcorner\;{}\<[14]%
\>[14]{}\mathrel{=}\;\ulcorner\;\Varid{i}\;\urcorner{}\<[E]%
\ColumnHook
\end{hscode}\resethooks
\end{minipage}
\begin{minipage}{0.45\textwidth}
\begin{hscode}\SaveRestoreHook
\column{B}{@{}>{\hspre}l<{\hspost}@{}}%
\column{13}{@{}>{\hspre}l<{\hspost}@{}}%
\column{E}{@{}>{\hspre}l<{\hspost}@{}}%
\>[B]{}\ulcorner\;\Varid{t}\;\cdot\;\Varid{u}\;\urcorner\;{}\<[13]%
\>[13]{}\mathrel{=}\;\ulcorner\;\Varid{t}\;\urcorner\;\cdot^{\mathrm{I}}\;\ulcorner\;\Varid{u}\;\urcorner{}\<[E]%
\\
\>[B]{}\ulcorner\;\lambda\;\Varid{t}\;\urcorner\;{}\<[13]%
\>[13]{}\mathrel{=}\;\lambda^{\mathrm{I}}\;\ulcorner\;\Varid{t}\;\urcorner{}\<[E]%
\\[\blanklineskip]%
\>[B]{}\ulcorner\;\varepsilon\;\urcorner\Varid{*}\;{}\<[13]%
\>[13]{}\mathrel{=}\;\varepsilon^{\mathrm{I}}{}\<[E]%
\\
\>[B]{}\ulcorner\;\delta\;\Varid{,}\;\Varid{x}\;\urcorner\Varid{*}\;{}\<[13]%
\>[13]{}\mathrel{=}\;\ulcorner\;\delta\;\urcorner\Varid{*}\;\Varid{,}^{\mathrm{I}}\;\ulcorner\;\Varid{x}\;\urcorner{}\<[E]%
\ColumnHook
\end{hscode}\resethooks
\end{minipage}

\subsection{Proving initiality}
\label{sec:proving-initiality}

We have implemented both directions of the isomorphism. Now to show this truly
is an isomorphism and not just a pair of functions between two types, we must 
prove that \ensuremath{\Varid{norm}} and \ensuremath{\ulcorner\_\urcorner} are mutual inverses - i.e. stability 
(\ensuremath{\Varid{norm}\;\ulcorner\;\Varid{t}\;\urcorner\;\equiv\;\Varid{t}}) and completeness (\ensuremath{\ulcorner\;\Varid{norm}\;\Varid{t}\;\urcorner\;\equiv\;\Varid{t}}).

We start with stability, as it is considerably easier. There are just a couple
details worth mentioning:
\begin{itemize}
  \item To deal with variables in the \ensuremath{\texttt{\textasciigrave}\_} case, we slightly generalise the
  inductive hypothesis, taking expressions of any sort and coercing them up 
  to sort \ensuremath{\Conid{T}} on the RHS.
  \item The case for variables relies on a bit of coercion manipulation and our 
  earlier lemma equating \ensuremath{\Varid{i}\;[\mskip1.5mu \;\Varid{id}\;^{+}\;\Conid{B}\;\mskip1.5mu]} and \ensuremath{\Varid{suc}\;\Varid{i}\;\Conid{B}}.
\end{itemize}

\begin{hscode}\SaveRestoreHook
\column{B}{@{}>{\hspre}l<{\hspost}@{}}%
\column{3}{@{}>{\hspre}l<{\hspost}@{}}%
\column{21}{@{}>{\hspre}l<{\hspost}@{}}%
\column{37}{@{}>{\hspre}l<{\hspost}@{}}%
\column{E}{@{}>{\hspre}l<{\hspost}@{}}%
\>[B]{}\Varid{stab}\;\mathbin{:}\;\Varid{norm}\;\ulcorner\;\Varid{x}\;\urcorner\;\equiv\;\Varid{tm}\!\sqsubseteq\;\sqsubseteq\!\Varid{t}\;\Varid{x}{}\<[E]%
\\
\>[B]{}\Varid{stab}\;\{\mskip1.5mu \Varid{x}\;\mathrel{=}\;\Varid{zero}\mskip1.5mu\}\;{}\<[21]%
\>[21]{}\mathrel{=}\;\Varid{refl}{}\<[E]%
\\
\>[B]{}\Varid{stab}\;\{\mskip1.5mu \Varid{x}\;\mathrel{=}\;\Varid{suc}\;\Varid{i}\;\Conid{B}\mskip1.5mu\}\;{}\<[21]%
\>[21]{}\mathrel{=}\;{}\<[E]%
\\
\>[B]{}\hsindent{3}{}\<[3]%
\>[3]{}\Varid{norm}\;\ulcorner\;\Varid{i}\;\urcorner\;[\mskip1.5mu \;\Varid{tm}\Varid{*}\!\sqsubseteq\;\Varid{v}\!\sqsubseteq\!\Varid{t}\;(\Varid{id}\;^{+}\;\Conid{B})\;\mskip1.5mu]\;{}\<[37]%
\>[37]{}\equiv\!\langle\;\Varid{t}[\sqsubseteq]\;\{\mskip1.5mu \Varid{t}\;\mathrel{=}\;\Varid{norm}\;\ulcorner\;\Varid{i}\;\urcorner\mskip1.5mu\}\;\Varid{⟩}\;{}\<[E]%
\\
\>[B]{}\hsindent{3}{}\<[3]%
\>[3]{}\Varid{norm}\;\ulcorner\;\Varid{i}\;\urcorner\;[\mskip1.5mu \;\Varid{id}\;^{+}\;\Conid{B}\;\mskip1.5mu]\;{}\<[37]%
\>[37]{}\equiv\!\langle\;\Varid{cong}\;(\lambda\;\Varid{j}\;\rightarrow\;\Varid{suc}[\;\anonymous \;\mskip1.5mu]\;\Varid{j}\;\Conid{B})\;(\Varid{stab}\;\{\mskip1.5mu \Varid{x}\;\mathrel{=}\;\Varid{i}\mskip1.5mu\})\;\Varid{⟩}\;{}\<[E]%
\\
\>[B]{}\hsindent{3}{}\<[3]%
\>[3]{}\texttt{\textasciigrave}\;\Varid{i}\;[\mskip1.5mu \;\Varid{id}\;^{+}\;\Conid{B}\;\mskip1.5mu]\;{}\<[37]%
\>[37]{}\equiv\!\langle\;\Varid{cong}\;\texttt{\textasciigrave}\_\;\Varid{suc}[\Varid{id}^{+}]\;\Varid{⟩}\;{}\<[E]%
\\
\>[B]{}\hsindent{3}{}\<[3]%
\>[3]{}\texttt{\textasciigrave}\;\Varid{suc}\;\Varid{i}\;\Conid{B}\;\blacksquare{}\<[E]%
\\
\>[B]{}\Varid{stab}\;\{\mskip1.5mu \Varid{x}\;\mathrel{=}\;\texttt{\textasciigrave}\;\Varid{i}\mskip1.5mu\}\;{}\<[21]%
\>[21]{}\mathrel{=}\;\Varid{stab}\;\{\mskip1.5mu \Varid{x}\;\mathrel{=}\;\Varid{i}\mskip1.5mu\}{}\<[E]%
\\
\>[B]{}\Varid{stab}\;\{\mskip1.5mu \Varid{x}\;\mathrel{=}\;\Varid{t}\;\cdot\;\Varid{u}\mskip1.5mu\}\;{}\<[21]%
\>[21]{}\mathrel{=}\;\Varid{cong}_{2}\;\_\cdot\_\;(\Varid{stab}\;\{\mskip1.5mu \Varid{x}\;\mathrel{=}\;\Varid{t}\mskip1.5mu\})\;(\Varid{stab}\;\{\mskip1.5mu \Varid{x}\;\mathrel{=}\;\Varid{u}\mskip1.5mu\}){}\<[E]%
\\
\>[B]{}\Varid{stab}\;\{\mskip1.5mu \Varid{x}\;\mathrel{=}\;\lambda\;\Varid{t}\mskip1.5mu\}\;{}\<[21]%
\>[21]{}\mathrel{=}\;\Varid{cong}\;\lambda\_\;(\Varid{stab}\;\{\mskip1.5mu \Varid{x}\;\mathrel{=}\;\Varid{t}\mskip1.5mu\}){}\<[E]%
\ColumnHook
\end{hscode}\resethooks

To prove completeness, we must instead induct on the initial CwF itself, which
means there are many more cases. We start with the motive:

\begin{hscode}\SaveRestoreHook
\column{B}{@{}>{\hspre}l<{\hspost}@{}}%
\column{E}{@{}>{\hspre}l<{\hspost}@{}}%
\>[B]{}\Varid{compl-}\mathbb{M}\;\mathbin{:}\;\Conid{Motive}{}\<[E]%
\ColumnHook
\end{hscode}\resethooks
\noindent
\begin{minipage}{0.6\textwidth}
\begin{hscode}\SaveRestoreHook
\column{B}{@{}>{\hspre}l<{\hspost}@{}}%
\column{23}{@{}>{\hspre}l<{\hspost}@{}}%
\column{E}{@{}>{\hspre}l<{\hspost}@{}}%
\>[B]{}\Varid{compl-}\mathbb{M}\;\Varid{.}\Conid{Tm}^{\mathrm{M}}\;\anonymous \;\anonymous \;\Varid{t}^{\mathrm{I}}\;{}\<[23]%
\>[23]{}\mathrel{=}\;\ulcorner\;\Varid{norm}\;\Varid{t}^{\mathrm{I}}\;\urcorner\;\equiv\;\Varid{t}^{\mathrm{I}}{}\<[E]%
\\
\>[B]{}\Varid{compl-}\mathbb{M}\;\Varid{.}\Conid{Tms}^{\mathrm{M}}\;\anonymous \;\anonymous \;\delta^{\mathrm{I}}\;{}\<[23]%
\>[23]{}\mathrel{=}\;\ulcorner\;\Varid{norm}\Varid{*}\;\delta^{\mathrm{I}}\;\urcorner\Varid{*}\;\equiv\;\delta^{\mathrm{I}}{}\<[E]%
\ColumnHook
\end{hscode}\resethooks
\end{minipage}
\begin{minipage}{0.35\textwidth}
\begin{hscode}\SaveRestoreHook
\column{B}{@{}>{\hspre}l<{\hspost}@{}}%
\column{15}{@{}>{\hspre}l<{\hspost}@{}}%
\column{18}{@{}>{\hspre}l<{\hspost}@{}}%
\column{E}{@{}>{\hspre}l<{\hspost}@{}}%
\>[B]{}\Varid{compl-}\mathbb{M}\;\Varid{.}\Conid{Con}^{\mathrm{M}}\;\anonymous \;{}\<[18]%
\>[18]{}\mathrel{=}\;\top{}\<[E]%
\\
\>[B]{}\Varid{compl-}\mathbb{M}\;\Varid{.}\Conid{Ty}^{\mathrm{M}}\;{}\<[15]%
\>[15]{}\anonymous \;{}\<[18]%
\>[18]{}\mathrel{=}\;\top{}\<[E]%
\ColumnHook
\end{hscode}\resethooks
\end{minipage}

To show these identities, we need to prove that our various recursively defined
syntax operations are preserved by \ensuremath{\ulcorner\_\urcorner}.

Preservation of \ensuremath{\Varid{zero}[\_]}, \ensuremath{\ulcorner\Varid{zero}\urcorner\;\mathbin{:}\;\ulcorner\;\Varid{zero}[\;\Varid{q}\;\mskip1.5mu]\;\urcorner\;\equiv\;\Varid{zero}^{\mathrm{I}}} reduces to
reflexivity after splitting on the sort.

Preservation of each of the projections out of sequences of terms 
(e.g. \ensuremath{\ulcorner\;\pi_0\;\delta\;\urcorner\Varid{*}\;\equiv\;\pi_0^{\mathrm{I}}\;\ulcorner\;\delta\;\urcorner\Varid{*}}) reduce to the 
associated $\beta$-laws of the initial CwF (e.g. \ensuremath{\rhd\!-\!\beta_0^{\mathrm{I}}}).

Preservation proofs for \ensuremath{\_[\_]}, \ensuremath{\_\uparrow\_}, \ensuremath{\_^{+}\_}, \ensuremath{\Varid{id}} and \ensuremath{\Varid{suc}[\_]} are all mutually 
inductive, mirroring their original recursive definitions. We must stay
polymorphic over sorts and again use our dummy \ensuremath{\Conid{Sort}} argument trick when
implementing \ensuremath{\ulcorner\Varid{id}\urcorner} to keep Agda's termination checker happy.

\noindent
\begin{minipage}{0.5\textwidth}
\begin{hscode}\SaveRestoreHook
\column{B}{@{}>{\hspre}l<{\hspost}@{}}%
\column{7}{@{}>{\hspre}l<{\hspost}@{}}%
\column{E}{@{}>{\hspre}l<{\hspost}@{}}%
\>[B]{}\ulcorner\Varid{[]}\urcorner\;{}\<[7]%
\>[7]{}\mathbin{:}\;\ulcorner\;\Varid{x}\;[\mskip1.5mu \;\Varid{ys}\;\mskip1.5mu]\;\urcorner\;\equiv\;\ulcorner\;\Varid{x}\;\urcorner\;[\mskip1.5mu \;\ulcorner\;\Varid{ys}\;\urcorner\Varid{*}\;\mskip1.5mu]^{\mathrm{I}}{}\<[E]%
\\
\>[B]{}\ulcorner\uparrow\urcorner\;{}\<[7]%
\>[7]{}\mathbin{:}\;\ulcorner\;\Varid{xs}\;\uparrow\;\Conid{A}\;\urcorner\Varid{*}\;\equiv\;\ulcorner\;\Varid{xs}\;\urcorner\Varid{*}\;\uparrow^{\mathrm{I}}\;\Conid{A}{}\<[E]%
\\
\>[B]{}\ulcorner^{+}\urcorner\;{}\<[7]%
\>[7]{}\mathbin{:}\;\ulcorner\;\Varid{xs}\;^{+}\;\Conid{A}\;\urcorner\Varid{*}\;\equiv\;\ulcorner\;\Varid{xs}\;\urcorner\Varid{*}\;\ensuremath{\mbox{$\circ$}}^{\mathrm{I}}\;\Varid{wk}^{\mathrm{I}}{}\<[E]%
\\
\>[B]{}\ulcorner\Varid{id}\urcorner\;{}\<[7]%
\>[7]{}\mathbin{:}\;\ulcorner\;\Varid{id}\;\{\mskip1.5mu \Gamma\;\mathrel{=}\;\Gamma\mskip1.5mu\}\;\urcorner\Varid{*}\;\equiv\;\Varid{id}^{\mathrm{I}}{}\<[E]%
\ColumnHook
\end{hscode}\resethooks
\end{minipage}
\begin{minipage}{0.45\textwidth}
\begin{hscode}\SaveRestoreHook
\column{B}{@{}>{\hspre}l<{\hspost}@{}}%
\column{E}{@{}>{\hspre}l<{\hspost}@{}}%
\>[B]{}\ulcorner\Varid{suc}\urcorner\;\mathbin{:}\;\ulcorner\;\Varid{suc}[\;\Varid{q}\;\mskip1.5mu]\;\Varid{x}\;\Conid{B}\;\urcorner\;\equiv\;\ulcorner\;\Varid{x}\;\urcorner\;[\mskip1.5mu \;\Varid{wk}^{\mathrm{I}}\;\mskip1.5mu]^{\mathrm{I}}{}\<[E]%
\\[\blanklineskip]%
\>[B]{}\ulcorner\Varid{id}\urcorner^{\prime}\;\mathbin{:}\;\Conid{Sort}\;\rightarrow\;\ulcorner\;\Varid{id}\;\{\mskip1.5mu \Gamma\;\mathrel{=}\;\Gamma\mskip1.5mu\}\;\urcorner\Varid{*}\;\equiv\;\Varid{id}^{\mathrm{I}}{}\<[E]%
\\
\>[B]{}\ulcorner\Varid{id}\urcorner\;\mathrel{=}\;\ulcorner\Varid{id}\urcorner^{\prime}\;\Conid{V}{}\<[E]%
\ColumnHook
\end{hscode}\resethooks
\end{minipage}


To complete these proofs, we also need $\beta$-laws for our initial CwF
substitutions, so we derive these now.

\noindent
\begin{minipage}{0.475\textwidth}
\noindent
\begin{hscode}\SaveRestoreHook
\column{B}{@{}>{\hspre}l<{\hspost}@{}}%
\column{3}{@{}>{\hspre}l<{\hspost}@{}}%
\column{28}{@{}>{\hspre}l<{\hspost}@{}}%
\column{E}{@{}>{\hspre}l<{\hspost}@{}}%
\>[B]{}\Varid{zero[]}^{\mathrm{I}}\;\mathbin{:}\;\Varid{zero}^{\mathrm{I}}\;[\mskip1.5mu \;\delta^{\mathrm{I}}\;\Varid{,}^{\mathrm{I}}\;\Varid{t}^{\mathrm{I}}\;\mskip1.5mu]^{\mathrm{I}}\;\equiv\;\Varid{t}^{\mathrm{I}}{}\<[E]%
\\
\>[B]{}\Varid{zero[]}^{\mathrm{I}}\;\{\mskip1.5mu \delta^{\mathrm{I}}\;\mathrel{=}\;\delta^{\mathrm{I}}\mskip1.5mu\}\;\{\mskip1.5mu \Varid{t}^{\mathrm{I}}\;\mathrel{=}\;\Varid{t}^{\mathrm{I}}\mskip1.5mu\}\;\mathrel{=}\;{}\<[E]%
\\
\>[B]{}\hsindent{3}{}\<[3]%
\>[3]{}\Varid{zero}^{\mathrm{I}}\;[\mskip1.5mu \;\delta^{\mathrm{I}}\;\Varid{,}^{\mathrm{I}}\;\Varid{t}^{\mathrm{I}}\;\mskip1.5mu]^{\mathrm{I}}\;{}\<[28]%
\>[28]{}\equiv\!\langle\;\Varid{sym}\;\pi_1\!\circ^{\mathrm{I}}\;\Varid{⟩}\;{}\<[E]%
\\
\>[B]{}\hsindent{3}{}\<[3]%
\>[3]{}\pi_1^{\mathrm{I}}\;(\Varid{id}^{\mathrm{I}}\;\ensuremath{\mbox{$\circ$}}^{\mathrm{I}}\;(\delta^{\mathrm{I}}\;\Varid{,}^{\mathrm{I}}\;\Varid{t}^{\mathrm{I}}))\;{}\<[28]%
\>[28]{}\equiv\!\langle\;\Varid{cong}\;\pi_1^{\mathrm{I}}\;\Varid{id}\circ^{\mathrm{I}}\;\Varid{⟩}\;{}\<[E]%
\\
\>[B]{}\hsindent{3}{}\<[3]%
\>[3]{}\pi_1^{\mathrm{I}}\;(\delta^{\mathrm{I}}\;\Varid{,}^{\mathrm{I}}\;\Varid{t}^{\mathrm{I}})\;{}\<[28]%
\>[28]{}\equiv\!\langle\;\rhd\!-\!\beta_1^{\mathrm{I}}\;\Varid{⟩}\;{}\<[E]%
\\
\>[B]{}\hsindent{3}{}\<[3]%
\>[3]{}\Varid{t}^{\mathrm{I}}\;{}\<[28]%
\>[28]{}\blacksquare{}\<[E]%
\ColumnHook
\end{hscode}\resethooks
\end{minipage}
\begin{minipage}{0.45\textwidth}
\noindent
\begin{hscode}\SaveRestoreHook
\column{B}{@{}>{\hspre}l<{\hspost}@{}}%
\column{E}{@{}>{\hspre}l<{\hspost}@{}}%
\>[B]{}\Varid{suc[]}^{\mathrm{I}}\;\mathbin{:}\;\Varid{suc}^{\mathrm{I}}\;\Varid{t}^{\mathrm{I}}\;\Conid{B}\;[\mskip1.5mu \;\delta^{\mathrm{I}}\;\Varid{,}^{\mathrm{I}}\;\Varid{u}^{\mathrm{I}}\;\mskip1.5mu]^{\mathrm{I}}\;\equiv\;\Varid{t}^{\mathrm{I}}\;[\mskip1.5mu \;\delta^{\mathrm{I}}\;\mskip1.5mu]^{\mathrm{I}}{}\<[E]%
\\
\>[B]{}\Varid{suc[]}^{\mathrm{I}}\;\mathrel{=}\;\Varid{...}{}\<[E]%
\ColumnHook
\end{hscode}\resethooks

\noindent
\begin{hscode}\SaveRestoreHook
\column{B}{@{}>{\hspre}l<{\hspost}@{}}%
\column{E}{@{}>{\hspre}l<{\hspost}@{}}%
\>[B]{}\Varid{,}\circ^{\mathrm{I}}\;\mathbin{:}\;(\delta^{\mathrm{I}}\;\Varid{,}^{\mathrm{I}}\;\Varid{t}^{\mathrm{I}})\;\ensuremath{\mbox{$\circ$}}^{\mathrm{I}}\;\sigma^{\mathrm{I}}\;\equiv\;(\delta^{\mathrm{I}}\;\ensuremath{\mbox{$\circ$}}^{\mathrm{I}}\;\sigma^{\mathrm{I}})\;\Varid{,}^{\mathrm{I}}\;(\Varid{t}^{\mathrm{I}}\;[\mskip1.5mu \;\sigma^{\mathrm{I}}\;\mskip1.5mu]^{\mathrm{I}}){}\<[E]%
\\
\>[B]{}\Varid{,}\circ^{\mathrm{I}}\;\mathrel{=}\;\Varid{...}{}\<[E]%
\ColumnHook
\end{hscode}\resethooks
\end{minipage}

We also need a couple lemmas about how \ensuremath{\ulcorner\_\urcorner} treats terms of different sorts
identically: \ensuremath{\ulcorner\sqsubseteq\urcorner\;\mathbin{:}\;\forall{}\;\{\mskip1.5mu \Varid{x}\;\mathbin{:}\;\Gamma\;\vdash\![\;\Varid{q}\;\mskip1.5mu]\;\Conid{A}\mskip1.5mu\}\;\rightarrow\;\ulcorner\;\Varid{tm}\!\sqsubseteq\;\sqsubseteq\!\Varid{t}\;\Varid{x}\;\urcorner\;\equiv\;\ulcorner\;\Varid{x}\;\urcorner} and
\ensuremath{\ulcorner\sqsubseteq\urcorner\Varid{*}\;\mathbin{:}\;\ulcorner\;\Varid{tm}\Varid{*}\!\sqsubseteq\;\sqsubseteq\!\Varid{t}\;\Varid{xs}\;\urcorner\Varid{*}\;\equiv\;\ulcorner\;\Varid{xs}\;\urcorner\Varid{*}}.

We can now proceed with the preservation proofs. There are quite a few
cases to cover, so for brevity we elide the proofs of \ensuremath{\ulcorner\Varid{[]}\urcorner} and \ensuremath{\ulcorner\Varid{suc}\urcorner}.

\begin{hscode}\SaveRestoreHook
\column{B}{@{}>{\hspre}l<{\hspost}@{}}%
\column{E}{@{}>{\hspre}l<{\hspost}@{}}%
\>[B]{}\ulcorner\uparrow\urcorner\;\{\mskip1.5mu \Varid{q}\;\mathrel{=}\;\Varid{q}\mskip1.5mu\}\;\mathrel{=}\;\Varid{cong}_{2}\;\_\Varid{,}^{\mathrm{I}}\_\;\ulcorner^{+}\urcorner\;(\ulcorner\Varid{zero}\urcorner\;\{\mskip1.5mu \Varid{q}\;\mathrel{=}\;\Varid{q}\mskip1.5mu\}){}\<[E]%
\ColumnHook
\end{hscode}\resethooks

\noindent
\begin{minipage}{0.45\textwidth}
\begin{hscode}\SaveRestoreHook
\column{B}{@{}>{\hspre}l<{\hspost}@{}}%
\column{3}{@{}>{\hspre}l<{\hspost}@{}}%
\column{28}{@{}>{\hspre}l<{\hspost}@{}}%
\column{E}{@{}>{\hspre}l<{\hspost}@{}}%
\>[B]{}\ulcorner^{+}\urcorner\;\{\mskip1.5mu \Varid{xs}\;\mathrel{=}\;\varepsilon\mskip1.5mu\}\;{}\<[28]%
\>[28]{}\mathrel{=}\;\Varid{sym}\;\Varid{•}\!-\!\eta^{\mathrm{I}}{}\<[E]%
\\
\>[B]{}\ulcorner^{+}\urcorner\;\{\mskip1.5mu \Varid{xs}\;\mathrel{=}\;\Varid{xs}\;\Varid{,}\;\Varid{x}\mskip1.5mu\}\;\{\mskip1.5mu \Conid{A}\;\mathrel{=}\;\Conid{A}\mskip1.5mu\}\;{}\<[28]%
\>[28]{}\mathrel{=}\;{}\<[E]%
\\
\>[B]{}\hsindent{3}{}\<[3]%
\>[3]{}\ulcorner\;\Varid{xs}\;^{+}\;\Conid{A}\;\urcorner\Varid{*}\;\Varid{,}^{\mathrm{I}}\;\ulcorner\;\Varid{suc}[\;\anonymous \;\mskip1.5mu]\;\Varid{x}\;\Conid{A}\;\urcorner\;{}\<[E]%
\\
\>[B]{}\hsindent{3}{}\<[3]%
\>[3]{}\equiv\!\langle\;\Varid{cong}_{2}\;\_\Varid{,}^{\mathrm{I}}\_\;\ulcorner^{+}\urcorner\;(\ulcorner\Varid{suc}\urcorner\;\{\mskip1.5mu \Varid{x}\;\mathrel{=}\;\Varid{x}\mskip1.5mu\})\;\Varid{⟩}\;{}\<[E]%
\\
\>[B]{}\hsindent{3}{}\<[3]%
\>[3]{}(\ulcorner\;\Varid{xs}\;\urcorner\Varid{*}\;\ensuremath{\mbox{$\circ$}}^{\mathrm{I}}\;\Varid{wk}^{\mathrm{I}})\;\Varid{,}^{\mathrm{I}}\;(\ulcorner\;\Varid{x}\;\urcorner\;[\mskip1.5mu \;\Varid{wk}^{\mathrm{I}}\;\mskip1.5mu]^{\mathrm{I}})\;{}\<[E]%
\\
\>[B]{}\hsindent{3}{}\<[3]%
\>[3]{}\equiv\!\langle\;\Varid{sym}\;\Varid{,}\circ^{\mathrm{I}}\;\Varid{⟩}\;{}\<[E]%
\\
\>[B]{}\hsindent{3}{}\<[3]%
\>[3]{}(\ulcorner\;\Varid{xs}\;\urcorner\Varid{*}\;\Varid{,}^{\mathrm{I}}\;\ulcorner\;\Varid{x}\;\urcorner)\;\ensuremath{\mbox{$\circ$}}^{\mathrm{I}}\;\Varid{wk}^{\mathrm{I}}\;\blacksquare{}\<[E]%
\ColumnHook
\end{hscode}\resethooks
\end{minipage}
\begin{minipage}{0.45\textwidth}
\begin{hscode}\SaveRestoreHook
\column{B}{@{}>{\hspre}l<{\hspost}@{}}%
\column{3}{@{}>{\hspre}l<{\hspost}@{}}%
\column{20}{@{}>{\hspre}l<{\hspost}@{}}%
\column{25}{@{}>{\hspre}l<{\hspost}@{}}%
\column{E}{@{}>{\hspre}l<{\hspost}@{}}%
\>[B]{}\ulcorner\Varid{id}\urcorner^{\prime}\;\{\mskip1.5mu \Gamma\;\mathrel{=}\;\Varid{•}\mskip1.5mu\}\;{}\<[20]%
\>[20]{}\anonymous \;\mathrel{=}\;\Varid{sym}\;\Varid{•}\!-\!\eta^{\mathrm{I}}{}\<[E]%
\\
\>[B]{}\ulcorner\Varid{id}\urcorner^{\prime}\;\{\mskip1.5mu \Gamma\;\mathrel{=}\;\Gamma\;\rhd\;\Conid{A}\mskip1.5mu\}\;{}\<[20]%
\>[20]{}\anonymous \;\mathrel{=}\;{}\<[E]%
\\
\>[B]{}\hsindent{3}{}\<[3]%
\>[3]{}\ulcorner\;\Varid{id}\;^{+}\;\Conid{A}\;\urcorner\Varid{*}\;\Varid{,}^{\mathrm{I}}\;\Varid{zero}^{\mathrm{I}}\;{}\<[25]%
\>[25]{}\equiv\!\langle\;\Varid{cong}\;(\_\Varid{,}^{\mathrm{I}}\;\Varid{zero}^{\mathrm{I}})\;\ulcorner^{+}\urcorner\;\Varid{⟩}\;{}\<[E]%
\\
\>[B]{}\hsindent{3}{}\<[3]%
\>[3]{}\ulcorner\;\Varid{id}\;\urcorner\Varid{*}\;\uparrow^{\mathrm{I}}\;\Conid{A}\;{}\<[25]%
\>[25]{}\equiv\!\langle\;\Varid{cong}\;(\_\uparrow^{\mathrm{I}}\;\Conid{A})\;\ulcorner\Varid{id}\urcorner\;\Varid{⟩}\;{}\<[E]%
\\
\>[B]{}\hsindent{3}{}\<[3]%
\>[3]{}\Varid{id}^{\mathrm{I}}\;\uparrow^{\mathrm{I}}\;\Conid{A}\;{}\<[25]%
\>[25]{}\equiv\!\langle\;\Varid{cong}\;(\_\Varid{,}^{\mathrm{I}}\;\Varid{zero}^{\mathrm{I}})\;\Varid{id}\circ^{\mathrm{I}}\;\Varid{⟩}\;{}\<[E]%
\\
\>[B]{}\hsindent{3}{}\<[3]%
\>[3]{}\Varid{wk}^{\mathrm{I}}\;\Varid{,}^{\mathrm{I}}\;\Varid{zero}^{\mathrm{I}}\;{}\<[25]%
\>[25]{}\equiv\!\langle\;\rhd\!-\!\eta^{\mathrm{I}}\;\Varid{⟩}\;{}\<[E]%
\\
\>[B]{}\hsindent{3}{}\<[3]%
\>[3]{}\Varid{id}^{\mathrm{I}}\;{}\<[25]%
\>[25]{}\blacksquare{}\<[E]%
\ColumnHook
\end{hscode}\resethooks
\end{minipage}

We also prove preservation of substitution composition 
\ensuremath{\ulcorner\ensuremath{\mbox{$\circ$}}\urcorner\;\mathbin{:}\;\ulcorner\;\Varid{xs}\;\ensuremath{\mbox{$\circ$}}\;\Varid{ys}\;\urcorner\Varid{*}\;\equiv\;\ulcorner\;\Varid{xs}\;\urcorner\Varid{*}\;\ensuremath{\mbox{$\circ$}}^{\mathrm{I}}\;\ulcorner\;\Varid{ys}\;\urcorner\Varid{*}} in similar fashion, folding \ensuremath{\ulcorner\Varid{[]}\urcorner}.
The main cases of \ensuremath{\Varid{compl-}\textbf{m}\;\mathbin{:}\;\Conid{Methods}\;\Varid{compl-}\mathbb{M}} can now be proved by just applying 
the preservation lemmas and inductive hypotheses, e.g:

\noindent
\begin{minipage}{0.335\textwidth}
\begin{hscode}\SaveRestoreHook
\column{B}{@{}>{\hspre}l<{\hspost}@{}}%
\column{3}{@{}>{\hspre}l<{\hspost}@{}}%
\column{21}{@{}>{\hspre}l<{\hspost}@{}}%
\column{E}{@{}>{\hspre}l<{\hspost}@{}}%
\>[B]{}\Varid{compl-}\textbf{m}\;\Varid{.}\Varid{id}^{\mathrm{M}}\;\mathrel{=}\;{}\<[E]%
\\
\>[B]{}\hsindent{3}{}\<[3]%
\>[3]{}\ulcorner\;\Varid{tm}\Varid{*}\!\sqsubseteq\;\Varid{v}\!\sqsubseteq\!\Varid{t}\;\Varid{id}\;\urcorner\Varid{*}\;{}\<[21]%
\>[21]{}\equiv\!\langle\;\ulcorner\sqsubseteq\urcorner\Varid{*}\;\Varid{⟩}\;{}\<[E]%
\\
\>[B]{}\hsindent{3}{}\<[3]%
\>[3]{}\ulcorner\;\Varid{id}\;\urcorner\Varid{*}\;{}\<[21]%
\>[21]{}\equiv\!\langle\;\ulcorner\Varid{id}\urcorner\;\Varid{⟩}\;{}\<[E]%
\\
\>[B]{}\hsindent{3}{}\<[3]%
\>[3]{}\Varid{id}^{\mathrm{I}}\;\blacksquare{}\<[E]%
\ColumnHook
\end{hscode}\resethooks
\end{minipage}
\begin{minipage}{0.6\textwidth}
\begin{hscode}\SaveRestoreHook
\column{B}{@{}>{\hspre}l<{\hspost}@{}}%
\column{3}{@{}>{\hspre}l<{\hspost}@{}}%
\column{35}{@{}>{\hspre}l<{\hspost}@{}}%
\column{E}{@{}>{\hspre}l<{\hspost}@{}}%
\>[B]{}\Varid{compl-}\textbf{m}\;\Varid{.}\_\ensuremath{\mbox{$\circ$}}^{\mathrm{M}}\_\;\{\mskip1.5mu \sigma^{\mathrm{I}}\;\mathrel{=}\;\sigma^{\mathrm{I}}\mskip1.5mu\}\;\{\mskip1.5mu \delta^{\mathrm{I}}\;\mathrel{=}\;\delta^{\mathrm{I}}\mskip1.5mu\}\;\sigma^{\mathrm{M}}\;\delta^{\mathrm{M}}\;\mathrel{=}\;{}\<[E]%
\\
\>[B]{}\hsindent{3}{}\<[3]%
\>[3]{}\ulcorner\;\Varid{norm}\Varid{*}\;\sigma^{\mathrm{I}}\;\ensuremath{\mbox{$\circ$}}\;\Varid{norm}\Varid{*}\;\delta^{\mathrm{I}}\;\urcorner\Varid{*}\;{}\<[35]%
\>[35]{}\equiv\!\langle\;\ulcorner\ensuremath{\mbox{$\circ$}}\urcorner\;\Varid{⟩}\;{}\<[E]%
\\
\>[B]{}\hsindent{3}{}\<[3]%
\>[3]{}\ulcorner\;\Varid{norm}\Varid{*}\;\sigma^{\mathrm{I}}\;\urcorner\Varid{*}\;\ensuremath{\mbox{$\circ$}}^{\mathrm{I}}\;\ulcorner\;\Varid{norm}\Varid{*}\;\delta^{\mathrm{I}}\;\urcorner\Varid{*}\;{}\<[35]%
\>[35]{}\equiv\!\langle\;\Varid{cong}_{2}\;\_\ensuremath{\mbox{$\circ$}}^{\mathrm{I}}\_\;\sigma^{\mathrm{M}}\;\delta^{\mathrm{M}}\;\Varid{⟩}\;{}\<[E]%
\\
\>[B]{}\hsindent{3}{}\<[3]%
\>[3]{}\sigma^{\mathrm{I}}\;\ensuremath{\mbox{$\circ$}}^{\mathrm{I}}\;\delta^{\mathrm{I}}\;\blacksquare{}\<[E]%
\ColumnHook
\end{hscode}\resethooks
\end{minipage}

The remaining cases correspond to the CwF laws, which must hold 
for whatever type family we eliminate into in order to retain congruence of 
\ensuremath{\_\equiv\_}. 
In our completeness proof, we are eliminating into equations, and so all of 
these cases are higher identities (demanding we equate different 
proof trees for completeness, instantiated with the LHS/RHS 
terms/substitutions). 

In a univalent type theory, we might try and carefully introduce additional 
coherences to our initial CwF to try and make these identities provable without 
the sledgehammer of set truncation (which prevents eliminating the initial 
CwF into any non-set).

As we are working in vanilla Agda, we'll take a simpler approach, and rely on 
dependent uniqueness of identity proofs (UIP)
\begin{hscode}\SaveRestoreHook
\column{B}{@{}>{\hspre}l<{\hspost}@{}}%
\column{E}{@{}>{\hspre}l<{\hspost}@{}}%
\>[B]{}\Varid{duip}\;\mathbin{:}\;\forall{}\;\{\mskip1.5mu \Varid{p}\;\mathbin{:}\;\Varid{x}\;\equiv\;\Varid{y}\mskip1.5mu\}\;\{\mskip1.5mu \Varid{q}\;\mathbin{:}\;\Varid{z}\;\equiv\;\Varid{w}\mskip1.5mu\}\;\rightarrow\;\Varid{p}\;\equiv\!\![\;\Varid{r}\;]\!\!\equiv\;\Varid{q}{}\<[E]%
\ColumnHook
\end{hscode}\resethooks
which enables, e.g., \ensuremath{\Varid{compl-}\textbf{m}\;\Varid{.}\Varid{id}\circ^{\mathrm{M}}\;\mathrel{=}\;\Varid{duip}}.
Note that proving this form of UIP relies 
on type constructor injectivity, specifically, injectivity of \ensuremath{\_\equiv\_}. 
We could use a weaker version, taking an additional proof of \ensuremath{\Varid{x}\;\equiv\;\Varid{z}}, 
but this would be clunkier to use as Agda has no hope of inferring such a
proof by unification.

Completeness is now just one call to the eliminator away.

\begin{hscode}\SaveRestoreHook
\column{B}{@{}>{\hspre}l<{\hspost}@{}}%
\column{E}{@{}>{\hspre}l<{\hspost}@{}}%
\>[B]{}\Varid{compl}\;\mathbin{:}\;\ulcorner\;\Varid{norm}\;\Varid{t}^{\mathrm{I}}\;\urcorner\;\equiv\;\Varid{t}^{\mathrm{I}}{}\<[E]%
\\
\>[B]{}\Varid{compl}\;\{\mskip1.5mu \Varid{t}^{\mathrm{I}}\;\mathrel{=}\;\Varid{t}^{\mathrm{I}}\mskip1.5mu\}\;\mathrel{=}\;\Varid{elim-cwf}\;\Varid{compl-}\textbf{m}\;\Varid{t}^{\mathrm{I}}{}\<[E]%
\ColumnHook
\end{hscode}\resethooks

\section{Conclusions and further work}
\label{sec:concl-furth-work}

The subject of the paper is a problem which we expect many 
people (including ourselves) 
would have thought trivial.
Theorem provers have made significant progress since the first POPLMark 
challenge \cite{aydemir2005poplmark} (which indeed focused on problems 
relating to binding and substitution), motivating a shifted focus
(onto logical relations proofs) in
newer benchmarks \cite{abel2019poplmark}. 
As it turns out, elegantly mechanising substitution
still requires some care, and we spent quite some time going
down alleys that didn't work (whilst getting to 
grips with the subtleties of Agda's termination checking).


The convenience of our solution relies on Agda's built-in 
support for lexicographic termination \cite{alti:jfp02}.
In contrast, Rocq's \ensuremath{\Conid{Fixpoint}} command merely supports structural recursion on a
single argument and Lean has only raw elimination principles as
primitive. Luckily, both of these proof assistants layer on additional
tactics to support more natural use of non-primitive 
induction, making our approach somewhat 
transferable. Indeed, Lean can be convinced that our substitution 
operations
terminate after specifying measures similar to those in
Section~\ref{sec:termination}, via the \ensuremath{\Varid{termination\char95 by}} tactic.

One reviewer asked about another alternative: since we are merging \ensuremath{\_\ni\_} and
\ensuremath{\_\vdash\_}
why not go further and merge them entirely? Instead of a separate type for
variables, one could have a term corresponding to de Bruijn index zero
(written \ensuremath{\Varid{•}\;\mathbin{:}\;\Gamma\;\rhd\;\Conid{A}\;\vdash\;\Conid{A}} and an explicit weakening operator on
terms (written
\ensuremath{\Varid{\char95 ↑}\;\mathbin{:}\;\Gamma\;\vdash\;\Conid{B}\;\rightarrow\;\Gamma\;\rhd\;\Conid{A}\;\vdash\;\Conid{B}}).
This has the unfortunate property that there is now more than one way to
write terms that used to be identical. For instance, the terms
\ensuremath{\Varid{•}\;\Varid{↑}\;\Varid{↑}\;\cdot\;\Varid{•}\;\Varid{↑}\;\cdot\;\Varid{•}} and \ensuremath{(\Varid{•}\;\Varid{↑}\;\cdot\;\Varid{•})\;\Varid{↑}\;\cdot\;\Varid{•}} are equivalent, where \ensuremath{\Varid{•}\;\Varid{↑}\;\Varid{↑}}
corresponds to the variable with de Bruijn index two. A development
along these lines is explored in \cite{wadler2024explicit}.


We see the current construction as a warmup for the
definition of substitution for dependent type theory
This is harder,
because then the typing of the constructors actually depends on the
substitution laws. Such a M{\"u}nchhausian \cite{altenkirch2023munchhausen} 
construction should be possible in Agda.
However, the theoretical underpinning of
inductive-inductive-recursive definitions is mostly unexplored, with
the exception of \cite{kaposi2023towards}.
There are
potential interesting applications: strictifying substitution laws is
essential to prove coherence of models of type theory in higher types,
in the sense of HoTT.

Hence an apparently trivial
problem isn't so easy after all, and it is a stepping stone to more
exciting open questions. But before you can run, you need to walk and
we believe that the construction here can be useful to others.





\bibliography{./local}

\end{document}